\documentclass[preprintnumbers,floatfix,twocolumn,superscriptaddress,a4paper,nofootinbib]{revtex4}

\usepackage{graphicx}
\usepackage{amsmath,amssymb,amsfonts}
\usepackage{hyperref,enumitem}
\usepackage{bbm}
\usepackage{xspace}
\usepackage{verbatim}
\usepackage{bm}
\usepackage{color}
\usepackage{cancel}
\usepackage{ulem}
\usepackage[T1]{fontenc} 
\usepackage{slashed}
\usepackage{soul}
\usepackage{booktabs}
\usepackage{enumitem}
\usepackage{booktabs}
\usepackage{siunitx}

\usepackage{makecell, tabularx}

\setcellgapes{3pt}

\usepackage[capitalise, english]{cleveref}
 \crefname{section}{Sec.}{Secs.}
  \crefname{table}{Tab.}{Tabs.}

\newcommand{\lam}{\lambda}
\newcommand{\Lam}{\Lambda}

\newcommand{\LamRpV}{\Lambda_{\not R_p}}
\newcommand{\ETmiss}{\not\!\!E_T}
\newcommand{\ETmissx}{\not\!\!\!E_T}

\newcommand{\Delphes}{\texttt{Delphes}\xspace}
\newcommand{\Madgraph}{\texttt{MadGraph}\xspace}
\newcommand{\Pythiaeight}{\texttt{Pythia 8}\xspace}

\newcommand{\Fastjet}{\texttt{FastJet}\xspace}
\newcommand{\Atlas}{\texttt{ATLAS}\xspace}
\newcommand{\Cms}{\texttt{CMS}\xspace}
\newcommand{\Checkmate}{\texttt{CheckMATE}\xspace}

\newcommand{\Sarah}{\texttt{SARAH}}
\newcommand{\Spheno}{\texttt{SPheno}}

\definecolor{weddingpurple}{rgb}{0.8,0, 0.8}

\newcommand{\AN}[1]{\textcolor{blue}}

\newcommand{\LamCMSSM}{$\Lam_{\not R_p}$-CMSSM}
\newcommand{\LamRPV}{\Lam_{\not R_p}}

\newcommand{\AddrHH}{
II. Institut f\"ur Theoretische Physik,
Universität Hamburg,
Luruper Chaussee 149,
22761 Hamburg, Germany
 }
\newcommand{\AddrBonn}{%
Bethe Center for Theoretical Physics \& Physikalisches Institut der 
Universit\"at Bonn, Nu{\ss}allee 12, 
 53115 Bonn, Germany
}
\begin{document}

\title{\huge
 R-Parity Violation at the LHC}
\preprint{BONN-TH-2017-03}

\author{Daniel Dercks} \email{dercks@desy.de}
\affiliation{\AddrBonn}\affiliation{\AddrHH}
\author{Herbi Dreiner} \email{dreiner@uni-bonn.de}
\affiliation{\AddrBonn}
\author{Manuel E. Krauss}\email{mkrauss@th.physik.uni-bonn.de}
\affiliation{\AddrBonn}
\author{Toby Opferkuch}\email{toby@th.physik.uni-bonn.de}
\affiliation{\AddrBonn}
\author{Annika Reinert}\email{areinert@th.physik.uni-bonn.de}
\affiliation{\AddrBonn}


\begin{abstract}
We investigate the phenomenology of the MSSM extended by a single $R$-parity violating coupling at the unification scale. For all 
$R$-parity violating couplings, we discuss the evolution of the particle spectra through the renormalization group equations and the 
nature of the lightest supersymmetric particle (LSP) within the CMSSM, as an example of a specific complete supersymmetric model. 
We use the nature of the LSP to classify the possible signatures. For each possible scenario we present in detail the current 
LHC bounds on the supersymmetric particle masses, typically obtained using simplified models. From this we determine the present 
coverage of $R$-parity violating models at the LHC. We find several gaps, in particular for a stau-LSP, which is easily obtained in 
$R$-parity violating models. Using the program \texttt{CheckMATE} we recast existing LHC 
searches to set limits on the parameters of all $R$-parity violating CMSSMs. We find that virtually all of them
are either more strongly constrained or similarly constrained in comparison to the $R$-parity conserving CMSSM, including 
the $\bar U\bar D\bar D$ models. For each $R$-parity violating CMSSM we then give the explicit lower mass bounds on all 
relevant supersymmetric particles. 
\end{abstract}
\maketitle

\section{Introduction}
Supersymmetry (SUSY) \cite{Golfand:1971iw,Volkov:1973ix,Wess:1973kz,Wess:1974tw} is a unique extension of the 
external symmetries of the Standard Model of particle physics (SM) \cite{Coleman:1967ad,Haag:1974qh}.\footnote{See also 
Ref.~\cite{Ramond:2016talk} on the early history of supersymmetry, 1967 - 1976,  and references therein.} As a solution to 
the hierarchy problem \cite{Gildener:1976ai,Veltman:1980mj}, the supersymmetry-breaking scale should be $\mathcal{O}
(\mathrm{TeV})$ and thus testable at the LHC. To-date no experimental sign of supersymmetry has been found 
\cite{Adam:2016ICHEPtalk}, pushing the lower mass limits for some supersymmetric particles into the TeV range, 
however with some clear model dependence, see for example Ref.~\cite{Drees:2015aeo} on discussing the impact of the 
$\sqrt{s}=8\,$TeV data.

Requiring supersymmetric invariance of the SM and imposing $R$-parity conservation yields the minimal 
supersymmetric SM (MSSM). Its 
superpotential is given by
\begin{eqnarray} 
W_{\mathrm{MSSM}} &=& \epsilon_{ab}\left[(Y_u)_{ij} {Q}_i^a {H}_u^b {\bar U}_j + (Y_d)_{ij} 
{Q}^a_i {H}_d^b {\bar D}_j \right.\nonumber \\
&&\left.+ (Y_e)_{ij} {L}^a_i {H}_d^b {\bar E}_j -  \mu
{H}_d^a {H}_u^b \right]\,,
\label{eq:W-RPC}
\end{eqnarray}
where we have explicitly included the $SU(2)$ indices,  while otherwise using standard notation 
\cite{AbdusSalam:2011fc}, for example $i,j=1,2,3$ are the generation indices. The above superpotential by construction 
conserves the discrete symmetry $R$-parity 
\begin{equation}
\mathbf{R}_{\mathbf{p}}=(-\mathbf{1})^{3B+L+2S},
\end{equation}
where $B$ denotes baryon number, $L$ lepton number and $S$ spin. This requires supersymmetric pair production 
in colliders, and often leads to missing transverse momentum signatures. These signatures arise as the lightest 
supersymmetric particle (LSP) is necessarily stable, guaranteed by conserved $\mathbf{R}_{\mathbf{p}}$, and in most cases electrically neutral thus evading experimental detection. For many 
years the constrained minimal supersymmetric SM (CMSSM) 
\cite{Dimopoulos:1981zb,Nilles:1982ik,Barbieri:1982eh,Chamseddine:1982jx,Kane:1993td} has been a benchmark 
for experimental supersymmetry searches. It is defined by five parameters,
\begin{equation}
M_0,\,M_{1/2},\,A_0,\,\tan\beta,\,\mathrm{sgn}(\mu)\,,
\label{eq:cmssm}
\end{equation}
at the unification scale, $M_X\simeq10^{16}\,$GeV, in comparison to the $\mathcal{O}(100)$ parameters present in 
the generic MSSM. Here $M_0$, $M_{1/2}$ are the universal scalar and gaugino masses, $A_0$ is the universal 
trilinear scalar interaction and $\tan\beta$ is the ratio of the two Higgs vacuum expectation values. $\mathrm{sgn}
(\mu)$ is the sign of the supersymmetric Higgsino mass term $\mu$.

The lack of a supersymmetric signal at the LHC puts increasing pressure on the CMSSM  in particular with respect to 
fine-tuning \cite{Cassel:2011tg,Bechtle:2012zk}. Several groups have performed combined frequentist fits of the CMSSM 
to all the relevant data, see for example \cite{Buchmueller:2013rsa,Bechtle:2013mda}. In Ref.~\cite{Bechtle:2015nua} it 
was shown that the CMSSM is experimentally excluded due to tension between the $(g-2)_\mu$ measurements and the 
LHC lower bounds on $M_0$. Several groups now instead investigate the phenomenological MSSM (pMSSM), which has 
a more extensive parameter set \cite{Djouadi:1998di,Berger:2008cq,Conley:2010du,deVries:2015hva,Bertone:2015tza}. In 
particular the slepton and squark masses at the unification scale are now given by separate parameters, decoupling the 
$(g-2)_\mu$ measurement and the LHC lower mass bound on the squarks \cite{deVries:2015hva}. 

Rather than relaxing the high-scale boundary conditions, we instead consider $R$-parity violating (RPV) supersymmetry 
\cite{Dreiner:1997uz,Barbier:2004ez,Escudero:2008jg}. Restricting ourselves to the minimal set of  fields as in the MSSM, the superpotential is extended 
to include the 48 terms  \cite{Weinberg:1981wj}
\begin{eqnarray} 
W &=& \epsilon_{ab}\left[\frac{1}{2} \lam_{ijk} {L}_i^a {L}_j^b {\bar E}_k + \lam'_{ijk} 
{L}^a_i {Q}_j^b {\bar D}_j -  \kappa_i
{L}_i^a {H}_u^b \right]\nonumber \\
&&+ \frac{1}{2} \epsilon_{xyz}\lam''_{ijk} {\bar U}^x_i {\bar D}_j^y {\bar D}_k^z \,. 
\label{eq:W-RPV}
\end{eqnarray}
Here $x,y$ and $z$ are $SU(3)$ color indices and the $\lam_{ijk},\,\lam'_{ijk},\,\lam''_{ijk}$ are dimensionless Yukawa couplings. As an orientation we present the most strictly bound Yukawas in each class of operators in \cref{tab:bounds}. A 
complete list is given in \cref{rpv-bounds}. The $\kappa_i$ are mass dimension-one mixing parameters. At a fixed energy 
scale they can be rotated away. This also holds for complex $\kappa_i$ and $\lam$'s \cite{Hall:1983id,Dreiner:2003hw}. 
Through the renormalization group equations RGEs they are in general, however, regenerated at other scales. As discussed in 
\cite{Allanach:2003eb}, supergravity models with universal breaking have alignment at the unification scale, and thus 
only radiatively generated $\kappa_i$ at the weak scale. The $\kappa_i$ are then very small and have no impact on the 
LHC phenomenology. We therefore discard them in the remainder of this paper.

\begin{table}[t!]
\renewcommand\arraystretch{1.3}
\begin{tabular}{cccccc}
\toprule
Couplings & $\lam_{ijk}$ & $\lam'_{1jk}$  & $\lam'_{2jk}$ & $\lam'_{3jk}$ & $\lam''_{ijk}$ \\ \midrule
Bound & \;0.49$^a$\; & \;$0.09^a$\; & \;0.59\; & \;1.1\; & \;0.5$^b$ \; \\ \bottomrule
\end{tabular}
\caption{The bounds on the most constrained coupling for each class of RPV operators at $M_W$. The bounds are given for sfermion 
masses $\tilde m$ of 1\,TeV and typically scale as $\tilde m^{-1}$. A complete list with the specific sfermion mass dependence 
is given in \cref{rpv-bounds}. $^a$\,We disregarded the stringent bound on $\lam_{133}$ and $\lam_{133}^\prime$ 
arising from upper bounds on neutrino masses \cite{Godbole:1992fb}. $^b$\,$\lam^{\prime\prime}_{112}$ and $\lam^{\prime
\prime}_{113}$ can be more strongly constrained under the assumption of a large hadronic scale for double nucleon decay 
\cite{Goity:1994dq}.} 
\label{tab:bounds}
\end{table}

$R$-parity was originally introduced to stabilize the proton. However, this is not a unique choice, with many viable alternatives
\cite{Ibanez:1991hv,Banks:1991xj,Dreiner:2005rd,Dreiner:2006xw,Dreiner:2011ft,Dreiner:2012ae,Perez:2013usa}. There are 
also a number of simple models which predict a subset of RPV couplings through other discrete gauge symmetries, see for 
example \cite{Dreiner:2003hw,Dreiner:2003yr,Csaki:2011ge}. In addition phenomenologically RPV supersymmetry models naturally 
accommodate light massive neutrinos, requiring neither 
right-handed neutrinos nor an additional heavy Majorana mass scale \cite{Hall:1983id,Davidson:2000uc,LopezFogliani:2005yw}.  
As a consequence of $R$-parity violation, the LSP can decay and is no longer a good dark matter candidate. However, others 
such as an axion, a sufficiently long-lived axino \cite{Kim:2001sh,Chun:1999cq,Chun:2006ss,Dreiner:2014eda} or even the 
gravitino \cite{Buchmuller:2007ui,Arcadi:2015ffa} can account for the measured dark matter relic density \cite{Ade:2013zuv}. 
 Furthermore, RPV can 
alleviate part of the light Higgs problem in supersymmetry \cite{Dreiner:2014lqa}, as it can lead to weaker lower mass 
bounds on the squark and gluino, see e.g. 
\cite{ATLAS:2013qla,ATLAS-CONF-2016-057,CaminalArmadans:2016hie} and the discussion below. We conclude that
RPV models are just as well motivated as $R$-parity conserving (RPC) models.

Throughout most of this paper we focus on the RPV--CMSSM, which we shall also denote $\Lam_{\not R_p}$--CMSSM, 
as defined in Ref.~\cite{Allanach:2003eb}. A given such model has \textit{one} additional non-zero trilinear RPV coupling 
$\Lam_{\not R_p}$ at the unification scale. We have the following parameters at $M_X$
\begin{eqnarray}
&&M_0,\,M_{1/2},\,A_0,\,\tan\beta,\,\mathrm{sgn}(\mu),\,\Lambda_{\not R_p}\,, \label{eq:rpv-cmssm} \\[2mm]
&& \mathrm{with}\;\Lambda_{\not R_p}\in\{\lam_{ijk},\lam'_{ijk},\lam''_{ijk}\} \,.\nonumber
\end{eqnarray}
Through the RGEs, several non-zero RPV couplings will be generated at the weak scale \cite{Allanach:1999mh}. As 
we see below, this can have an effect on the LHC phenomenology, in particular for $\tilde\tau$-LSPs.

It is the purpose of this paper to investigate the impact of the LHC on the allowed parameter ranges of the $\Lam_{\not R_p}$--CMSSM. 
At the LHC supersymmetric production is dominated by  the squarks and gluinos, but electroweak gaugino production can also have an 
impact. For not too large RPV-couplings, the produced sparticles will cascade to the LSP via standard MSSM operators. As the 
cosmological constraint no longer applies for an unstable LSP, it need not be the lightest neutralino. The LSP then decays in the
detector via $R$-parity violating couplings, usually through the dominant coupling given at the unification scale. It is thus the nature 
of the LSP, as well as its decay which mainly determines the resulting signatures. We summarize the possibilities as follows:
\begin{equation}
\mathrm{sig.}=\left(\begin{array}{c}
\tilde q\tilde q \\[0.8mm]
\tilde q \tilde g\\[0.8mm]
\tilde g\tilde g\\[0.8mm]
\tilde \ell^+ \tilde \ell^- \\[0.8mm]
 \tilde \nu \tilde \nu \\[0.8mm]
 \tilde\chi^0\tilde\chi^\pm
\end{array}\right)_{\mathrm{prod}}
\!\!\!\otimes
\left(\begin{array}{c}
\tilde\chi^0_1\\[0.6mm]
\tilde\chi^\pm_1\\[0.6mm]
\tilde\nu_i\\[0.6mm]
\tilde \ell^\pm_i\\[0.6mm]
\tilde\tau \\[0.6mm]
\tilde q \\[0.6mm]
\tilde b \\[0.6mm]
\tilde t \\[0.6mm]
\tilde g
\end{array}\right)_{\stackrel{\mathrm{possible}}{\mathrm{LSP}}}
\!\!\!\!\!\!\!
\otimes
\left(\begin{array}{c}
 L_1 L_2 {\bar E}_1\\
\ldots\\
 L_1  Q_1 {\bar D}_1\\
\ldots\\
 {\bar U}_3 {\bar D}_2  {\bar D}_3
\end{array}\right)_{\stackrel{\mathrm{LSP}}{\mathrm{decay}}} \!\!\!\!\!\!
\label{eq:rpv-signatures}
\end{equation}
Here $\tilde q$ refers to any squark and the last array represents all forty-five new trilinear RPV operators, as given in 
Eq.~(\ref{eq:W-RPV}). A first systematic analysis of these bewildering possibilities for a neutralino LSP was presented in 
\cite{Dreiner:1991pe}. A more general classification was presented in \cite{Dreiner:2012wm}, allowing for all possible LSPs and also 
all supersymmetric mass orderings. This is presently beyond a systematic comparison with LHC data. Here we instead address the 
case of the $\LamRPV$--CMSSM, Eq.~(\ref{eq:rpv-cmssm}). With the smaller number of parameters this is feasible. In particular in 
this paper we investigate the following points:
\begin{enumerate}
\item[\textbf{(1)}] In \cref{sec:rges}, we analyze the possible LSPs in the $\LamRPV$--CMSSM. This depends on the type and size of the dominant 
RPV coupling, as well as the CMSSM parameters. We employ the supersymmetric RGEs \cite{Allanach:1999mh} and go beyond previous 
work \cite{Dreiner:2008ca}, to take into account the recent Higgs boson discovery.
\item[\textbf{(2)}] In \cref{sec:neutralinoLSP,sec:stauLSP,sec:nonneutralinoLSP}, 
we review in detail the RPV LHC signatures, 
summarize the experimental lower mass bounds and thus determine the current LHC coverage of the $\LamRPV$--MSSM. The
experiments typically set limits on the parameters of simplified models, with no interpretation in the $\LamRPV$--CMSSM. These searches
can however be applied to a wide range of RPV models.
\item[\textbf{(3)}] In \cref{sec:checkmate-tests} we investigate the $\LamRPV$--CMSSM as a complete supersymmetric model. We use the program 
\texttt{CheckMATE} \cite{Drees:2013wra,Kim:2015wza,Dercks:2016npn} to determine the LHC bounds on the various versions of this model 
and compare it to the \texttt{CheckMATE} constraints in the RPC--CMSSM.
\item[\textbf{(4)}] In \cref{sec:checkmateboundstable} we use the results from \cref{sec:checkmate-tests} to determine absolute lower mass bounds
on the supersymmetric particles for a $\tilde\chi^0_1$-LSP and for a $\tilde\tau_1$-LSP scenario in the $\LamRPV$--CMSSM, respectively. We 
compare our bounds to the simplified models experimental bounds in \cref{sec:neutralinoLSP,sec:stauLSP,sec:nonneutralinoLSP}.
\end{enumerate}
In \cref{sec:summary} we summarize and conclude. In \cref{rpv-bounds} we collate the current  weak-scale bounds 
on the $R$-parity violating trilinear couplings. Most bounds are proportional to the mass of a supersymmetric scalar fermion (sfermion). For heavy 
sfermion masses above a TeV many bounds are weak to non-existent.

\section{The Renormalization Group Evolution of the RPV--CMSSM}
\label{sec:rges}
The renormalization group equations  (RGEs) of supersymmetric $R$-parity violation have previously been studied in 
Refs.~\cite{Brahmachari:1994wd,Goity:1994dq,Hempfling:1995wj,Barger:1995qe,deCarlos:1996ecd,Nardi:1996iy}. 
The full two-loop equations are given in Ref.~\cite{Allanach:1999mh}. Through the interface with {\tt SARAH}  
\cite{Staub:2008uz,Staub:2009bi,Staub:2010jh,Staub:2012pb,Staub:2013tta,Staub:2015kfa}, they have been 
implemented in the numerical program {\tt  SPheno} \cite{Porod:2003um,Porod:2011nf}, which we employ here.
\subsection{General Considerations}
As we saw in Eq.~(\ref{eq:rpv-signatures}), the nature of the LSP plays an important role in determining all 
possible LHC signatures. Within the $\Lambda_{\not R_p}$--CMSSM, the LSP is determined dynamically as 
a function of  the input parameters given in Eq.~(\ref{eq:rpv-cmssm}). The universal gaugino masses at 
$M_X$ of the CMSSM  imply that the lightest {\it gaugino} is always the neutralino at the weak scale. Regions 
with the chargino as the LSP are only possible for very small chargino masses and are hence excluded by LEP searches, while the gluino is always heavier. Whether or 
not a given {\it sfermion} could be the 
LSP depends on the corresponding soft SUSY-breaking scalar and gaugino masses at the weak scale. Therefore, apart from the initial choice of 
$M_0$, their RGE evolution from $M_X$ to the TeV scale is crucial. To a good approximation, it is sufficient to consider the one-loop RGEs and 
neglect the contributions from the 1st and 2nd generation Yukawa couplings of the $R$-parity conserving MSSM (RPC-MSSM). Given these 
assumptions the RGEs for the scalar masses squared can be parametrized as \cite{Allanach:2003eb,Dreiner:2008ca}
\begin{equation}
16\pi^2 \frac{d(\tilde m^2_{Y})}{dt} = -a_i g_i |M_i|^2 - b g_1 \mathcal S + \Lambda_{\not R_p}^2 \mathcal F + c T_\Lambda^2\,,
\label{eq:rges_generic}
\end{equation}
where $t=\log (Q/M)$, $Q$ is the renormalization scale and $M$ is the reference scale. $M_i$, $i=1,2,3$, are the soft breaking gaugino masses, 
and $g_i$ the corresponding gauge couplings for $U(1)_Y$, $SU(2)_L$ and $SU(3)_c$, respectively. $\Lambda_{\not R_p}$ denotes the non-zero
RPV coupling and $T_\Lambda$ are the corresponding  RPV trilinear soft supersymmetry breaking sfermion interactions, namely $T_\Lam = \Lam_{\not R_p} A_0$ at $M_X$.
In the $LL\bar E$ case we have: $(T_\lam^k)_{ij}\equiv (T_\lam)_{ijk}\equiv \lam_{ijk} A_0$. The coefficients $a_i,b,$ and $c$ depend on ${Y}=
\{E,L,Q,D,U\}$. $\mathcal F$ is a linear function of the soft SUSY-breaking squared scalar masses and is positive if the latter are all positive, and 
$\mathcal S$ is given by
\begin{align}
\mathcal S &= m_{H_u}^2 + m_{H_d}^2 \notag\\
&\qquad+ {\rm Tr}\left( \tilde m_Q^2-\tilde m_L^2-2 \tilde m_U^2 + \tilde m_D^2 + \tilde m_E^2 \right)\,.
\end{align}

For the sleptons and sneutrinos, the  RGEs for the diagonal $\tilde m^2_Y$ entries read
\begin{align}
16\pi^2 \frac{d(\tilde m^2_E)_{ii}}{dt} &= - \frac{24}{5} g_1^2 |M_1|^2 + \frac{6}{5} g_1^2 \mathcal S  + 2{\rm Tr} (\lam^{i\dagger} 
\lam^i) (\tilde m_E^2)_{ii} \notag \\ &\qquad+ 4 {\rm Tr} (\tilde m_L^2 \lam^{i\dagger} \lam^i) + 2 {\rm Tr} (T_\lam^{i\dagger} 
T_\lam^i)\,,
\label{eq:rges_SER}
\end{align}
and
\begin{align} \label{eq:rges_SEL}
16&\pi^2 \frac{d(\tilde m^2_L)_{ii}}{dt} = - \frac{6}{5} g_1^2 |M_1|^2 -6 g_2^2|M_2|^2 -\frac{3}{5} g_1^2 \mathcal S  \\
&+ \sum_r \big[ 2 (\tilde m^2_L)_{ii} ( \lam^r \lam^{r\dagger})_{ii}  
 + 2 (\lam^r  (\tilde m_L^2)^T \lam^{r \dagger})_{ii} 
 \notag \\
&
+ 2 (\tilde m_E^2)_{rq} (\lam^r \lam^{q\dagger})_{ii} 
+ 6 (\tilde m_L^2  \lam'^r \lam'^{r\dagger})_{ii} \notag \\
&+ 6  (\tilde m_D^2)_{rq} (\lam'^r \lam'^{q\dagger})_{ii} + 6(\lam'^k (\tilde m_Q^2)^T \lam'^{k\dagger})_{ii} \notag \\
&+ 2 ( T_\lam^r T_\lam^{r\dagger})_{ii} + 
6 (T_{\lam'}^rT_{\lam'}^{r\dagger})_{ii} 
 \big]\,. \notag
\end{align}
Here we have used the notation $\lam_{ijk} = (\lam^k)_{ij},\, \lam'_{ijk} = (\lam'^k)_{ij}$ so that, e.g., ${\rm Tr} (\lam^{i\dagger} \lam^i) = \sum_{k,l}\lam^*_{kli} \lam_{kli} $ and $\sum_r ( T_\lam^r T_\lam^{r\dagger})_{ii} \equiv \sum_{k,r} (T_{\lam})_{ikr} (T_{\lam}^*)_{ikr}$. The complete contributions to all soft-masses including all Yukawa couplings can be found in Ref.~\cite{Allanach:2003eb}.

As is known from the RPC case, large gaugino masses contribute with a negative slope to the RGEs, thus raising 
the sfermion masses when running from the high to the low scale. This gaugino mass effect typically dominates 
over the reverse effect due to the soft-mass contribution $\Lam_{\not R_p}^2 \mathcal F$ to the RGEs, even for 
large values of $M_0(M_X)$. Thus small $M_0$ is in general favorable for obtaining light sfermions. The most 
important effect  in the RGEs comes  from the $A$-terms, \textit{i.e.} $T_\Lam$, which always decrease the 
soft-masses at the low scale. As is well-known for the RPC--CMSSM, large $A_0$ values of several TeV are 
needed in order to explain the Higgs mass, if at the same time the stops are at the TeV scale. See, e.g., 
Ref.~\cite{Bechtle:2015nua} for a recent global fit to the RPC--CMSSM. Therefore, in a constrained SUSY model, 
it is natural to have large trilinear SUSY-breaking terms, leading to  sizable effects in the weak-scale sfermion 
masses.

\begin{figure}[t]
\includegraphics[width=\linewidth]{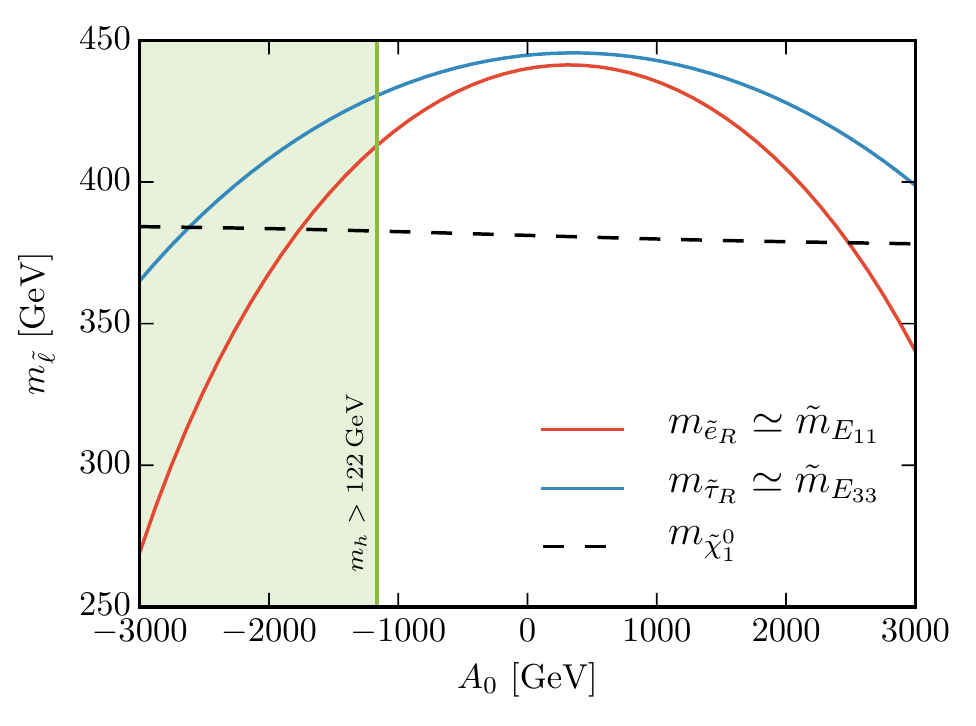}
\caption{Dependence at one-loop of the stau, selectron and lightest neutralino mass on the initial choice of $A_0$ at 
$M_X$ for $\lam_{231}|_{M_X}=0.1$, as well as $M_0=300~$GeV, $M_{1/2}=900~$GeV and $\tan \beta=10$. The 
green shaded area indicates the region where the Higgs mass is sufficiently heavy, $m_h>\SI{122}{\GeV}$, see text.}
\label{fig:selectron_A0}
\end{figure}

Therefore sfermion LSPs are most easily obtained for small $M_0$ and large $|A_0|$. For illustration purposes we 
show in \cref{fig:selectron_A0} the dependence of the right-handed selectron mass on $A_0$ for an RPV coupling 
$\lam_{231}=0.1$ at $M_X$.\footnote{This large value of the RPV coupling is consistent with the low-energy bounds, 
\textit{cf.} \cref{tab:bounds} and \cref{rpv-bounds}.} Here we show as a solid red line the mass of the 
right-handed selectron, as a solid blue line the mass of the right-handed stau and as a black dashed line the mass of 
the lightest neutralino. We see that the lightest neutralino mass is largely unaffected by the initial value of $A_0$.
Due to the strong $T_\lam$ dependence of the soft mass $(\tilde m^2_E)_{11}$, the right-handed selectron becomes 
the LSP for $A_0\lesssim -1.8~$TeV or $A_0\gtrsim 2.4~$TeV in this scenario. The dominantly right-handed stau is 
also strongly affected by the choice of $A_0$ and becomes the next-to-lightest sparticle (NLSP) for even larger $|A_0|$.
The $A_0$ dependence of $m_{\tilde\tau_R}$ is due to the terms in the RGEs proportional to the (RPC) $\tau$ Yukawa, 
which are not included in Eqs.~(\ref{eq:rges_generic})-(\ref{eq:rges_SEL}), but have been used in the full numerical 
evaluation. For $A_0\lesssim -1.1~$TeV, corresponding to the green shaded region, the mass of the SM-like Higgs is 
in the correct range, 122\,GeV$<m_h<128\,$GeV \cite{Aad:2015zhl,Athron:2016fuq}, in accordance with the $\pm3\,$GeV uncertainties
in the numerical programs we employ. Whereas it is too small in the rest of the plot.

We note that large trilinear sfermion interactions tend to make the electroweak vacuum unstable, \textit{i.e.} the scalar potential can develop additional
minima, which for large $|T_i|$ can be deeper than the electroweak minimum \cite{Frere:1983ag,Casas:1995pd}. 
The latter can then tunnel to the 
energetically preferred configuration at possibly unacceptably large rates. However, using the numerical program {\tt Vevacious} 
\cite{Camargo-Molina:2013qva}, we have verified that possible new minima induced by the RPV operators 
$T_\Lambda$ are, for the scenarios we 
consider here, never deeper than the vacua we already find in the RPC--CMSSM. Therefore, the findings of 
Ref.~\cite{Camargo-Molina:2013sta} 
concerning the vacuum stability of the RPC--CMSSM also apply here. All scenarios we present in the following feature an electroweak 
vacuum which 
is either stable or metastable long-lived, meaning that the tunnelling time to the global minimum is longer than the age of the Universe.


\subsection{Determining the LSP}
\label{sec:determ-lsp}

The different possible LSP scenarios in the RPV--CMSSM have been first explored in Ref.~\cite{Dreiner:2008ca}. 
This analysis was centered around comparatively small values of $M_0,M_{1/2}$ and $A_0$. With the recent 
measurement of the Higgs mass, these small values of the input parameters lead to an unacceptably light Higgs 
and are hence excluded. Here we reevaluate the possible LSPs in the RPV--CMSSM, taking into account the new 
Higgs measurements \cite{Aad:2015zhl}. Due to the necessity of quite heavy universal soft-breaking parameters, 
the supersymmetric contributions to the muon anomalous magnetic moment can only alleviate but not solve the 
well-known discrepancy \cite{Davier:2010nc,Stockinger:2007pe,Jegerlehner:2017lbd}. We first consider the case 
of small couplings and then large couplings which can strongly affect the RGEs.

\subsubsection{Small $\Lambda_{\not R_p}$}
\begin{figure}[!t]
\includegraphics[width=0.47 \textwidth]{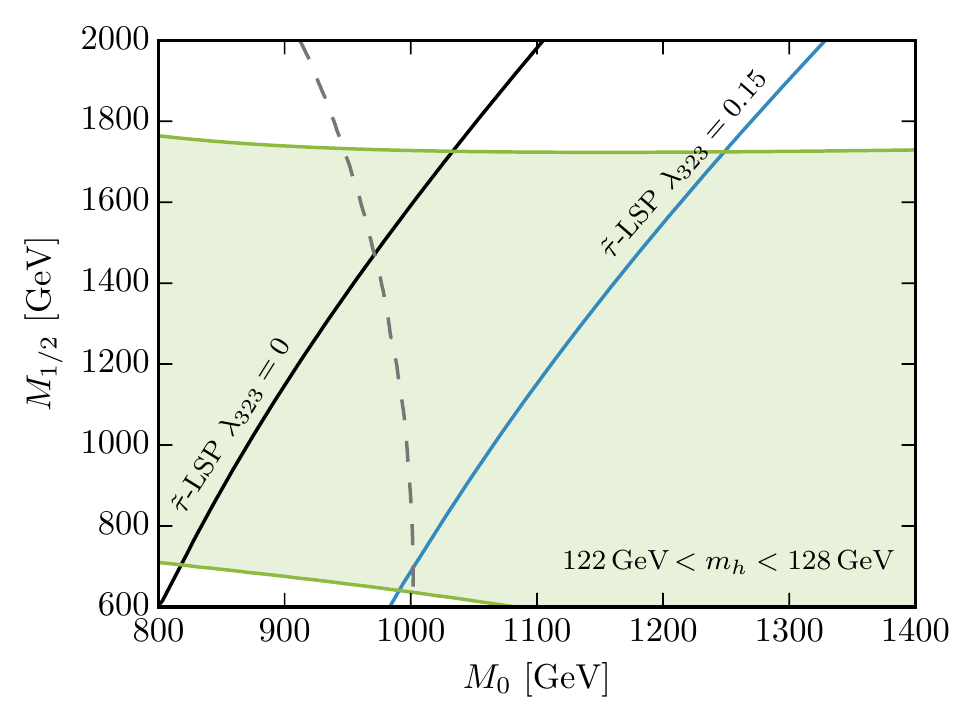}
\caption{LSP Plot for both $\lam_{ijk}=0$ (black) and $\lam_{323}|_{M_X}=0.15$ (blue). The parameter space left 
of the respective contours features a $\tilde \tau$ LSP whereas on the right the $\tilde \chi^0_1$ is the LSP. The 
parameter space on the left of the dashed gray line is excluded in the case of $\lam_{323}|_{M_X}=0.15$ from the bounds on the RPV coupling, see 
\cref{tab:bounds1}. We see how the stau-LSP region expands as we turn on, in this case, the $\lam_{323}$ coupling. 
The green shaded region corresponds to parameter space where the Higgs mass lies in the range $122\,\mathrm
{GeV} < m_h < 128\,$GeV \cite{Aad:2015zhl,Athron:2016fuq}.}
\label{fig:stau-LSP}
\end{figure}

For small values of the RPV couplings, $\LamRpV\lesssim0.05$, the RGE running of the soft supersymmetry 
breaking masses is largely unaffected by the $R$-parity violating interactions. As in the RPC--CMSSM, we thus 
obtain wide ranges of parameter space with a neutralino LSP. However, for small values of $M_0$ and larger 
values of $M_{1/2}$, $A_0$ and $\tan\beta$, the lightest stau, $\tilde\tau_1$, is the LSP. For conserved $R$-parity 
the LSP is stable and these regions of parameter space are excluded on astrophysical and cosmological grounds 
\cite{Gould:1989gw}. However, for RPV these regions are viable as the LSP decays. An example of such a 
parameter region for $A_0=-3\,$TeV and $\tan\beta=30$ is given in \cref{fig:stau-LSP} to the left of the black 
contour. The Higgs mass measurement restricts the larger $M_{1/2}$ regions compared to 
Refs.~\cite{Allanach:2003eb,Dreiner:2008ca}. 

\subsubsection{Large $\lam_{ijk}$} 

For a large $\LamRPV$ the RGEs are significantly modified and the low-energy mass spectrum must be re-evaluated 
\cite{Allanach:1999mh}. Specifically for $\LamRPV=\lam_{ijk}$ and large, the running of the soft-breaking masses  
results in $(\tilde m_E)_{kk} < (\tilde m_L)_{ii},\,(\tilde m_L)_{jj}$ at the weak scale, where $\tilde m_L$ includes both 
$\tilde m_{\tilde\ell_L}$ and $\tilde m_{\nu_L}$.
As already known from the RPC case, the gaugino contribution is larger for the doublet than for the singlet sleptons. 
Furthermore from Eqs.~(\ref{eq:rges_SER}) and (\ref{eq:rges_SEL}) we see that $(\tilde m_E^2)_{kk}$ receives twice the contribution from the $A$-term 
compared to $(\tilde m_{L}^2)_{ii}$, $(\tilde m_{L}^2)_{jj}$. Explicitly, consider $\Lam_{\not R_p} =\lam_{ijk}=-\lam_{jik}\not=0, \lam_{m n \ell}=0$ otherwise. Defining $T_\Lam = 
A\,\Lam_{\not R_p}$, we have $2 \,{\rm Tr}(T^{k\dagger}_\lam T^k_\lam) = 4\,A^2\,\Lam_{\not R_p}^2$, due to the antisymmetry in the indices, while $\sum
_r 2\,(T^r_\lam T^{r\dagger}_\lam)_{ii,jj} = 2\,A^2\,\Lam_{\not R_p}^2$. Therefore, in general, large $\lam_{ijk}$ can lead to a right-handed slepton LSP of 
flavor $k$. Using a 1-step integration a very rough estimate for the soft-breaking mass squared is:
 \begin{equation}
  (\tilde  m_E^2)_{kk} \simeq  (\tilde  m_E^2)^{\rm RPC}_{kk}- 0.76 \, |\lam_{ijk}|^2 (A_0^2+3 M_0^2)\,,
  \label{eq:LLE_rule_of_thumb}
 \end{equation}
where $(\tilde  m_E^2)^{\rm RPC}_{kk} \simeq M_0^2+0.15\, M_{1/2}^2 -\frac{2}{3} X_k \delta_{k3}$ \cite{Ibanez:1983di,Ibanez:1984vq,Drees:1995hj} 
and $X_k$ includes the effects from the third-generation RPC soft-breaking trilinear interactions. Neglecting the $D$-term contributions, for the stau, it reads
  \begin{equation}
  X_3 \simeq \frac{(1+\tan^2\beta)}{10^4} \left( M_0^2 + 0.15 M_{1/2}^2 + 0.33 A_0^2 \right)\,.
  \label{eq:3rd-soft}
  \end{equation}
For various $\lam_{ijk}$ choices, one can obtain the analogue of \cref{fig:selectron_A0} for both a $\tilde \mu$ and a $\tilde \tau$. The latter case is 
qualitatively different because of the  large RPC $\tau$ Yukawa coupling, which is $\tan\beta$-enhanced, \textit{cf.} Eq.~(\ref{eq:3rd-soft}). In 
\cref{fig:stau-LSP} we show the $\tilde \tau$ LSP regions in the $M_0-M_{1/2}$ plane for both $\lam_{ijk}=0$ (solid black curve) and $\lam_{323}=0.15$ 
(solid blue curve) at $M_{X}$. The $\tilde\tau$ region is to the left of the respective curves and is significantly enlarged for a non-zero $\lam_{ij3}$.
The green region represents the allowed Higgs mass and the region to the left of the gray dashed line is excluded by the bounds on the RPV coupling for the case
$\lam_{323}|_{M_X}=0.15$.


\subsubsection{Large $\lam'_{ijk}$}
For large  $\lam'_{ijk}$, the only possible non-neutralino LSP candidate is a slepton or sneutrino of flavor $i$,  as the squarks are always heavier due to the 
large RGE contribution from the gluino. The RPC contribution to the approximately integrated RGEs for the squarks are 
\cite{Ibanez:1983di,Ibanez:1984vq,Drees:1995hj,Martin:1997ns} 
\begin{eqnarray}
(\tilde m_{Q}^2)_{kk} &\simeq& M_0^2+  5.2\, M_{1/2}^2 -\frac{1}{3}(X_b + X_t)\delta_{k 3}\,, \\
(\tilde m_{D}^2)_{kk} &\simeq& M_0^2+  4.8\, M_{1/2}^2 -\frac{2}{3}X_b \delta_{k3}\,,
\end{eqnarray}
whereas for the SU(2) doublet sleptons we have
\begin{equation}
(\tilde m_{L}^2)_{kk} \simeq M_0^2+  0.52\, M_{1/2}^2-\frac{1}{3} X_\tau \delta_{k3}\,. 
\end{equation}
In the latter case the $M_{1/2}$ coefficient is about an order of magnitude smaller. When including the RPV effects, a one-step integration for the slepton 
soft-breaking masses is not sensible. This is because the $\lam'$ coupling  increases by a factor of $\sim 3$ when running from the high to the low scale, 
see, \textit{e.g.},  Ref.~\cite{Dreiner:2008rv}, thus requiring a numerical treatment.\footnote{We found large discrepancies between the full treatment and 
the approximate one-step integration. This is also the case using the approximations in Refs.~\cite{deCarlos:1996ecd,deCarlos:1996yh}.}

As the $D$-term contributions to the sparticle masses slightly suppress $m_{\tilde \nu}$ w.r.t. $m_{\tilde \ell_L}$, only the sneutrino can become the LSP for 
large $\lam'_{ijk}$ and $i=1,2$. For $i=3$ the effect of the left-right-mixing in the stau sector reduces the lightest stau mass below $m_{\tilde \nu_\tau}$. 
The non-neutralino LSP candidates for the large-$\lam'$ scenario are thus $\tilde \nu_e,\,\tilde \nu_\mu$ and $\tilde \tau_1$, where the latter is mainly a 
$\tilde\tau_L$, unlike $\tilde \tau_R$ within RPC models. However, the parameter space for a $\lam'$-induced non-neutralino and non-stau LSP is small 
because of (i) the already comparably large $\tilde  m_L^2$ from the RPC RGEs alone and (ii) the smaller effect of a large $A_0$ when compared with 
the $\lam_{ijk}\neq0$ case. 

\begin{figure}
\includegraphics[width=1.04\linewidth]{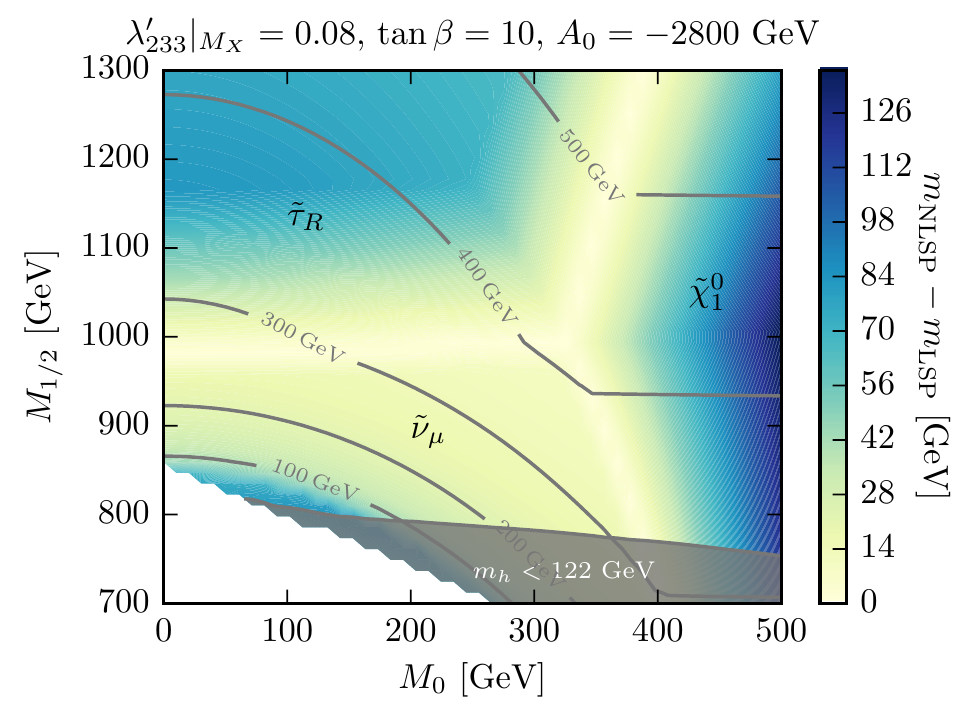}
\caption{Regions in the $M_0$--$M_{1/2}$ plane with different LSPs for $\lam'_{233}|_{M_X}=0.08$. The other parameters 
are $\tan\beta=10$ and $A_0=-2.8~$TeV. The gray contours are LSP iso-mass curves, which are labeled in black for the 
different LSP regions. In most of the $\tilde \nu_\mu$ LSP region, the NLSP is the smuon. The gray shaded region 
corresponds to a too small Higgs mass below \SI{122}{\GeV}. Using 
the scale on the right, the color regions show the mass difference $m_{\mathrm{NLSP}}-m_{\mathrm{LSP}}$ in GeV. 
}
\label{fig:sneutrino-LSP}
\end{figure}

In \cref{fig:sneutrino-LSP}, as an example we show the nature of the LSP in the $M_0$--$M_{1/2}$ plane for the 
case $\lam'_{233}|_{M_X}=0.08$, where $\tan\beta=10$ and $A_0=-2.8\,$TeV.  The gray contours denote the LSP iso-mass 
curves. In the gray shaded region at the bottom the Higgs mass is too small  \cite{Aad:2015zhl,Athron:2016fuq}. 
Here and in the following figures, all shown regions satisfy the bounds on single RPV couplings as given in \cref{rpv-bounds}.
The 
color scale on the right is given in GeV and denotes the mass difference $m_{\mathrm{NLSP}}-m_{\mathrm{LSP}}$. 
The LSP name is given in black and the boundary of the mass cross over is shown in beige. For large $M_0$ 
 the lightest neutralino is the LSP. For $M_0\lesssim350\,$GeV and $M_{1/2}\gtrsim980\,$GeV, $\tilde\tau_1$ is the 
LSP. Only in the small remaining region is a $\tilde\nu_\tau$-LSP obtained. The white region in the lower left corner 
results in tachyons.

\subsubsection{Large $\lam''_{ijk}$}

\begin{figure}
\includegraphics[width=1.04\linewidth]{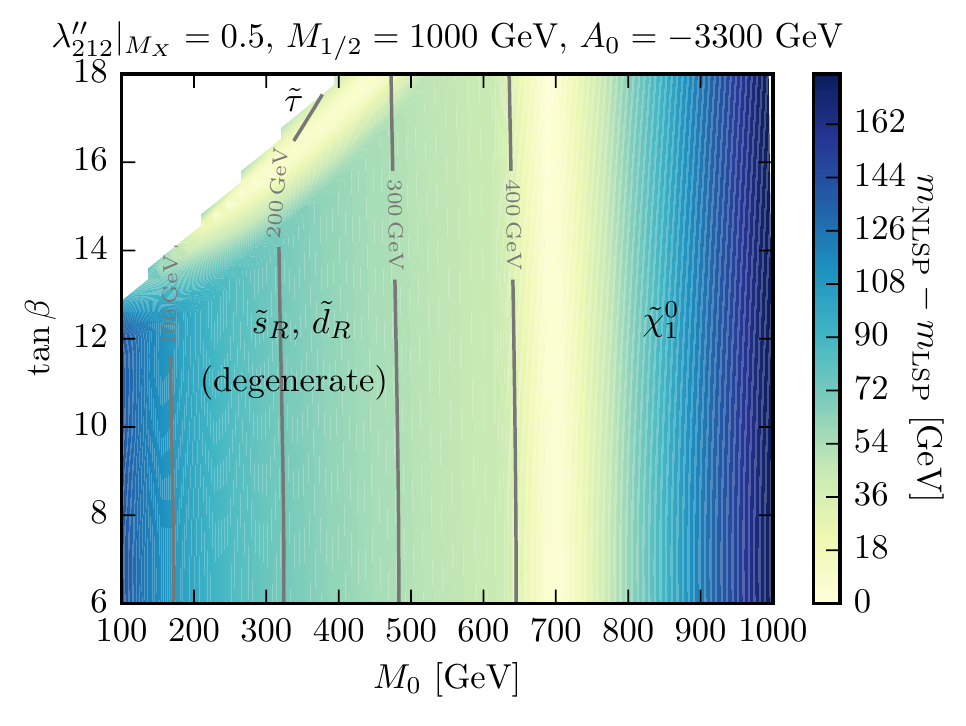}
\caption{Different LSPs in the $M_0$--$\tan\beta$ plane for $\lam_{212}^{\prime\prime}|_{M_X}=0.5$. 
The other parameters are $M_{1/2}
=1~$TeV and $A_0=-3.3~$TeV.  The white region in the upper left corner corresponds to a tachyonic stau and/or squarks. The other labels are 
as in \cref{fig:sneutrino-LSP}.}
\label{fig:lampp212-LSP}
\end{figure}

The constraints on the $\lam''$ couplings, \textit{cf.} \cref{tab:bounds1}, are typically rather loose, leaving more room to have strong RPV effects on the RGE 
squark running. An exception are the strong bounds on $\lam''_{112,113}$, \textit{cf.} \cref{rpv-bounds}. Therefore the only possibility with an operator solely coupled to the $1^{\rm st}$ 
and/or $2^{\rm nd}$ generation squarks is $\lam''_{212}$. In \cref{fig:lampp212-LSP} we show as an example the LSP nature for the case $\lam''_{212}|_
{M_X}=0.5$ in the $M_0$--$\tan\beta$ plane. This is to also show the $\tan\beta$ dependence of the LSP nature for the case of two particle species ($\tilde 
\chi^0_1$ and $\tilde s_R,\tilde d_R$) whose mass is largely independent of $\tan\beta$, and the lightest stau, whose mass depends strongly on $\tan\beta$.
The other parameters are $M_{1/2}=1~$TeV and $A_0=-3.3~$TeV. Here, for large $M_{1/2}$, the $\tilde d_R$ 
and $\tilde s_R$ are almost degenerate and can be the joint LSPs. In the figure we have disregarded the $\tilde{s}_R$--$\tilde{d}_R$ mass-splitting, as it is 
below 1~GeV. Throughout the figure the neutralino and gluino masses do not vary significantly as $M_{1/2}$ is fixed. Their masses are $m_{\tilde{\chi}_1^0} 
\simeq \SI{430}{\GeV}$ and $m_{\tilde g} \simeq \SI{2.2}{\TeV}$. The remaining labels are as in \cref{fig:sneutrino-LSP}. In the center of the figure we have
a large $\tilde{s}_R$/$\tilde{d}_R$-LSP region. To the right, for large values of $M_0$, we again have a $\tilde\chi^0_1$-LSP. In the far upper left corner the 
white area indicates a tachyonic stau and/or squarks. Just below that is a small region with a $\tilde\tau$-LSP. In the $\lam''_{212}$-scenario the charm-squarks, 
in turn, cannot become the LSP because of their slightly heavier soft-breaking masses at the low scale, $\tilde m_{U}^2-\tilde m_{D}^2 \simeq 0.05\,M_{1/2}^2$ 
\cite{Ibanez:1983di,Ibanez:1984vq,Martin:1997ns}. 

For $\lam''_{ij3},\,i\neq 3$ the $\tilde b_R$ couples directly to the leading RPV operator, allowing for a sbottom LSP. We always have $m_{\tilde b_R}<m_{\tilde 
u_{i=1,2}},\, m_{\tilde d_{i=1,2}}$ due to the larger RPC bottom Yukawa coupling.

\begin{figure}
\includegraphics[width=1.04\linewidth]{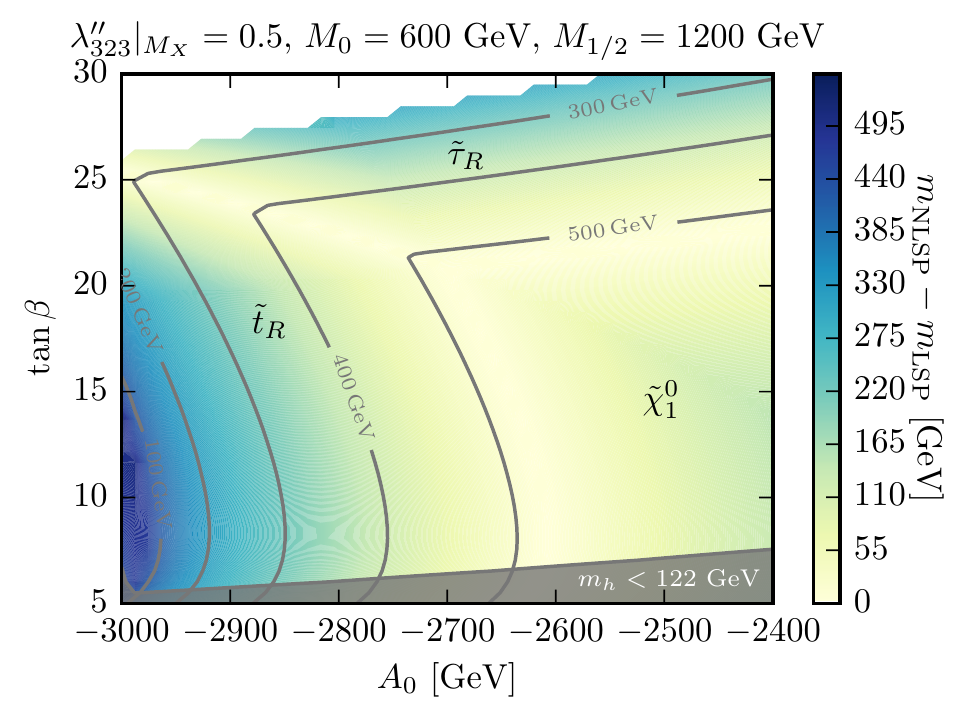}
\caption{Regions in the $A_0$-$\tan\beta$ plane with different LSPs for $\lam''_{323}|_{M_X}=0.5$. The other parameters are $M_0=0.6~$TeV and $M_{1/2}
=1.2~$TeV. In the top left white region the staus become tachyonic. As in \cref{fig:lampp212-LSP}, the neutralino and gluino masses do not vary significantly as $M
_{1/2}$ is fixed. Their masses are $m_{\tilde{\chi}_1^0} \simeq \SI{520}{\GeV}$ and $m_{\tilde g} \simeq \SI{2.6}{\TeV}$. The other labels are as in \cref{fig:sneutrino-LSP}. }
\label{fig:stop-LSP}
\end{figure}

Similarly for $\lam''_{3jk}$, we can only get a stop LSP as a novel scenario, even for $k=3$, as the RPC 
top-Yukawa dominates. To demonstrate this, we show in \cref{fig:stop-LSP} the LSP nature in the 
$A_0$--$\tan\beta$ plane, for the case $\lam''_{323}|_{M_X}=0.5$. The other parameters are fixed as $M_0=600\,
$GeV and $M_{1/2}=1200\,$GeV. This results in a gluino mass fixed around $m_{\tilde g}\simeq2.6\,$TeV 
and a lightest neutralino mass of about $m_{\tilde\chi^0_1}\simeq520\,$GeV. The labelling is otherwise as in 
\cref{fig:sneutrino-LSP}. Note the scaling of the $x$-axis is fairly fine. Besides the usual $\tilde\chi^0
_1$- and $\tilde\tau$-LSP regions we have an extended $\tilde t_R$-LSP region for $A_0\lesssim -2.65\,$TeV. 
No $\tilde b$-LSP region is obtained.

\subsection{Summarizing the LSP Scenarios}
In \cref{tab:LSPs} we summarize all possible LSP scenarios we have found, stating the required (large) RPV 
coupling, as appropriate. This is the first main result of this paper. For small values of $\Lam_{\not R_p}\ll1$ we 
reproduce the results of the RPC--CMSSM, \textit{i.e.} large regions of parameter space give a neutralino 
LSP. However significant regions of parameter space result in a stau LSP, which is predominantly a right-handed stau. This occurs for small values of $M_0$ and 
moderate to large values of $M_{1/2}$.  In \cref{fig:stau-LSP} we show as an example regions of parameter space with a stau LSP. Both the neutralino and the stau 
LSP scenarios are special, in the sense that any $R$-parity violating operator can be dominant. When discussing the phenomenology at the LHC, we thus in principle have 
to consider all 45 different possibilities for a dominant RPV operator. We do this for the neutralino in \cref{sec:neutralinoLSP}, and for the stau in \cref{sec:stauLSP}.
The other LSP cases for large couplings are discussed in \cref{sec:nonneutralinoLSP}.

\begin{table}[t]
\begin{tabular}{c l}
\toprule
\;LSP\; & Required Couplings \\ \midrule
$\tilde\chi^0_1$ & $\LamRPV\ll1$ or large $M_0$ 
\\[2mm]
$\tilde\tau_1$ & $\LamRPV\ll1$, small $M_0$ and large $M_{1/2}$\\[2mm]
$\tilde\tau_1$ & $\lam_{ij3}$ (dominantly $\tilde\tau_R$), $\lam'_{3jk}$ ($\tilde\tau_L$)\\[2mm]
$\tilde e_R$ & $\lam_{ij1} $ \\[2mm]
$\tilde \mu_R$ & $\lam_{ij2}$ \\[2mm]
$\tilde \nu_e$ & $\lam'_{1jk}\,,~\{j,k\} \neq \{1,1\}{}^\ddagger$ \\[2mm]
$\tilde \nu_\mu$ & $\lam'_{2jk}$ \\[2mm]
$\;\tilde s_R,\tilde d_R\;\;$& $\lam''_{212}$ (degenerate LSPs)\\[2mm]
$\tilde b_1$& $\lam''_{123},\,\lam''_{213},\,\lam''_{223}{}^\ddagger$ (dominantly $\tilde b_R$)\\[2mm]
$\tilde t_1$& $\lam''_{3jk}$ (dominantly $\tilde t_R$) \\ \bottomrule
\end{tabular}
\caption{Summary of the various LSP scenarios in the $\LamRPV$--CMSSM as a function of the dominant necessary RPV coupling 
at $M_X$.${}^\ddagger$ Note that the couplings on the right column are \emph{required} but their presence is not necessarily \emph{sufficient} to get the corresponding LSP. The couplings $\lam'_{111},\,\lam''_{112}$ and $\lam''_{113}$ are too constrained to produce 
an LSP of that kind, see \cref{tab:bounds1}.}
\label{tab:LSPs}
\end{table}


\subsection{RGE-Induced Operators}
\label{sec:RGEinduced}

\begin{figure*}
\centering
\includegraphics{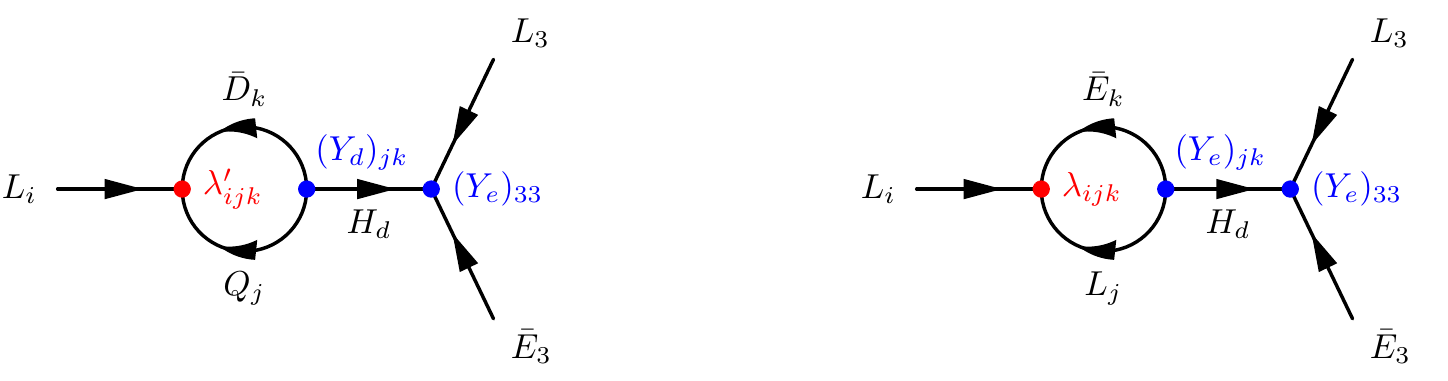}
\caption{Superfield Feynman diagrams corresponding to the one-loop RGE-induced $\lam_{i33}$ operators. The diagrams assuming non-zero $\lam_{ijk}^\prime$ and $\lam_{ijk}$ are shown on 
the left- and right-hand side respectively.}
\label{fig:rge-induced-couplings}
\end{figure*}
\begin{figure}[htbp]
\includegraphics[width=\linewidth]{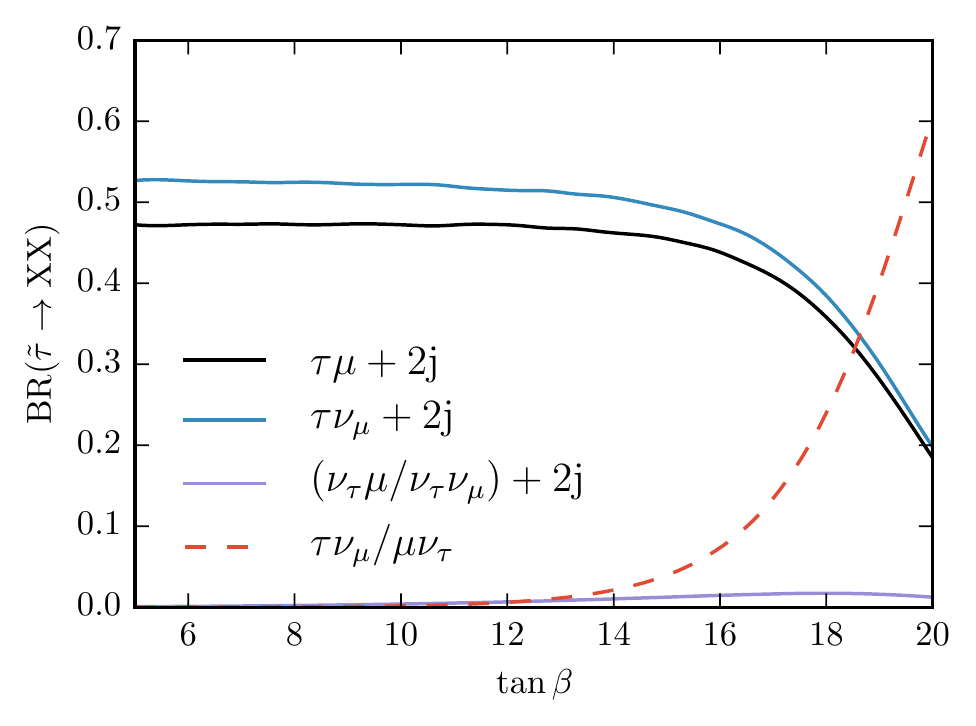}
\caption{Branching ratios of the decaying $\tilde \tau$ LSP as a function of $\tan\beta$ using $\lam'_{211}|_{\rm GUT}=
0.07$ as well as $M_0=0.2~$TeV, $M_{1/2}=1~$TeV and $A_0=-1.75~$TeV. The solid black and blue curves indicate the
four-body decay via the neutralino. In solid lavender we show the branching ratio for the four-body decay  via the chargino. 
The dashed red curve denotes the branching ratio for the two-body decay via the RGE-induced coupling
$\lam_{233}$.}
\label{fig:staudecayTB}
\end{figure}
\begin{figure*}
\includegraphics[width=\linewidth]{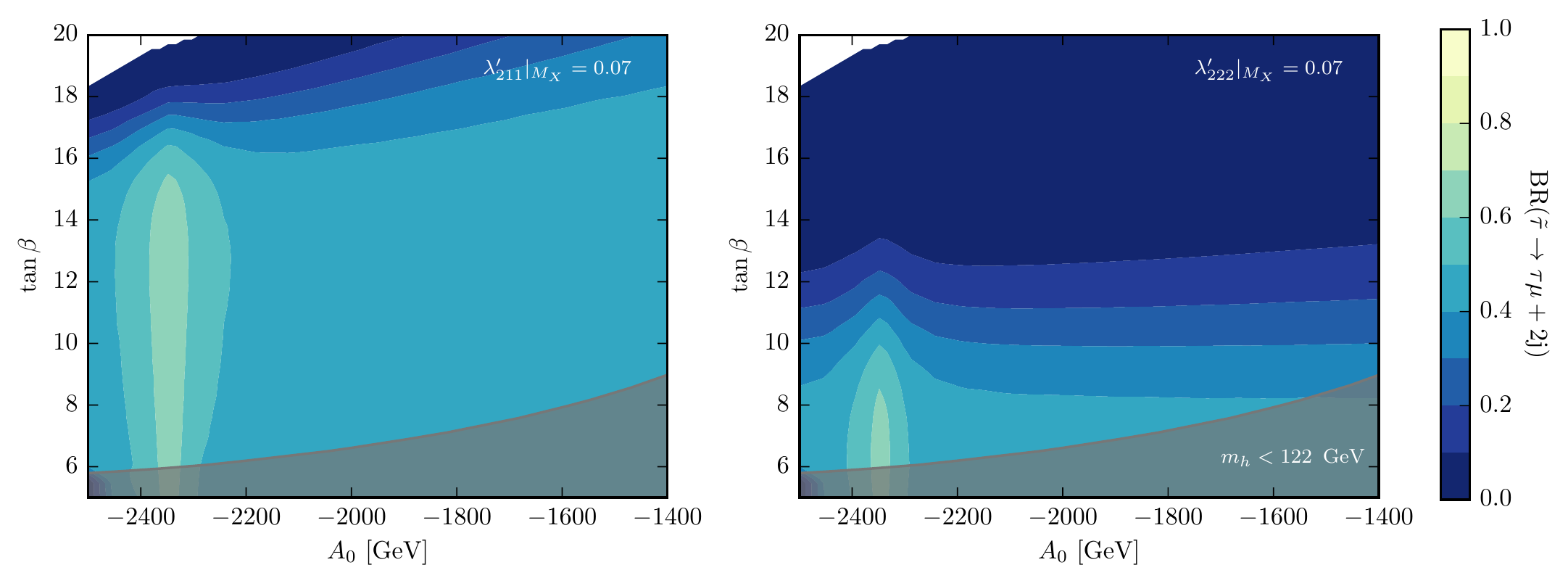}
\caption{
Branching ratios of the stau four-body decay $\tilde \tau \to \tau \mu j j$ as a function of $A_0$ and $\tan\beta$ 
using $\lam'_{211}|_{M_X}=0.07$ (left) and $\lam'_{222}|_{M_X}=0.07$ (right). The other parameter values are 
$M_0=0.2~$TeV and $M_{1/2}=1~$TeV. The gray parameter space is excluded due to a too light Higgs mass, 
whereas the white upper-left corner features a tachyonic stau.}
\label{fig:staudecayA0TB}
\end{figure*}
Once lepton (baryon) number is violated by a non-zero  $\Lambda_{\not R_p}|_{M_X}$,  other lepton 
(baryon) number-violating operators $\LamRPV^{\rm ind} \not=0$ are induced at $M_W$ via RGE effects. 
This occurs via diagrams of the type shown in \cref{fig:rge-induced-couplings}. The explicit RGEs for 
$\lam$ and $\lam'$ are, for instance, given in Ref.~\cite{Dreiner:2008rv}. These RGE-generated operators 
are typically phenomenologically irrelevant as they are loop suppressed and $\LamRPV^{\rm ind} \ll \Lam
_{\not R_p}$. However, they become relevant once new decay channels open which would otherwise be absent
\cite{Allanach:2003eb,Allanach:2006st,Desch:2010gi}. 

Consider a stau LSP and a single non-zero RPV coupling $\lam'_{ijk}$ with $i\neq 3$. Then at tree-level the $\tilde \tau_1$ 
cannot decay directly to an $R$-parity even two-body final state but rather decays via the chain
\begin{align}
\tilde \tau_1 \to \left\{
\begin{array}{lcl}
\tau+\tilde \chi^{0\,(*)}&\to&\tau +(\ell_iu_j d_k,\,\nu_id_j d_k)\,,\\[2mm]
\nu_\tau +\tilde \chi^{\pm\,(*)}&\to&\nu_\tau+(\ell_id_j  d_k,\,\nu_iu_j d_k)\,,
\end{array}
\right.
\label{eq:4-body-stau}
\end{align}
to a four-body final state. Here we have neglected charge-conjugations and assumed the chargino to be dominantly wino. $X^{(*)}$ indicates 
that the respective particle $X$ need not necessarily be on--shell. There is also a similar scenario for $\lam_
{ijk} $ with $\{i,j,k\}\neq 3$.

However, through RGE running, in both cases the couplings $\lam^{(\prime)}_{ijk}|_{M_X}$ generate a non-zero 
$\lam_{i33}|_{M_W}$, enabling the two-body decays $\tilde \tau \to \ell_i \nu_\tau/\tau \nu_i$. The RGEs for this 
case read 
\begin{align}\label{eq:taurge}
16 \pi^2 \frac{d}{dt} \lam_{i33} &= \lam_{i33} \Big[ -\frac{9}{5} g_1^2 - 3 g_2^2 + 4 (Y_e)^2_{33} \Big] \\ \notag
&+ 3\lam'_{ijk} (Y_e)_{33} (Y_d)_{jk} + \lam_{ijk}  (Y_e)_{33}  (Y_{e})_{jk}\,.
\end{align}
The RGE-induced RPV operators scale as a function of the down-type Yukawas, which are themselves a function
of $\tan\beta$. This means that the size of $\tan\beta$ has a strong effect on the relative magnitude of the
initial and the induced coupling and thus on the branching ratio of a 
stau LSP into  four- and  two-body final states. This is depicted in \cref{fig:staudecayTB} where we show various
stau LSP decay branching ratios as a function of $\tan\beta$, assuming a non-zero $\Lam_{\not R_p} = \lam'_
{211}=0.07$ at $M_X$. In solid black and blue we show the two four-body decays via the neutralino, corresponding to
the top line in Eq.~(\ref{eq:4-body-stau}), respectively. Note that the two actually differ, unlike the assumptions in
many experimental analyses, \textit{cf.} Sec.~\ref{sec:neutralinoLSP}. The solid lavender curve shows the negligible
decay via the chargino, corresponding to the second line in Eq.~(\ref{eq:4-body-stau}). The dashed red curve shows 
the branching ratio for the RGE-generated two-body decay via the operator $\lam_{233}$. This becomes significant for
$\tan\beta>16$ and dominant for $\tan\beta\gtrsim18$. 

The four-body decay branching ratio $\tilde\tau\to\tau\mu+2\,$jets for $\lam'_{211}|_{M_X}=0.07$ is shown in the left plot 
of \cref{fig:staudecayA0TB} as a function of $\tan\beta$ and $A_0$. For $A_0\simeq -2300\,$GeV, at the finger 
shaped region, there is a small resonance in the partial width $\tilde{\tau}\to\tau\mu+2$j. This occurs due to a 
level-crossing in the mixing between the left- and right-handed gauge eigenstates of the staus as $|A_0|$ increases. 
Subsequently, the largest branching ratio occurs where the stau left-right mixing is maximal as the RPV operator $L\bar 
Q \bar D$ involves only left-handed sleptons. One should note that this level crossing only appears as a result of the large 
RPV coupling. For smaller RPV couplings, the right smuon is always the lighter one. However, for the relatively large RPV 
value used in \cref{fig:staudecayA0TB}, the RGE effects of this coupling in conjunction with $A_0$ drive the 
left-handed slepton soft-mass towards smaller values, leading to a level-crossing at a particular value of $A_0$. Once 
again, the gray parameter region is excluded as the Higgs mass is too small.

If we consider the analogous scenario for $\lam'_{222}|_{M_X}=0.07$ instead of $\lam'_{211}$, shown in the left plot of \cref{fig:staudecayA0TB},  the partial widths of 
the four-body decays do not change. On the other hand the RGE-induced coupling $\lam_{233}$ is larger because of the 
significantly larger strange-quark Yukawa coupling $(Y_d)_{22}\gg(Y_d)_{11}$. Therefore the corresponding two-body 
partial width is much larger. 
%
%
%
%
For a more 
detailed discussion of the effect of the RGE-induced operators and the impact of four-body decays, we refer to 
Ref.~\cite{Dreiner:2008rv}.

\section{LHC Coverage of RPV-induced Neutralino LSP Decay Scenarios}
\label{sec:neutralinoLSP}
As we saw in the previous section, throughout wide ranges of the RPV--CMSSM parameter space, and in particular 
for $\Lam_{\not R_p}\ll1$, the LSP is given by the lightest neutralino. We first discuss in this section the decay 
lifetime of the neutralino, to see what ranges of parameter space we probe when restricting ourselves to prompt 
decays. We then discuss in detail the LHC final state signatures, depending on the dominant RPV operator at $M_
X$, and their coverage by existing LHC searches.

\subsection{Neutralino Lifetime}
For a given dominant RPV operator, if kinematically allowed, the neutralino LSP will decay via a three-body mode 
to the $R$-parity even particles of the operator, \textit{e.g.}
\begin{eqnarray}
 L_i L_j {\bar E}_k: \quad&\tilde\chi^0_1&\to \{\ell_i^-\nu_j\ell^+_k,\,\nu_i\ell_j^-\ell^+_k\}+c.c. 
\label{eq:neutralino-LLE-decay}
\end{eqnarray}
Note that in $R$-parity violating models the neutralino LSP can in principle be very light or even massless 
\cite{Choudhury:1999tn,Dreiner:2003wh,Dreiner:2009ic,Dreiner:2011fp}. A very light neutralino would decay for example as $\tilde\chi_1^0\to 
\gamma\nu_{i,j}$ via $ L_i L_j{\bar E}_i$ or $ L_i Q_j{\bar D}_j$ operators. However, neutralino masses below 
$\sim50\,$GeV \cite{Olive:2016xmw} only arise with non-universal gaugino masses at the unification scale, which 
is outside of the RPV--CMSSM which we investigate here. Thus we shall only consider the three-body neutralino 
decay modes for the $\tilde\chi^0_1$-LSP.

Here we are interested in the effects of the $R$-parity violating neutralino decay on LHC physics. In this paper we 
restrict ourselves to neutralinos decaying promptly in the detector, \textit{i.e.} 
\begin{equation}
c\tau_{\tilde\chi^0_1}\lesssim 10^{-4}\,\mathrm{m}\,. 
\end{equation}
We shall consider the long-lived case with detached vertices
\begin{equation}
10^{-4}\,\mathrm{m}<    
c\tau_{\tilde\chi^0_1}<5\,\mathrm{m}\,,
\end{equation}
elsewhere. 

As an example, for a pure photino the partial neutralino decay width via  $ L_1 Q_2{\bar D}_1$ is given by \cite{Dawson:1985vr}
\begin{equation}
\Gamma(\tilde\gamma\to\nu_e s \bar d)=\frac{3\alpha e_{\tilde d}^2 \lam'^2_{121}}{128\pi^2} \frac{M_{\tilde\chi^0_1}^5}
{M_{\tilde f}^4} \,, \quad M_{\tilde\chi^0_1}\ll M_{\tilde f}\,, 
\label{eq:lifetime}
\end{equation}
assuming the final state fermion masses are negligible. $\tilde f$ represents the virtual squarks/sleptons in the 
propagator of the decay, all assumed to be degenerate. The neutralino lifetime is thus inversely proportional to 
the $R$-parity violating coupling squared and depends sensitively on the neutralino and the sfermion masses. 
In the general case it depends on the neutralino admixture \cite{Dreiner:1994tj}. 

\begin{figure*}
\includegraphics[width=\linewidth]{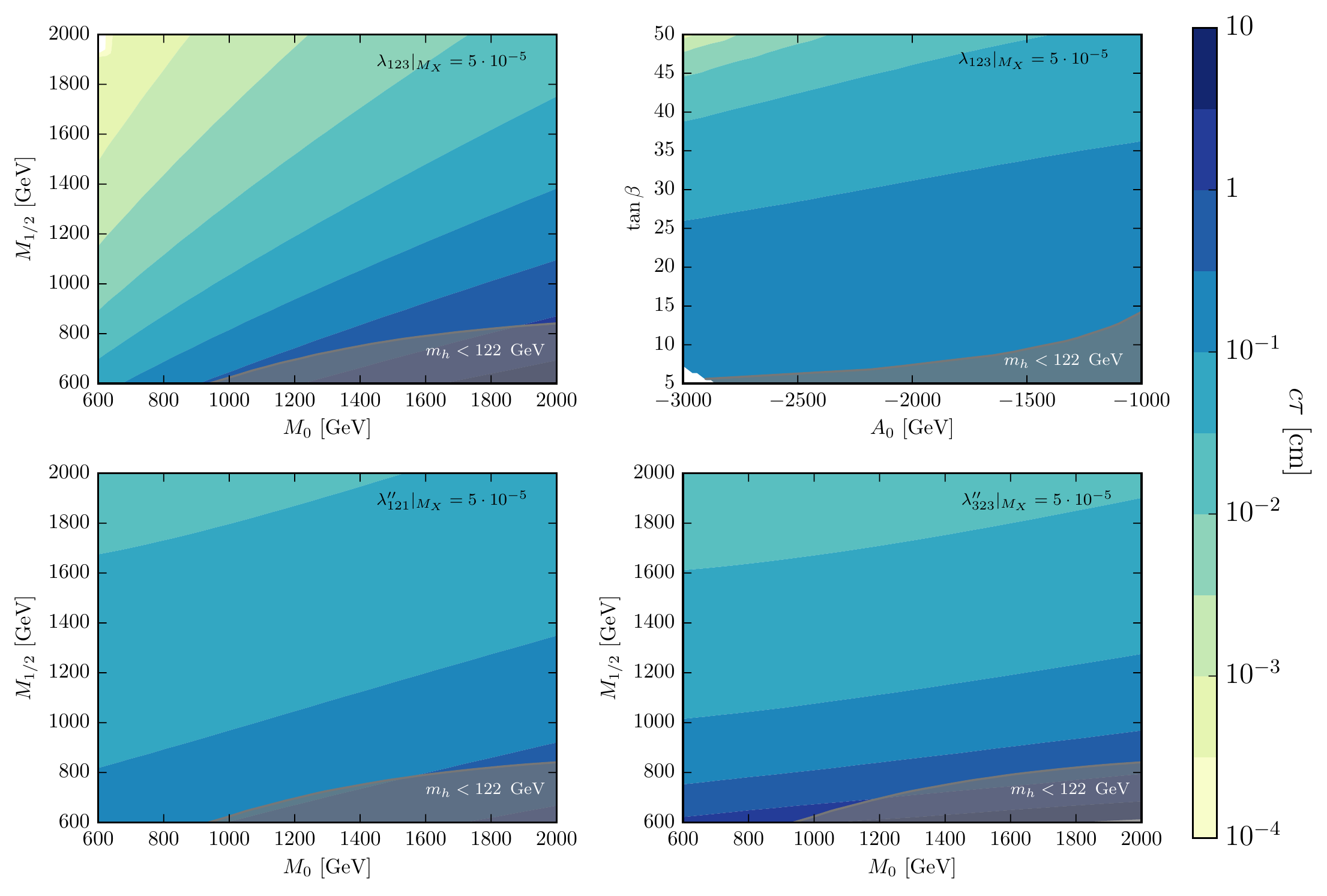}
\caption{Decay length $c\tau$ of the lightest neutralino for $\Lam_{\not R_p}=5\cdot 10^{-5}$ at the GUT-scale 
as a function of $\{M_0,M_{1/2}\}$ using $\tan\beta=10$ and $A_0=-2~$TeV (left upper panel) or $\{A_0,\tan\beta\}$ 
using $M_0=1.5~$TeV and $M_{1/2}=1~$TeV (right upper panel). For both upper panels:  $\Lam_{\not R_p}=\lam_
{123}=-\lam_{213}$. The parameter space in the gray shaded region is excluded due to a too small Higgs mass. In 
the upper left corner of the figure, the $\tilde \tau$ is the LSP, so that the two-body decay $\tilde \chi^0_1\to \tau \tilde 
\tau_1$ is possible, therefore drastically increasing the neutralino width. For the  lower figures, we have used $\Lam_
{\not R_p}=\lam''_{112}=-\lam''_{121}$ (left) and $\Lam_{\not R_p}=\lam''_{323}=-\lam''_{332}$ (right).}
\label{fig:ctau_LLE}
\end{figure*}

In Fig.~\ref{fig:ctau_LLE} we show in the $\Lam_{\not R_p}$--CMSSM, for fixed $\Lam_{\not R_p}=5\cdot 10^
{-5}$ at the unification scale, how the neutralino decay length $c\tau$ (color scale on the right) depends on 
$M_0$ and $M_{1/2}$ for the cases $\Lam_{\not R_p} = \lam_{123}$ (upper-left panel), $\lam''_{112}$ (lower-left 
panel) and $\lam''_{323}$ (lower-right panel), respectively. We also show the $A_0$, $\tan\beta$ dependence 
of the decay length for $\lam_{123}$ in the upper-right panel. In the upper-left panel, we see that we get decay 
lengths ranging from 10\,cm, which is readily observable as a detached vertex, down to $1~\mu$m.  

Besides being proportional to the $R$-parity violating coupling squared, as we see in Eq.~(\ref{eq:lifetime}), the decay 
width scales as the fifth power of the neutralino mass, which is strongly connected to $M_{1/2}$, and the fourth 
inverse power of the scalar fermion propagator mass, which is not only strongly correlated with $M_0$, but also 
depends on $M_{1/2}$ via the RGEs. When comparing the top left with the lower panels in Fig.~\ref{fig:ctau_LLE}, 
we see that the lifetime can be quite different for equally sized $\Lam_{\not R_p}$ at the unification scale.  $\lam''$ is 
almost a factor three larger than $\lam$ at the low-scale, due to the RGEs. However for $\lam$ the slepton masses 
in the propagator are much lighter than the squark masses for $\lam''$.  The propagator effect dominates, as $\Gamma
\propto (\lam^{(\prime\prime)})^2M^{-4}_{\tilde f}$.  Furthermore, when comparing the $M_{1/2}$ dependence for $\Lam
_{\not R_p}=\lam$ versus $\Lam_{\not R_p}=\lam''$, there is a much slower decrease of $c\tau$ with increasing $M_
{1/2}$ for the $\lam''$ scenarios, since the squark masses also increase with $M_{1/2}$. In the lower panels we also see
a marked difference in the decay lengths for the two couplings, $\lam''_{323}$ and $\lam''_{122}$. For $\lam''_{323}$ the 
final state top quark leads to a phase space suppression if $M_{1/2}$ and therefore $m_{\tilde \chi^0_1}$ is small. For 
larger $M_{1/2}$ as well as large $M_{0}$ (corresponding to large $|A_0|$ in the setup at hand), a 
lighter virtual top squark, in comparison to the first-/second-generation squarks, is more important. This lighter top squark 
leads to smaller decay lengths for $\lambda''_{323}$ compared to $\lambda''_{121}$ in these regions.

When $m_{\tilde\chi_1^0}<m_t+m_b$ (not shown in the figure), the three body decay $\tilde\chi^0_1\to tbs$ is kinematically 
forbidden. We found that the corresponding four-body decay $\tilde\chi^0_1 \to t^{(*)} bs \to (b W^+) b s$ can be prompt for $
\lam_{323}''\sim \mathcal O(0.1)$ and $m_t+m_b-m_{\tilde\chi_1^0} \lesssim10~$GeV. In addition, CKM-effects in the RGE 
evolution of the $\lam''$ coupling lead to a nonzero $\lam^{\prime\prime}_{223}$ at the scale of the decaying particle and 
therefore open the alternate three-body decay $\tilde \chi^0_1 \to cbs$. However, we found the 4-body decay remains the 
dominant channel.\footnote{The numerical details depend on assumptions where the CKM mixing takes place 
\cite{Agashe:1995qm,Allanach:1999ic}.} The RGE-generated coupling is loop suppressed and thus if only the decay $\tilde
\chi^0_1\to cbs$ is kinematically accessible, the lightest neutralino is stable on detector scales. Note, however, that a 
neutralino as light as a top quark or lighter  requires $M_{1/2} \lesssim 410~$GeV, resulting in $m_{\tilde g} \lesssim 1~$TeV, 
and can therefore be safely regarded as excluded, as we see below.


\subsection{Neutralino LSP Decay via an $LL\bar E$ Operator}
\label{sec:neutralino-LSP-LLE}
In the case of $LL\bar E$ operators, the neutralino decays to two charged leptons and a neutrino, \textit{cf.} 
Eq.~(\ref{eq:neutralino-LLE-decay}).  At the LHC with the pair or associated production of squarks and/or gluinos, 
we expect cascade decays to two neutralinos. Such a process would therefore contain at least four charged leptons 
and some jets in the final state. The possible flavor and charged combinations depend on the dominant 
operator.\footnote{As discussed above, for example an operator $L_1L_2\bar E_1$ can generate at 1-loop $L_2Q_j
\bar D_k$ operators. The corresponding neutralino decay branching ratios are however suppressed, as there is no 
kinematic or other suppression of the leading operators.} We have summarized the charged lepton part of the leading 
order final state signatures in Table~\ref{tab:lle-signatures}. These are always accompanied by (at least) two neutrinos, 
resulting in additional missing transverse momentum in the signature.

\begin{table}
\begin{tabular}{ccc}
\toprule
Scenario\; & \; Charged Lepton Signatures\; & \; RPV Operators \\ \midrule 
Ia & $e^+ e^- e^+ e^-$ &$\lam_{121,131}$\\[1.5mm]
Ib & $\mu^+ \mu^- \mu^+ \mu^-$ &$\lam_{122,232}$\\[1.5mm]
Ic & $\tau^+ \tau^- \tau^+ \tau^-$ &$\lam_{133,233}$\\[1.5mm]
Id&$e^+ e^-e^\pm\mu^\mp$ &$\lam_{121}$\\[1.5mm]
Ie&$e^+ e^-e^\pm\tau^\mp$ &$\lam_{131}$\\[1.5mm]
If&$\mu^+ \mu^-\mu^\pm e^\mp$ &$\lam_{122}$\\[1.5mm]
Ig&$\mu^+ \mu^-\mu^\pm\tau^\mp$ &$\lam_{232}$\\[1.5mm]
Ih&$\tau^+ \tau^-\tau^\pm e^\mp$ &$\lam_{133}$\\[1.5mm]
Ii&$\tau^+ \tau^-\tau^\pm\mu^\mp$ &$\lam_{233}$\\[1.5mm]
Ij&$e^+\mu^-e^\pm\mu^\mp$ &$\lam_{121,231,122,132}$\\[1.5mm]
Ik&$e^+\tau^-e^\pm\tau^\mp$ &$\lam_{131,231,123,133}$\\[1.5mm]
I$\ell$&$\mu^+\tau^-\mu^\pm\tau^\mp$ &$\lam_{132,232,123,233}$\\[1.5mm]
Im &$e^-\tau^+\mu^\pm\tau^\mp$ &$\lam_{123}$\\[1.5mm]
In &$e^-\mu^+\tau^\pm\mu^\mp$ &$\lam_{132}$\\[1.5mm]
Io &$e^-\mu^+e^\pm\tau^\mp$ &$\lam_{231}$ \\ \bottomrule
\end{tabular}
\caption{Possible charged lepton final states in the $LL\bar E$ case for a pair of LSP neutralinos resulting 
from the cascade decays of pair/associated produced SUSY particles. In each case, if distinct, the charged 
conjugate final state is also possible. The various charge combinations have equal branching ratios, due to 
the Majorana nature of the $\tilde\chi^0_1$-LSP. All final states are accompanied by (at least) two neutrinos 
typically leading to some missing transverse momentum.}
\label{tab:lle-signatures}
\end{table}

As can be seen, for each dominant RPV operator, we can get SFOS (same flavor, opposite sign) lepton pairs. In all 
cases we can also get SFSS (same flavor, same sign) lepton pairs. This includes the somewhat exotic signatures 
(Im) $\tau^-\tau^-e^+\mu^+$, (In) $\mu^-\mu^-e^+\tau^+$, and (Io) $e^-e^-\mu^+\tau^+$. For each dominant operator
one should check which final state leads to the optimal experimental sensitivity. In the case of a discovery, we see that 
each operator has two alternate channels, with definitive branching ratios, which should give a good experimental cross 
check.

Both \Atlas \cite{ATLAS:2012kr,Aad:2014iza,Aad:2014hja} and \Cms \cite{CMS:2013qda,Khachatryan:2016iqn,CMS-PAS-SUS-16-022} have searched for supersymmetry with RPV in four lepton events. Typically they have investigated simplified 
models where the supersymmetric particles are pair-produced and then directly decay to the neutralino LSP. The latter 
decays to a 3-body final state via the dominant RPV operator. We thus have the simplest cascades for the various 
produced supersymmetric particles
\begin{eqnarray}
\left.
\begin{array}{rcl}
\tilde\chi^\pm_1&\to& W^\pm\tilde\chi^0_1 \\[2mm]
\tilde\ell^\pm&\to&\ell^\pm\tilde\chi^0_1\\[2mm]
\tilde\nu&\to&\nu\tilde\chi^0_1\\[2mm]
\tilde q&\to&q\tilde\chi^0_1\\[2mm]
\tilde g&\to&q\bar q\tilde\chi^0_1\\[2mm]
\end{array}\right\}\;\;\mathrm{and}\;\;
\tilde\chi^0_1\to\ell^\pm\ell^\mp\nu\;.
\label{eq:cascade-lle}
\end{eqnarray}
The best resulting bounds with the appropriate reference are given in \cref{tab:best-bounds-lle} together with 
the couplings probed by the given experiments. We see that for each type of supersymmetric particle only a small 
subset of couplings has explicitly been probed. We note however, that the analysis considering the chargino limit, 
which looks for $N_{e,\mu}\geq4$ (here $N_{e,\mu}$ refers to the number of first two generation charged leptons), 
can easily be extended to the operators $\lam_{131,132, 231,232}$ by computing the appropriate branching ratios 
of the neutralino decays and adjusting the signal rate accordingly. This holds for all searches focusing on $(e^\pm,
\mu^\pm)$. 

Similarly in the case of the squark and gluino limits from Ref.~\cite{CMS-PAS-SUS-12-027}, \Cms searched for 
separate signatures with: a) $N_{e,\mu}=4$, b) $N_{e,\mu}=2$, $N_\tau=2$, and c)~$N_\tau=4$. As we can see from 
\cref{tab:lle-signatures}, this also covers the six operators not explicitly listed in the analysis. Thus again, adjusting 
the neutralino decay branching ratios allows, in this case, for complete coverage of all possible scenarios, when employing
this existing search.

\begin{table*}
\renewcommand\arraystretch{1.40}
\begin{tabular}{cccccc}
\toprule
Particle & Lower Bound [GeV] & $LL\bar E$ Coupling & Simplified Model & Comment & Reference \\ \midrule
$\tilde\chi_1^0$ & 900 (740) & $\lam_{122}$ ($\lam_{123,233}$) &  $m_{\tilde \chi^\pm_1} = m_{\tilde \chi^0_1}+1\,$GeV & Wino production & \cite{Khachatryan:2016iqn}
 \\
$\tilde\chi_1^0$ & 900 (560) [260] & $\lam_{122}$ ($\lam_{123}$) [$\lam_{233}$] & $m_{\tilde \chi^\pm_1} = m_{\tilde \chi^0_1}+1\,$GeV & Higgsino production & \cite{Khachatryan:2016iqn}
 \\
 \midrule
$\tilde\chi^\pm_1$ &  up to 750 (470) & $\lam_{121}$ ($\lam_{133}$) & $\tilde\chi^\pm_1\to W^\pm\tilde\chi^0_1$ & $\tilde W^- \tilde W^+$ production & \cite{ATLAS:2013qla}
 \\ 
$\tilde\chi_1^\pm$ & up to 1100 & $\lam_{121,122}$& $\tilde\chi^\pm_1\to W^\pm\tilde\chi^0_1$ & 13\,TeV update of \cite{ATLAS:2013qla} &\cite{ATLAS:2016soo}
 \\ \midrule
$\tilde\ell^\pm_L$ & 500 (425) & $\lam_{121,122}$ ($\lam_{133,233}$)& $\tilde\ell^\pm\to \ell^\pm\tilde\chi^0_1$ 
 &$N_\ell\geq4$, 8\,TeV&\cite{Aad:2014iza} \\ 
$\tilde\ell^\pm_R$ & 425 (325) & $\lam_{121,122}$ ($\lam_{133,233}$)& $\tilde\ell^\pm\to \ell^\pm\tilde\chi^0_1$ 
 &$N_\ell\geq4$, 8\,TeV&\cite{Aad:2014iza} \\ 
$\tilde\nu_L$ & 450 & $\lam_{121,122}$ & $\tilde\nu \to \nu\tilde\chi^0_1$ 
 &$N_\ell\geq4$, 8\,TeV&\cite{Aad:2014iza} \\ \midrule
$\tilde q$ & \; 1850 (1750) [1600] \;& $\lam_{122}$ ($\lam_{123}$) [$\lam_{233}$]& $\tilde q\to q\tilde\chi^0_1$ & $N_\ell\geq3$,  
  8\,TeV&\cite{CMS-PAS-SUS-12-027}\\ 
$\tilde t_R$ & 950 (900) [900]& \;$\lam_{122}$ ($\lam_{123}$) [$\lam_{233}$]\;& $\tilde t_R\to t\tilde\chi^0_1$ &
$m_{\tilde\chi^0_1}=300\,$GeV in \cite{CMS-PAS-SUS-12-027}  &  \cite{CMS-PAS-SUS-12-027,Chatrchyan:2013xsw,Khachatryan:2016iqn}\\
$\tilde g$ & \;1450 (1270) [1200] \{1050\}\;& \; $\lam_{121,122}$ ($\lam_{123}$) [$\lam_{233}$] \{$\lam_{133}$\}\;& $\tilde g\to q\bar q\tilde\chi^0_1$ & 
 $N_\ell\geq3$, 8\,TeV&\cite{Aad:2014iza,CMS-PAS-SUS-12-027}\\\bottomrule
\end{tabular}
\caption{Best limits in RPV searches using simplified models and $LL\bar E$ operators. The pair-produced SUSY particles 
are assumed to decay down to the neutralino LSP which itself always decays as $\tilde\chi^0_1\to\ell_i^\pm\ell_k^\mp\nu_j,\,
\ell_j^\pm\ell_k^\mp\nu_i,\,$ for $L_iL_j\bar E_k$. The bounds are only estimates, as they have been read off the relevant 
plots. The first column shows the particle on which a bound is set. The second and third columns show the lower mass bounds 
and the $LL\bar E$ operator which has been assumed for the respective scenario.
}
\label{tab:best-bounds-lle}
\end{table*}

In Ref.~\cite{Aad:2014iza}, \Atlas designed three signal regions specifically for $LL\bar E$ RPV searches
\begin{equation}
\!\!\!\!\!(N_{e,\mu},\, N_\tau,\ETmiss[\SI{}{GeV}])=\left\{\!
\begin{array}{cl}
(i) &(\geq4,\geq0,\geq75)\,,\\[2mm] 
(ii) &(=3,\geq1,\geq100)\,,\\[2mm]
(iii) &(=2,\geq2,\geq100)\,. 
\end{array}\right.\!\!\!\!\!\!\!\!\!
\label{eq:4-leptons}
\end{equation}
Here, $\ETmiss$ [GeV] gives the missing transverse energy in \SI{}{GeV}. \Atlas always employed SFOS signatures. 
They also considered several simplified models with chargino, slepton, sneutrino or gluino pair production, respectively, 
followed by the decays as in Eq.~(\ref{eq:cascade-lle}). This resulted in several of the bounds listed in 
\cref{tab:best-bounds-lle}. Comparing Eq.~(\ref{eq:4-leptons}) with \cref{tab:lle-signatures} we see that these 
searches cover all possible scenarios. Thus, again possibly adjusting for the decay branching ratios of the neutralinos, 
these searches can be employed to constrain all RPV $LL{\bar E}$ models. 

Employing Refs.~\cite{Aad:2014iza,CMS-PAS-SUS-12-027} we see that at least at the level of simplified models the 
neutralino LSP model with a dominant $L_iL_j\bar E_k$ operator has been tested at the LHC, setting lower mass 
bounds. When going to the full $\Lam_{\not R_p}$--CMSSM we expect these bounds to be weaker, as the rates will be 
degraded through additional decay modes. However, several distinct decay chains will contribute to a signal rate, 
possibly compensating the above degradation. In \cref{sec:checkmate-tests} we set bounds on the $\Lam_{\not 
R_p}$--CMSSM parameter space using the LHC searches which have been implemented in the program {\tt CheckMATE} 
\cite{Drees:2013wra,Kim:2015wza}, but which have not necessarily been designed for RPV searches. We note that a fit 
similar to \cite{Bechtle:2015nua} could possibly exclude these models, even though the dark matter constraint does not 
apply. In \cref{sec:checkmateboundstable} we give the explicit resulting lower mass bounds for the individual 
supersymmetric particles.

\Atlas has performed an RPV--mSUGRA/CMSSM search \cite{Aad:2014iza} for the fixed parameters $M_0=A_0=0$ 
excluding $M_{1/2}<800\,$GeV. This corresponds roughly to a gluino mass of 1.8~TeV. \Cms has also performed an 
RPV--CMSSM analysis using 9.2\,fb$^{-1}$ of data at $\sqrt{8}\,$TeV for the specific coupling $\lam_{122}$  
\cite{CMS-PAS-SUS-12-027}. They obtain a lower bound of $M_{1/2}\gtrsim1200\,$GeV for $M_0=1000\,$GeV, $\tan
\beta=40$ and $A_0=0$. This corresponds roughly to a lower gluino mass bound of 2.6\,TeV, and a lower squark 
mass bound of 1.9\,TeV, at this benchmark point.

We also note that in scenarios Ij-I$\ell$, we have with equal rates the special signatures SFSS-SF'SS, \textit{i.e.} for 
two distinct lepton flavors they have same flavor, same sign. These should have an even lower background and in 
particular for the $\tau$ scenarios could lead to improved bounds. 

The special case of stop pair production followed by the cascade decay to neutralinos which then decay via 
$LL\bar E$ operators was also investigated in Ref.~\cite{CMS-PAS-SUS-12-027}. In 
Ref.~\cite{CMS-PAS-SUS-13-010,Khachatryan:2016iqn}, simplified models of squark, gluino and stop pair production have been considered. For not too light neutralino masses they obtain lower mass bounds of about $\SI{1750}{GeV}$ for the squarks,
$\SI{1500}{GeV}$ for the gluinos and $\SI{950}{GeV}$ for the top squark, when considering $\lam_{122}$. 
Similar results are obtained for $\lam_{121}$.

In addition to colored production, the electroweak production of wino- or higgsino-like neutralinos has been considered
\cite{Khachatryan:2016iqn}. In Ref.~\cite{ATLAS:2013qla}, the pair-production of wino-like charginos is considered. It is 
assumed that the charginos are the NLSPs which decay to $W\tilde \chi^0_1$. The results are then interpreted in terms 
of $\lam_{121}\neq 0$ and $\lam_{133}\neq 0$, with the chargino mass bound depending on the mass difference $m_{
\tilde \chi^\pm_1}-m_{\tilde \chi^0_1}$. The respective lower mass bounds range up to 750\,GeV for $\lam_{121}$ and 
470\,GeV for $\lam_{133}$. The analysis has been updated using 13\,TeV data \cite{ATLAS:2016soo}, yielding bounds 
which range up to 1140\,GeV, when considering  $\lam_{12a}\neq 0,~a=1,2$.

However, in a realistic model, a wino-like chargino is accompanied by a wino-like neutralino. The relevant associated 
production $pp\to \tilde \chi^0 \tilde \chi^\pm$ has a larger cross-section than both the neutralino or chargino 
pair-production alone. This has been taken into account in Ref.~\cite{Khachatryan:2016iqn} where wino- and higgsino-LSPs 
are treated separately in a simplified model setup, always assuming the associated chargino state to be $\sim 1~$GeV 
heavier. However, it is further assumed that the bino is sufficiently heavy to play no role in either the production or decay. 
The resulting bounds for winos are about $\SI{900}{GeV}$ for $\lam_{122}$ and $\SI{740}{GeV}$ for both $\lam_{123}$ 
and $\lam_{233}$. For a higgsino-like neutralino LSP the range of the bounds is much larger, constraining the mass 
below $260-900$ GeV, where the strongest bound is obtained for $\lam_{122}$ and the weakest one for $\lam_{233}$.

In \cref{sec:checkmate:lle}, we will compare this simplified model approach directly with the bounds we obtain for a more 
realistic scenario with CMSSM boundary conditions. In that case, the bino is always the LSP whose production cross 
section is, however, negligible compared to the wino and higgsino production. This results in a decay of the produced 
wino/higgino state down to a bino first which then itself decays via the $LL\bar E$-operator-induced three-body decay. 
We will see that, using the search strategy of Ref.~\cite{ATLAS:2013qla} and also taking into account electroweak 
neutralino-chargino production, we can improve upon the bounds which have been obtained in Ref.~\cite{ATLAS:2013qla}.


\subsection{Neutralino LSP Decay via an $LQ\bar D$ Operator}
\label{sec:NeutralinoLQD}
At leading order a neutralino decays via an $ L Q{\bar D}$ operator to one lepton and two jets:
\begin{equation}
L_i Q_j{\bar D}_k:\qquad \tilde\chi^0_1\to \{\ell^-_i  u_j \bar d_k,\,\nu_i d_j \bar d_k\} + c.c.
\label{eq:neutralino-decay-LQD}
\end{equation}
When discussing the decay signatures in the following  `\textit{j}' shall denote a first- or second-generation quark jet, 
and `$\ell$'  a first or second generation charged lepton. Top, bottom and tau are treated separately, as they can be 
identified by tagging-algorithms. The neutrino in the final state leads to missing transverse energy, $\ETmiss$. When
collectively describing states: $a,b,c=1,2$ denote the first two generations and $i,j,k=1,2,3$ refer to all three generations.

The possible signatures arising from the decay of a pair of neutralinos together with the couplings they probe are 
summarized in \cref{tab:lqd-signatures}. Here we assume that squarks or gluinos are pair-produced 
at the collider and cascade-decay to two neutralino LSPs. We see that we obtain at most two charged leptons, as 
well as various combinations of jets, $b$-,\,$t$-quarks, $\tau$-leptons and $\ETmiss$, depending on the dominant coupling.
\begin{table}
 \renewcommand\arraystretch{1.40}
\begin{tabular}{ccc}
\toprule
Scenario\; & \; Signature\; & \; $LQ\bar D$ Operator\\ \midrule 
IIa & $[\ell^+_a\ell^\pm_a,\,\ell^+_a \!\ETmiss]\; 4j$ &$\lam^\prime_{abc}$\\
IIb & $[\ell^+_a\ell^\pm_a,\,\ell^+_a\! \ETmiss]\; 2b\,2j$ &$\lam^\prime_{ab3}$\\
IIc & $[\ell^+_a\ell^\pm_a \bar t \stackrel{{\tiny(-)}}{t},\,\ell^+_a\bar t\,b\ETmiss]\;2j$ &$\lam^\prime_{a3c}$\\
IId & $[\ell^+_a\ell^\pm_a \bar t \stackrel{{\tiny(-)}}{t},\,\ell^+_a\bar t\,b\ETmiss]\;2b$ &$\lam^\prime_{a33}$\\
IIe & $[\tau^+\tau^\pm,\,\tau^+ \!\ETmiss]\;4j$  &$\lam^\prime_{3bc}$\\
IIf & $[\tau^+\tau^\pm,\,\tau^+\! \ETmiss]\; 2b\,2j$  &$\lam^\prime_{3b3}$\\
IIg & $[\tau^+\tau^\pm \bar t \stackrel{{\tiny(-)}}{t},\,\tau^+\bar t\,b\ETmiss]\;2j$ &$\lam^\prime_{33c}$\\
IIh & $[\tau^+\tau^\pm \bar t \stackrel{{\tiny(-)}}{t},\tau^+\bar t\,b\ETmiss]\;2b$ &$\lam^\prime_{333}$\\
IIi & $4j\ETmiss$ &$\lam^\prime_{abc},\lam'_{3bc}$\\
IIj & $2b\,2j\ETmiss$ &$\lam^\prime_{ab3},\lam'_{3b3},\lam'_{a3c},\lam'_{33c}$\\
IIk & $4b\ETmiss$ &$\lam'_{a33},\lam^\prime_{333}$\\
\bottomrule
\end{tabular}
\caption{Possible final states in the $LQ\bar D$ case for a pair of neutralino LSPs decaying via the same operator. There 
will be further accompanying particles from the cascade decay of the originally produced particles, e.g. for squark 
pair-production $\tilde q\tilde q^*\to q\bar q\tilde\chi^0_1\tilde\chi^0_1$, giving two extra jets. $\ell$ denotes a charged 
lepton and the indices $a,b,c = 1,2$ denote leptons or quarks from the first or second generation. We have separated out 
the signatures with no charged lepton. For each listed signature there is a corresponding charge-conjugate signature due 
to the Majorana nature of the neutralino.}
\label{tab:lqd-signatures}
\end{table}

In the simplified models considered by the experimental collaborations, the gluinos or squarks cascade-decay as in Eq.~(\ref{eq:cascade-lle}), 
however with the neutralino decay replaced by that in Eq.~(\ref{eq:neutralino-decay-LQD}). 
Thus in order to obtain the total signature, those in \cref{tab:lqd-signatures} should be supplemented by 4 (2) jets 
in the case of gluino (squark) pair production. For example, for the dominant coupling $\lam'_{123}$, cases IIb and IIj, 
assuming squark pair production we obtain the signatures:
\begin{eqnarray}
\tilde q\tilde q^*+\mathrm{IIb:} &&[\ell^+_a\ell^\pm_a,\,\ell^+_a\! \ETmiss]\; 2b\,4j\,, \\
\tilde q\tilde q^*+\mathrm{IIj:} && 2b\,4j \ETmiss \,.
\end{eqnarray}
A promising signature would then be two same-sign charged leptons, and two $b$-jets. This has been searched for in 
Ref.~\cite{ATLAS-CONF-2015-018} yielding
\begin{eqnarray}
m_{\tilde q}&\geq& 1160\,(1360)\,\mathrm{GeV}, \quad m_{\tilde\chi^0_1}=0.5\,(0.9)\, m_{\tilde q}\,.
\end{eqnarray}
Note that Ref.~\cite{ATLAS-CONF-2015-018} assumed BR$(\tilde\chi^0_1\to\tau^\pm+2j)=0.5$, with the remaining 
50\% being decays to neutrinos and two jets. Even in these simplified models this is not true in general, as we saw
in Fig.~\ref{fig:staudecayTB}. The exact number depends on the admixture of the neutralino LSP \cite{Dreiner:1994tj},
as well as on the masses of the involved off-shell stau and sneutrino propagators. 

In \cref{tab:lqd-signatures} we see that each coupling leads to three distinct signatures, modulo lepton charge  assignments. In the case of a discovery these should be cross-checked against each other. Presently, the most sensitive mode should be chosen, most likely same-sign di-leptons, together with possible $b$-quarks for $\lam'_{ij3}$.

\begin{table*}
 \renewcommand\arraystretch{1.40}
\begin{tabular}{cccccc}
\toprule
Particle & Lower Bound [GeV] & $LQ\bar D$ Coupling & \; Simplified Model\; & Comment & Reference \\ \midrule
$\tilde \chi^0_1 $ & 720 (620) [660] \{500\}& $\lam'_{131}$ ($\lam'_{131}$) [$\lam'_{233}$] \{$\lam'_{233}$\}& & $\tan\beta= 2$ (40) [2] \{40\}  & \cite{Khachatryan:2016iqn}\\ \midrule
$\tilde\mu$ & 440 (825) [1290]  & $\lam'_{211}=$0.003 (0.01) [0.04] & $\tilde\mu\to\mu\tilde\chi^0_1$& res. $\tilde\mu$ prod, $m_{\tilde\chi^0_1}=200\,$GeV  & \cite{CMS-PAS-SUS-13-005}\\ \midrule
$\tilde q$ & 1160 (1090) [1065] & $\lam'_{abc,ab3}$ ($\lam'_{3bc}$) [$\lam'_{3b3}$]& $\tilde q\to q\tilde\chi^0_1$& $m_{\tilde\chi^0_1}
=0.5\, m_{\tilde q}$ & \cite{ATLAS-CONF-2015-018}\\
& 1315 (1360) [1225] \{1215\}& $\lam'_{abc}$ ($\lam'_{ab3}$) [$\lam'_{3bc}$] \{$\lam'_{3b3}$\}&& 
$m_{\tilde\chi^0_1}=0.9\, m_{\tilde q}$ & \cite{ATLAS-CONF-2015-018}\\ 
%
& 1310 (1400) [2000] & $\lam'_{23c,233}$  & & $m_{\tilde g}\lesssim 2000\, (1500) \,[1000]\,$GeV 
& \cite{CMS-PAS-SUS-12-027} \\ \midrule
$\tilde g$ & 1010 (970) [1070] \{1050\}& \;$\lam'_{abc}$ ($\lam'_{ab3}$) [$\lam'_{3bc}$] \{$\lam'_{3b3}$\}\;& $\tilde g\to q\bar q\tilde
\chi^0_1$& $m_{\tilde\chi^0_1}=0.1\, m_{\tilde g}$ & \cite{ATLAS-CONF-2015-018}\\ 
& 1135 (1085) [1220] & $\lam'_{abc}$ ($\lam'_{ab3}$) [$\lam'_{3bc,3b3}$]& & 
$m_{\tilde\chi^0_1}=0.5\, m_{\tilde g}$ & \cite{ATLAS-CONF-2015-018}\\ 
& 1285 (1260) [1200] & $\lam'_{abc}$ ($\lam'_{ab3}$) [$\lam'_{3bc,3b3}$] & & 
$m_{\tilde\chi^0_1}=0.9\, m_{\tilde g}$ & \cite{ATLAS-CONF-2015-018}\\ 
& 2000 (1500) [1000] & $\lam'_{23c,233}$  & & $m_{\tilde q}\lesssim 1310\, (1400) \,[2000]\,$GeV 
& \cite{CMS-PAS-SUS-12-027} \\
& 1520 (1770) [1820] & $\lam'_{abc}$  & & $m_{\tilde\chi^0_1}= 100\,(500)\,[890]\,$GeV & \cite{Aaboud:2017dmy} \cite{Aaboud:2017faq}
\\ \midrule
$\tilde t$ & 890 (1000)  & $\lam'_{1bc}$ ($\lam'_{2bc}$)  & $\;\tilde t\to b(\ell^+2j)_{\tilde\chi^+_1}\;$ & 
$m_{\tilde\chi^+_1}=100\,$GeV & \cite{Khachatryan:2016ycy}\\ 
 & 580  & $\lam'_{3bc}$   & $\;\tilde t\to b(\tau^+2j)_{\tilde\chi^+_1}\;$ & $m_{\tilde\chi^+_1}=100\,$GeV & \cite{Khachatryan:2014ura}\\
 & 710 (860)  & $\lam'_{132}$ ($\lam'_{232}$)  & $\;\tilde t\to 2b(\ell^+j)_{\tilde\chi^+_1}\;$ & $m_{\tilde\chi^+_1}=m_{\tilde t}-(100\,$GeV) 
 & \cite{Khachatryan:2015vaa}\\
\bottomrule
\end{tabular}
\caption{Best limits in RPV searches using simplified models and $LQ\bar D$ operators. Here $a,b,c\in\{1,2\}$. The 
pair-produced supersymmetric particles are assumed to decay directly to the neutralino LSP. The bounds are only 
estimates, as they have been read off the relevant plots. The neutralino LSP always decays as $\tilde\chi^0_1\to(
\ell_i^-u_j\bar d_k,\;\nu_i d_j\bar d_k)+c.c.$ for $L_iQ_j\bar D_k$. We have included the analysis of a scalar top 
decaying via a chargino and not the neutralino LSP in the last line, since it is similar. The neutralino decay can be 
blocked due to the heavy top quark. The search in Ref.~\cite{CMS-PAS-SUS-12-027} allows for simultaneously 
non-decoupled squarks and gluinos. In Ref.~\cite{Chatrchyan:2013xsw} \Cms was able for a given $m_{\tilde\chi^0_1}
\in[200,800]$GeV to exclude a range of scalar top squark masses, without a fixed lower bound. We have included 
the bound on the smuon mass, which is from a search for resonant production, since it also decays via the neutralino 
LSP. Each mass bound is for a fixed value of the RPV coupling. }
\label{tab:best-bounds-lqd}
\end{table*}

A summary of the experimental lower mass bounds on the supersymmetric particles for a given dominant coupling is 
given in \cref{tab:best-bounds-lqd}. For squarks we have lower mass bounds ranging from about 1 to 1.4 TeV. In 
one special case for light gluino masses there is a bound of 2 TeV. The lower gluino mass bounds are similar, ranging 
from about 1 to 1.3~TeV, with some stricter bounds achieved in special scenarios with light squarks or heavy neutralino 
LSPs. The pair-production cross section of neutralino LSPs is typically much smaller than for squarks and gluinos 
resulting in the correspondingly weaker lower limits ranging from 500 to 720 GeV. Unlike the $LL\bar E$ case there are 
no RPV searches for charginos or sleptons here. In the last three lines of the table we have included lower limits on the 
top squark, which in the case at hand decays via a chargino instead of a neutralino while the chargino decays to a 
charged lepton and two jets. The lower mass bounds range from 580 GeV to 1100 GeV depending on the flavor of the 
charged lepton and the jets. This decay is not in the spirit of this section, but we considered it similar enough to include, 
as the neutralino decay could be blocked by the heavy top quark.

Regarding the coverage of the $LQ\bar D$ $R$-parity violating signatures by explicit $R$-parity violating searches, we 
see that Ref.~\cite{ATLAS-CONF-2015-018} focused on leptons plus jets signatures. They consider $\tau$ leptons and 
$b$ quarks, but explicitly omit top quarks, and also do not consider the neutrino $\ETmissx$ cases. They thus cover the 
signatures IIa, IIb, IIe, and IIf. Comparing with \cref{tab:best-bounds-lqd} we see that \cite{ATLAS-CONF-2015-018} 
covers a wide range of possible $LQ\bar D$ couplings, however it omits the couplings $\lam'_{i3k}$, since in that case 
the charged leptons are accompanied by a heavy top quark. Ref.~\cite{CMS-PAS-SUS-12-027} explicitly looked for the 
cases $\lam'_{23k}$, partially covering the signatures IIc and IId. The signatures IIg-IIk either involve charged leptons 
with top quarks, or have $\ETmissx$ signatures instead of the charged leptons. None of these have been covered by 
explicit RPV searches at the LHC. In particular the couplings $\lam'_{33k}$ have not been looked for. Note that in 
Ref.~\cite{Khachatryan:2016iqn}, \Cms did search for electroweak gaugino production decaying via $\lam'_{331,333}$, 
however the sensitivity was insufficient to lead to any bound. In Ref.~\cite{ATLAS-CONF-2015-018} explicitly looked for 
$\tau$'s and $b$-quarks, giving specific sensitivity to the cases $\lam'_{3jk}$ and $\lam'_{ij3}$. However, they did not 
look for top quarks together with charged leptons and therefore explicitly omitted $\lam'_{i3j}$. 

We make a special mention of Ref.~\cite{CMS-PAS-SUS-13-005}, where \Cms analyzed resonant smuon production, 
with the smuon decaying via the neutralino LSP \cite{Dreiner:2000vf,Dreiner:2006sv}. The production cross section is 
proportional to the RPV coupling squared and in order to get an appreciable rate requires $\lam'_{211}\gtrsim0.003$. The 
lower mass limit then depends strongly on the assumed coupling value, as can be seen in \cref{tab:best-bounds-lqd}. 
We point out that \Cms have interpreted this search also in terms of the RPV--CMSSM.

We note that some of the signatures listed in \cref{tab:lqd-signatures} are also covered by RPC searches. The first 
two scenarios, IIa and IIb, involve only light leptons and jets, possibly $b$-jets.  They always include an option also with 
$\ETmiss$. The scenarios IIi-k involve multijet events with missing transverse momentum, but zero leptons. These 
correspond to the standard RPC supersymmetry searches, see for instance 
Refs.~\cite{Aad:2015mia,CMS-PAS-SUS-16-042,CMS-PAS-SUS-16-037} for isolated leptons and jets, and 
\cite{Aad:2013wta,Aad:2014wea,Aaboud:2016zdn,Sirunyan:2017cwe,CMS-PAS-SUS-16-036} for zero leptons and 
multijets accompanied by missing energy. The signatures including a $\tau$ lepton, \textit{i.e.} IIe-IIh are in principle also 
covered by these multijet analyses. In  this case, a hadronically decaying tau lepton is not explicitly tagged but handled 
as a hadronic object.  Thus the couplings $\lam'_{i3k}$ have in principle been probed via the 
scenarios IIi-IIk. The exact sensitivity will only be known once the corresponding experimental searches have been 
interpreted in terms of these RPV models.

All of the searches listed in \cref{tab:best-bounds-lqd}, except the smuon search, employ minimal or next-to-minimal 
simplified models, with an intermediate neutralino LSP state in the decay chain. Thus it is difficult to see how the bounds in 
\cref{tab:best-bounds-lqd} are modified in the case of realistic cascade decay branching ratios, as for example in the 
CMSSM. In \cref{sec:checkmate-tests}, we shall use LHC analyses implemented in the computer program 
\Checkmate{} to obtain realistic limits on the parameters of the RPV--CMSSM models. We shall see for example that the 
 search of Ref.~\cite{Aad:2013wta} which looks for RPC as well as $\bar U \bar D \bar D$-RPV is very sensitive to the signatures arising from the $\lam'_{3ij}$ operators within 
CMSSM boundary conditions. In Ref.~\cite{ATLAS-CONF-2015-018}, a re-interpretation of the RPC searches of 
Refs.~\cite{Aad:2015mia,Aad:2014pda,Aad:2014wea,Aad:2013wta} in terms of lepton-number-violating SUSY is 
presented. The resulting lower bounds on the gluino mass are around $\SI{1}{TeV}$, for some cases similar limits 
are obtained for squark masses. In a CMSSM context, the bounds are even stronger as we see in
\cref{sec:checkmateboundstable}.


\subsection{Neutralino LSP Decay via an $\bar U \bar D\bar D$ Operator}\label{sec:neutralinoLSP_UDD}

If the RPV operator $\bar U_i \bar D_j\bar D_k$ is non-zero, the neutralino LSP decays via an intermediate squark to three 
jets: 
\begin{equation}
\bar U_i \bar D_j\bar D_k: \qquad \tilde\chi^0_1\to u_id_jd_k+c.c.
\end{equation}
Thus, for neutralino pair production we get six jets, possibly accompanied by further jets from intermediate cascade decays. 
The detailed jet flavor listings are given in \cref{tab:udd-signatures}. As we see, out of the six jets up to two can be 
bottom quark jets and up to two can be top quarks. Most searches for these scenarios are focused on multijet events with 
different numbers of b-tags 
\cite{ATLAS:2012dp,ATLAS-CONF-2016-057,Aad:2015lea,Aad:2016kww,ATLAS:2016yhq,Khachatryan:2016xim,Khachatryan:2014lpa,Chatrchyan:2013gia,Chatrchyan:2013fea,ATLAS-CONF-2013-091}. 
Some analyses also account for leptons, which could arise from leptonic top decays 
\cite{ATLAS-CONF-2015-018,CMS-PAS-SUS-16-013,Khachatryan:2016iqn,Chatrchyan:2013fea,ATLAS:2016mnt,ATLAS:2017wgj}. 
\begin{table}
 \renewcommand\arraystretch{1.40}
\begin{tabular}{ccc}
\toprule
Scenario\; & \; Signature\; & \; $\bar U\bar D\bar D$ Operator\\ \midrule 
IIIa & $6j$  &$\lam^{\prime\prime}_{a12}$\\
IIIb & $(2b)(4j)$  &$\lam^{\prime\prime}_{ab3}$\\
IIIc & $(2t)(4j)$  &$\lam^{\prime\prime}_{312}$\\
IIId & $(2t)(2b)(2j)$  &$\lam^{\prime\prime}_{3b3}$\\
\bottomrule
\end{tabular}
\caption{Possible final states in the $\bar U\bar D\bar D$ case for a pair of neutralino LSPs decaying via the same operator. 
There will be further accompanying particles from the cascade decay of the originally produced particles, e.g. for squark pair 
production $\tilde q\tilde q^*\to q\bar q\tilde\chi^0_1\tilde\chi^0_1$, giving two extra jets. The indices $a,b = 1,2$ denote 
quarks from the first or second generation.}
\label{tab:udd-signatures}
\end{table}
The top quarks are not necessarily produced  via the RPV operator but can originate from cascade decays, \textit{e.g.}\ in 
stop pair production.

\begin{table*}
 \renewcommand\arraystretch{1.40}
\begin{tabular}{ccclcc}
\toprule
Particle & \; Lower Bound [GeV] \;& \;$\bar U\bar D\bar D$ Coupling\; & Simpl. Model & Comment & Reference \\ 
\midrule
$\tilde q$ & 1725  (1900) [2800] & $\lam''_{112}$ & \;$\tilde q_R\to j(\ell\ell \tilde\chi^0_1)_{\tilde\chi^0_2}$ \;
& $m_{\tilde g}\leq2400\, (1500)\, [1200]\,$GeV& \cite{CMS-PAS-SUS-12-027}\\
\midrule
$\tilde t$ & 950 (980) & $\lam''_{3b3}$ & $\tilde t\to t\tilde\chi^0_{1,2}/b\tilde\chi^+_1$ & $\tilde\chi^0_1=\tilde H$ ($\tilde B$),& 
\cite{Aaboud:2017faq} \\[-1mm]
& &&  &  $m_{\tilde\chi^0_1} =300\,$GeV & \\  
 & 1090 (1260) & $\lam''_{3b3}$ & $\tilde t\to t\tilde\chi^0_{1,2}/b\tilde\chi^+_1$ & $\tilde\chi^0_1=\tilde H$ ($\tilde B$),& 
 \cite{Aaboud:2017faq} \\[-1mm]
& &&  &  $m_{\tilde\chi^0_1} =800\,$GeV & \\  
\midrule
$\tilde g$ & 1200  (1500) [2400] & $\lam''_{112}$ & 
$\tilde g\to jj(\ell\ell \tilde\chi^0_1)_{\tilde\chi^0_2}$ & \;$m_{\tilde q}\leq2800\, (1900)\, [1725]\,$GeV\;& \cite{CMS-PAS-SUS-12-027}\\
  &1850 (2100) & $\lam''_{112}$ & $\tilde g\to t\bar t\tilde\chi^0_1$ & $m_{\tilde\chi^0_1}=100\, (800)\,$GeV & \cite{Aaboud:2017faq}\\
  &&&&$m_{\tilde g}>2m_t+m_{\tilde\chi^0_1}$ & \cite{Aaboud:2017faq}\\
 & 650  (950) [1020] & $\lam''_{212}$ & $\tilde g\to q\bar q \tilde H^0_1$ & $m_{\tilde q}\leq 100\, (500)\, [900]\,
 $GeV& \cite{Khachatryan:2016xim}\\ &  &  &  & $m_{\tilde H^0_1}=\frac{3}{4} m_{\tilde q},\,m_{\tilde c}<m_{\tilde g}$& \\
 & 675  (1020) [1075] & $\lam''_{212}$ & $\tilde g\to q\bar q \tilde H^0_1$ & $m_{\tilde q}\leq 100\, (500)\, [900]\,
 $GeV& \cite{Khachatryan:2016xim}\\ &  &  &  & $m_{\tilde H^0_1}=\frac{3}{4} m_{\tilde q};\;m_{\tilde b}<m_{\tilde g}$& \\
 & 650  (1020) [1100] & $\lam''_{213}$ & $\tilde g\to q\bar q \tilde H^0_1$ & $m_{\tilde q}\leq 100\, (500)\, [900]\,
 $GeV& \cite{Khachatryan:2016xim}\\ &  &  &  & $m_{\tilde H^0_1}=\frac{3}{4} m_{\tilde q},\,m_{\tilde b}<m_{\tilde g}$& \\
 & 650  (990) [1075] & $\lam''_{213}$ & $\tilde g\to q\bar q \tilde H^0_1$ & $m_{\tilde q}\leq 100\, (500)\, [900]\,
 $GeV& \cite{Khachatryan:2016xim}\\ &  &  &  & $m_{\tilde H^0_1}=\frac{3}{4} m_{\tilde q};\;m_{\tilde c}<m_{\tilde g}$& \\
 & 1040 (1555)  & $\lam''_{ijk}$ &$\tilde g\to q\bar q\tilde\chi^0_1$  & $m_{\tilde\chi^0_1}=100\,(900)\,$GeV & 
 \cite{ATLAS-CONF-2016-057}\\ &&&&all $\lam''_{ijk}\not=0$& \\
 & 800\, (1050) & $\lam''_{abc}$ &$\tilde g\to 5q$& $m_{\tilde\chi^0_1}=50\,(600)\,$GeV & \cite{ATLAS-CONF-2013-091}\\
\bottomrule
\end{tabular}
\caption{Best limits in RPV searches using simplified models and $\bar U\bar D\bar D$ operators. Here $a,b,c\in\{1,2\}$. 
The pair-produced supersymmetric particles are assumed to decay directly to the neutralino LSP if not stated otherwise. 
The bounds are only estimates, as they have been read off the relevant plots. The neutralino LSP always decays as $
\tilde\chi^0_1\to\bar u_i\bar d_j\bar d_k +c.c.$ for $\bar U_i\bar D_j\bar D_k$. Each mass bound is for a fixed value of 
the RPV coupling.}
\label{tab:best-bounds-udd}
\end{table*}

In \cref{tab:best-bounds-udd} we have collected the best LHC lower mass bounds on the squark mass (treating the 
stop separately) and the gluino mass for various dominant couplings. We see that most searches have been performed 
for the case of gluino pair production. The mass bounds range from 650 GeV up to 2400 GeV depending on the scenario. 
For the case of a simplified model with only a light gluino and a neutralino mass of $m_{\tilde\chi^0_1}=100\,$GeV, \Atlas 
obtained a lower bound of 840 GeV. The search employed the  ``total jet mass of large-radius" \cite{Aad:2015lea}. 
Compared to a jet-counting analysis, it was shown that this technique allows for slightly higher sensitivity for light jets, 
whereas the jet-counting analysis provides the better bounds in the case of $b$-tagging 
requirements. The strictest bound of 2400 GeV is achieved in a \Cms search for a simplified model which also contains 
kinematically accessible squarks \cite{CMS-PAS-SUS-12-027}.

For squarks we found only one direct search \cite{CMS-PAS-SUS-12-027} by \Cms. They utilized an extended simplified 
model involving accessible squarks, gluinos {\it and} the second neutralino $\tilde\chi^0_2$. The pair production of $SU(2)$ 
singlet squarks is considered, followed by the cascade decay via 
\begin{equation}
\tilde q_R\to q \tilde\chi^0_2\to q [\ell^+\tilde \ell^{(*)-}]\to q[\ell^+(\ell^-\tilde\chi^0_1)]\,.
\end{equation}
The intermediate sleptons need not be on-shell but cannot be too heavy.

In the case of the stop we have an \Atlas search, which also assumes an intermediate chargino \cite{Aaboud:2017faq}, 
giving rise to an additional decay mode. The resulting lower mass bounds are of the order 1\,TeV.

The most pertinent question for us is how well the $\bar U\bar D\bar D$ models with a neutralino LSP are covered by 
searches at the LHC. Looking at \cref{tab:best-bounds-udd} we see that the search from Ref.~\cite{ATLAS-CONF-2016-057} 
in the last line seems to cover all possible $\lam''_{ijk}$. However, \Atlas here explicitly assumed that {\it all} $\lam''_{ijk}
\not=0$ simultaneously, with every coupling taking the same value. Thus the sensitivity of the search could rely unduly 
on bottom quarks from $\lam''_{ij3}$ and/or from top quarks from $\lam''_{3jk}$ couplings. If we look at explicit searches 
based on the single coupling dominance assumption, then we have gluino searches for $\lam''_{112,212,213}$ 
\cite{CMS-PAS-SUS-12-027,Khachatryan:2016xim,Aaboud:2017faq,Aaboud:2017dmy}. All such searches make additional assumptions on 
the squark and/or neutralino masses. We would expect the search for $\lam''_{213}$ to be equally sensitive to $\lam''_{
223}$. Similarly we would expect the $\lam''_{213}$ search to apply equally to $\lam''_{113}$, provided $m_{\tilde u,\tilde 
d}<m_{\tilde g}$, and to $\lam''_{123}$, provided $m_{\tilde u,\tilde s}<m_{\tilde g}$. Thus for the case of gluino production 
there is no coverage only of the three couplings $\lam''_{3jk}$. This should be performed by the LHC experimental groups. 
As we see from \cref{tab:udd-signatures}, the final states involve either two top quarks and four jets or two top quarks, 
two bottom quarks and two jets, plus of course the accompanying jets from the cascade decay. 

In \cref{sec:checkmate-tests}, we also take into account dominant $\lam''_{3jk}$ couplings in the context of CMSSM 
boundary conditions and show that we can set bounds on this scenario using multijet analyses as well as searches for 
same-sign leptons. We do however expect a boost in discovery potential when designing a dedicated search for the 
signature outlined above.

As we see in rows two and three of \cref{tab:best-bounds-udd}, the searches for top squarks cover at least two of 
these missing scenarios, namely $\lam''_{3b3}$. Overall then the case $\lam''_{312}$ is missing for the scenarios 
considered here. 

Once again, none of these models have been interpreted in terms of the CMSSM, thus it is difficult to see how 
realistic these mass bounds are. We shall come back to this question in \cref{sec:checkmate-tests} and with explicit 
mass bounds in \cref{sec:checkmateboundstable}.

\section{LHC Coverage of RPV-induced Stau LSP Decay Scenarios}
\label{sec:stauLSP}
As we saw in \cref{sec:rges}, even for small RPV couplings, we have substantial regions of CMSSM parameter 
space, where the LSP is the lightest stau, $\tilde\tau_1$. The question is: how does the phenomenology change 
compared to a neutralino LSP, and what are the experimental constraints? The stau LSP case has been discussed 
from the theoretical perspective in some detail in the literature 
\cite{Allanach:2003eb,Allanach:2006st,deCampos:2007bn,Dreiner:2008rv,Desch:2010gi}. In \cite{Allanach:2006st} a 
set of appropriate LHC benchmarks was defined. Here we are interested in the case of small, but not too small 
$\Lam_{\not R_p}$. Thus the stau is the LSP but it still decays promptly. We shall consider the long-lived case where 
the stau leads to detached vertices, or is even stable on detector scales, elsewhere.

In order to discuss the LHC coverage for the stau LSP scenario, we distinguish two cases\footnote{Here again: $a,b,c
\in\{1,2\}$ and $i,j,k\in\{1,2,3\}$.}
\begin{enumerate}
\item[(a)] $\tilde \tau$ LSP with the dominant operator directly coupling to a tau/stau: $\Lam_{\not R_p} \in
\{\lam_{aj3,a3c},\,\lam'_{3jk}\}$.
\item[(b)]  $\tilde \tau$ LSP with the dominant operator {\it not} coupling to tau/stau $\Lam_{\not R_p} \in
\{\lam_{12c},\,\lam'_{ajk},\,\lam''_{ijk}\}$.
\end{enumerate}
In case (a) we do not expect significant changes compared to the neutralino phenomenology. Consider the small 
$\Lam_{\not R_p}$ case in the CMSSM, where we get a $\tilde\tau$ LSP for small $M_0$ and larger $M_{1/2}$, 
as discussed in \cref{sec:determ-lsp}. There the NLSP is typically the lightest neutralino $\tilde\chi^0_1$. We 
then have the usual cascade decays of the strongly produced sparticles down to the neutralino, as shown in the 
bottom two rows on the left in Eq.~(\ref{eq:cascade-lle}). This is followed by the decay of the neutralino to the 
$\tilde\tau$. For example for the gluino 
we could have
\begin{equation}
\tilde g\to q \bar q \tilde\chi^0_1\to q\bar q(\tau^\pm\tilde\tau^\mp)\,.
\label{eq:stau-lsp-cascade}
\end{equation}
This is followed by the direct RPV stau decay (plus charge conjugate)
\begin{equation}
\tilde\tau^-\to \left\{
\begin{array}{ll} 
\nu_a\ell_c^-\,, & \;\;\;L_aL_3\bar E_c\,,\\[2mm]
\!\!(\nu_1\ell_2^-,\ell_1^-\nu_2)\,,\;\; & \;\;\; L_1L_2\bar E_3\,,\\[2mm]
\!\!(\nu_a\tau^-,\ell_a^-\nu_\tau)\,, & \;\;\; L_aL_3\bar E_3\,,\\[2mm]
\bar u_j d_k\,, & \;\;\; L_3Q_j\bar D_k\,,
\end{array}
\right.
\label{eq:2bdystau}
\end{equation}
depending on the dominant operator. These final states with the same couplings were also obtained for the 
neutralino LSP scenarios. However, there are slight differences here for the stau. If the RPV operator involves an $L_3$ chiral superfield, 
the neutralino decay to the tau neutrino is strongly disfavored in comparison to the on-shell decay mode. For 
example:
\begin{eqnarray}
\tilde\chi^0_1\mathrm{-LSP:} & \tilde\chi^0_1\to \{ \nu_\tau e^\pm e^\mp,\,\tau^\pm\nu_e e^\mp \}; & L_1L_3\bar E_1\,, \\[2mm]
\tilde\tau_1\mathrm{-LSP:} &\!\!\!\! \tilde\chi^0_1\to \tau^\pm\tilde\tau^\mp \to \tau^\pm(\nu_e e^\mp);& L_1L_3\bar E_1 \,.
\end{eqnarray}
In the $L_1L_3\bar E_1$ case one thus loses the $e^\pm e^\mp$ signature, instead having a tau lepton with 
100\% branching ratio. This could dilute the experimental sensitivity. Similarly for $L_3Q_j\bar D_k$ operators, 
where the branching ratio for the charged tau is enhanced, possibly increasing the sensitivity compared to the 
neutralino LSP scenario.

In case (b), the stau does not couple to the dominant operator and the phenomenology changes, as the two-body 
decay channels in Eq.~(\ref{eq:2bdystau}) are absent. The leading stau decay
is four-body. In the case of a $\lam''_{ijk}$ operator, for example, the decay chains would be
\begin{align} \label{eq:staudec} 
\tilde \tau &\to \tau \tilde \chi^0{}^{(*)} \qquad \qquad \quad \text{or}  \;\; \to \nu_\tau \chi^\pm{}^{(*)} \\  \notag 
 & \hookrightarrow \tau u_i \tilde u_i^{(*)}/\tau d_{j,k} \tilde d_{j,k}^{(*)} \quad    \quad ~\hookrightarrow \nu_\tau u_{j,k} 
 \tilde d_{j,k}^{(*)}/\nu_\tau d_i \tilde u_i^{(*)}\\ \notag & ~~\hookrightarrow \tau u_i d_j d_k
 \qquad  \qquad \quad \quad ~\hookrightarrow \nu_\tau u_{j,k} u_i d_{k,j}/\nu_\tau d_i d_j d_k
\end{align}
where we have neglected charge-conjugations and denote virtuality by  `${}^{(*)}$'. We have allowed for decays via a 
chargino, as the intermediate states are virtual. 

Whether the four-body decay is prompt or not depends heavily on the sparticle mass spectrum. 
The stau decay length, assuming a four-body decay only, scales roughly as
\cite{Allanach:2003eb}
\begin{align}
%
c\tau_{\tilde \tau_1} \simeq\, & 6.2\cdot 10^{-6}\,{\rm m} \left( \frac{10^{-3}}{\LamRpV}\right)^2  \\ \notag &\times \left
(\frac{m_{\tilde \chi}}{100\,{\rm GeV}}\right)^2 \left(\frac{m_{\tilde f}}{100\,{\rm GeV}}\right)^4 \left( \frac{100\,{\rm GeV}}{m_{\tilde \tau_1}} \right)^7\,,
\end{align}
where $\tilde \chi$ refers to the virtual neutralino/chargino and $\tilde f$ to the virtual sfermion in the decay.

In the stau LSP parameter regions, the lightest stau is mostly a $\tilde\tau_R$, so that the coupling to the wino is 
reduced w.r.t. the bino. In conjunction with the hierarchy in the CMSSM of $m_{\tilde B}<m_{\tilde W}$, this means 
that the decays via charginos, \textit{i.e.} the ones resulting in $\nu_\tau+X$ are suppressed and the final states 
including a $\tau^\pm$ dominate. However, for $\lam''_{3jk}$ this is not true. In this case, all neutralino-mediated 
channels end up in a top quark, which is kinematically suppressed w.r.t. the chargino mediated $\nu_\tau b\,d_j d_k$ 
mode, unless the stau is very heavy.

As discussed in \cref{sec:RGEinduced}, even if the non-zero RPV operators at a given scale only involve 
couplings to first- and/or second-generation (s)leptons, operators coupling to (s)taus will be induced through the RGE 
evolution. Thus scenarios (a) and (b) are not strictly separated. This is particularly important in the case of large 
$\tan \beta$ for any $\lam_{ijk}$ operators where $i,j,k\neq3$ as well as all $\lam'_{ijj}$ operators with $i\neq 3$, 
see  Eq.~(\ref{eq:taurge}) and the corresponding \cref{fig:staudecayTB,fig:staudecayA0TB} in 
\cref{sec:RGEinduced}. Thus in complete models, such as the CMSSM, this must be taken into account.

\begin{table}
\begin{tabular}{lcl}\toprule
Operators & \;\;LHC Signatures\;\; & Couplings\\\midrule
$LL\bar E$ & 2$\tau$\,4$\ell\,\ETmiss$ & $\lam_{12c} $\\[1mm]
 & 2$\ell\,\ETmiss$ & $\lam_{a3b,ab3,a33}$\\[1mm]
 & $\tau\,\ell$\,$\ETmiss$ & $\lam_{a33}$\\[1mm]
 & 2$\tau$\,$\ETmiss$ & $\lam_{a33}$ \\\midrule
$LQ\bar D$  & 4j\,2$\tau$\,2$\ell$ & $\lam'_{aij}$\\[1mm]
& 4j\,2$\tau$\,1$\ell\,\ETmiss$ & $\lam'_{aij}$\\[1mm]
& 4j\,2$\tau$\,$\ETmiss$ & $\lam'_{aij}$\\[1mm]
& 4j & $\lam'_{3ij}$\\
\midrule
$\bar U\bar D\bar D$  & 6j\,2$\tau$ & $\lambda''_{ijk}$\\[1mm]
& 6j\,$\ETmiss$ & $\lam''_{3jk}$~${}^\ddagger$\\[1mm]
 \bottomrule
\end{tabular}
\caption{LHC signatures for a stau LSP from the decay of two staus. $a,b,c=1,2$; $i,j,k=1,2,3$. We did not distinguish 
between top quarks, bottom quarks and the jets arising from the first two generations. These are analogous to 
\cref{tab:lqd-signatures,tab:udd-signatures}. The special case in the last line denoted with 
${}^\ddagger$ refers to scenarios where the stau LSP is light so that the decay into a top quark is kinematically 
disfavoured, leading to the dominance of the chargino-mediated final state $\nu_\tau d_3 d_j d_k$ as in \cref{eq:staudec}. } 
\label{tab:stau-LSP-LHC}
\end{table}

The complete listings of the stau-LSP LHC signatures for all RPV operators, including both $LL\bar E$ and $LQ\bar D$ 
as well as the two- and four-body stau decays, are given in Ref.~\cite{Desch:2010gi}, assuming a cascade originating 
from a squark. See Tables~II-IV, therein. We summarize the signatures of two decaying staus in \cref{tab:stau-LSP-LHC}.
Depending on the production mechanism of the two staus at the LHC, the signature will be accompanied by extra taus, missing $\ETmiss$ and jets.
We do not distinguish 
between top quarks, bottom quarks and the jets arising from the first two generations. These are dependent on the 
generation indices of the couplings and are analogous to \cref{tab:lqd-signatures,tab:udd-signatures}. 
We see that these scenarios always involve multiple $\tau's$ in the final state. For all $LL\bar E$ and $LQ\bar D$ couplings 
we can also have $\ETmiss$ arising from neutrinos. In the $LL\bar E$ case it is possible to have 1st and 
2nd generation charged leptons. Contrastingly, the $\bar U\bar D\bar D$ scenario contains challenging final states with 
only two taus and jets.

LHC analyses explicitly looking for $\tilde \tau$ LSPs are rarely performed. The only one we are aware of with prompt 
stau decays is Ref.~\cite{ATLAS:2012kr}, where the CMSSM with $\lam_{121}=0.032$ at $M_X$ is considered. It is 
assumed there that the stau always decays via a four-body decay into $\tau e\mu \nu_e$ and $\tau e e \nu_\mu$ with 
equal probability and the results are interpreted in the $M_{1/2}-\tan\beta$ plane, assuming $M_0=A_0=0$. Values of 
$M_{1/2}\lesssim 820~$GeV could be excluded for most values of $\tan\beta$. Note however that the RGE-induced 
operator $\lam_{233}$ and the corresponding partial two-body decay widths, $\Gamma_2$, of $\tilde \tau_1\to (\mu\nu_
\tau,\,\tau \nu_\mu)$ have not been taken into account. While there is only a mild dependence of the ratio of the total four-body 
versus two-body decay width, $\Gamma_4/\Gamma_2$, on $M_{1/2}$ \cite{Dreiner:2008rv}, the ratio, however, 
scales with $1/\tan^2\beta$. Therefore, only taking into account $\Gamma_4$ is a well-justified assumption in the low $
\tan\beta$ region, but the $\tilde \tau$-LSP results of Ref.~\cite{ATLAS:2012kr} for large $\tan\beta$ can, unfortunately, 
not be trusted. Note also that Ref.~\cite{ATLAS:2012kr} uses data from the 7~TeV run of the LHC. Because of the high 
occurrence  of $\tilde \tau$ LSPs in RPV models, we want to encourage the LHC collaborations to reanalyze these 
scenarios with more recent data.

In conclusion these models are basically not covered at all by LHC searches. Individual final states listed in 
\cref{tab:stau-LSP-LHC} have been searched for in other contexts, however. In particular, we would like to mention a 
recent analysis looking for multi-lepton final states and including up to two hadronic tau tags, with various signal regions requiring 
different amounts of $\ETmissx$ \cite{CMS:2017fdz}. This search
is dedicated to RPC models where pair-produced electroweak gauginos decay via an intermediate slepton or sneutrino down
to the lightest neutralino. Including the leptonically decaying tau modes, this search is therefore in principle sensitive to the 
$LL\bar E$ and $LQ\bar D$ signatures of \cref{tab:stau-LSP-LHC}. Unfortunately, so far no interpretation in terms of 
RPV models has been performed. Thus the corresponding mass sensitivity is also unknown. In \cref{sec:checkmate-tests}, 
we discuss the impact of $\tilde\tau$ LSPs on the LHC sensitivity, making use of non-dedicated searches and 
taking into account RGE-generated operators as well as four-body decays. In \cref{sec:checkmateboundstable}
we present our best estimate of the LHC mass bounds in these scenarios.

\section{LHC Coverage of RPV-induced Non-Standard LSP Decay Scenarios}
\label{sec:nonneutralinoLSP}

The standard scenario in the $\Lam_{\not R_p}$--CMSSM with small RPV couplings is the neutralino or stau LSP. 
Instead, as we saw in \cref{sec:determ-lsp}, for various large RPV couplings, we can also have the following 
LSPs: $\tilde e_R,\,\tilde\mu_R,\,\tilde\nu_e,\,\tilde\nu_\mu,\,(\tilde s_R,\tilde d_R),\,\tilde b_1,$ and $\tilde t_1$, 
\textit{cf.} \cref{tab:LSPs}. Here we discuss the phenomenology of these models and how they are 
covered by LHC searches. We also briefly summarize other related searches at the LHC with non-standard LSPs,
which do not occur in the $\LamRPV$--CMSSM.


\subsection{Slepton LSPs}

\subsubsection{Selectron or Smuon LSP}
For large $\lam_{ijc},\;c=1,2$, we can have either an $\tilde e_R$ or $\tilde\mu_R$ LSP, respectively. Since by construction the 
LSP couples to the dominant operators, we have the decays
\begin{eqnarray}
\tilde e_R^-&\to& \{\ell_i^-\nu_j,\,\nu_i\ell^-_j\},\;L_iL_j\bar E_1\,,\\[1mm]
\tilde \mu_R^-&\to& \{\ell_i^-\nu_j,\,\nu_i\ell^-_j\},\;L_iL_j\bar E_2\,.
\end{eqnarray}
At the LHC, where strong production usually dominates,\footnote{In the case of $LL\bar E$ operators, however, the 
most stringent bounds may arise from electroweakino pair-production, as we shall see in \cref{sec:checkmate-tests}. } 
the cascade decay will proceed as before to the neutralino, and the latter will decay to an on-shell slepton LSP
\begin{equation}
\tilde g\to q \tilde q^* \to q[\bar q \tilde\chi^0_1]\to q[\bar q (\tilde e_R^- e^+)]\,,
\label{eq:e-LSP-cascade}
\end{equation}
and analogously for the $\tilde\mu_R$ case. This model has not been explicitly constrained by existing analyses 
of LHC searches. However, we see from Eq.~(\ref{eq:e-LSP-cascade}) that the signature is identical to the case 
discussed in \cref{sec:neutralino-LSP-LLE}, with a neutralino LSP and a dominant $LL\bar E$ coupling. The 
only difference is that here the intermediate $\tilde e_R$ is on-shell. Thus the signatures are covered by the 
corresponding $L_iL_j\bar E_{1,2}$ searches discussed in \cref{sec:neutralino-LSP-LLE}, respetively. The 
experimental lower mass bounds can differ from \cref{tab:best-bounds-lle}, since the on-shell $\tilde e_R/
\tilde\mu_R$-LSP can lead to differing kinematic distributions.

\begin{table*}
\centering
\renewcommand\arraystretch{1.3}
\begin{tabular*}{0.9\textwidth}{l @{\extracolsep{\fill}} l cccc}
\toprule
Particle & Lower Bound [GeV] & $\bar U \bar D \bar D$ Coupling & Simpl. Model & Comment & Reference \\ 
\midrule 
$\tilde t$ & 405 & $\lam^{\prime\prime}_{312}$ & $\tilde{t}\rightarrow qq$ &445-510 GeV also excluded & 
\cite{ATLAS-CONF-2015-026}\\
	& $385$ & $\lam^{\prime\prime}_{3b3}$ & $\tilde{t}\rightarrow \bar d_b\bar b $ &  & \cite{Khachatryan:2014lpa}\\ \midrule
$\tilde g$ & 1440 & $\lambda''_{312}$ & $\tilde g \to t \tilde t$ &  $m_{\tilde t}=800\,$GeV	& \cite{Aaboud:2017dmy}\\
 & 1460 & $\lambda''_{3b3}$ & $\tilde g \to t \tilde t$ &  $m_{\tilde t}=700\,$GeV	& \cite{Aaboud:2017dmy}\\
\bottomrule
\end{tabular*}
\caption{Lower mass bounds for the case of a stop LSP discussed in \cref{sec:squark-LSPs-CMSSM}. Here $b=1,2$.}
\label{tab:Squark-LSP-searches}
\end{table*}

\subsubsection{Sneutrino LSP}
A $\tilde\nu_e$-LSP is obtained in the $\Lam_{\not R_p}$--CMSSM for large $\lam'_{1jk}$, $\{j,k\}\not=\{1,1\}$. A $\tilde\nu_\mu$-LSP 
is obtained for large $\lam'_{2jk}$. The LSP decays are then given as (plus charge conjugate)
\begin{eqnarray}
\tilde \nu_e^*&\to& d_j\bar d_k,\;\;\;L_1Q_j\bar D_k,\;\{j,k\}\not=\{1,1\}\,,\\[1mm]
\tilde \nu_\mu^* &\to& d_j\bar d_k,\;\;\;L_2Q_j\bar D_k\,.
\end{eqnarray}
The cascade decay at the LHC would proceed from for example a gluino down to the on-shell sneutrino, but through a 
neutralino, as with the charged sleptons above
\begin{equation}
\tilde g\to q \tilde q^* \to q[\bar q \tilde\chi^0_1]\to q[\bar q (\tilde \nu_{e,\mu}^* \nu_{e,\mu})]
\to q[\bar q (\{d_j\bar d_k\} \nu_{e,\mu})]
\,.
\label{eq:nu-LSP-cascade}
\end{equation}
Again for this specific model there is \textit{no} interpretation of LHC data within or outside of the $\Lam_{\not R_p}
$--CMSSM. However, in principle $L_{1,2}Q_j\bar D_k$ models have been searched for, as discussed in 
\cref{sec:NeutralinoLQD}. There the intermediate neutralino could also decay to a charged lepton:  $\tilde\chi^
0_1\to (\tilde \ell^+_i)^*\ell^-_i\to\ell^-_iu_j\bar d_k$, with comparable probability. The charged lepton in the final
state leads to a higher sensitivity than the diluted~$\ETmissx$ from the neutrino in Eq.~(\ref{eq:nu-LSP-cascade}).
For the sneutrino LSP scenario one might think the slepton three-body decay is suppressed. However in the 
$\LamRPV$--CMSSM the charged slepton and the sneutrino are nearly mass degenerate, as discussed in 
\cref{sec:determ-lsp}. Thus for the sneutrino LSP the associated charged slepton is the NLSP and the neutralino
the NNLSP. Thus the decay via the charged slepton is in fact two-body and if at all only marginally suppressed.
Therefore, the searches in \cref{sec:NeutralinoLQD} are approximately applicable, with the further difference that
the on-shell slepton and sneutrino will modify the kinematic distributions. Overall, these scenarios still need to be 
checked at the LHC, presumably via $\ETmiss$ searches.

Note, that other interesting scenarios can occur if we go beyond the CMSSM boundary conditions or/and consider more than 
one RPV operator at $M_X$. We thus mention searches for an $s$-channel production of tau-sneutrino LSPs, $\tilde\nu_\tau$, 
which decay further into leptons, assuming both $\lam'_{311}$ and one of $\{\lam_{132},\,\lam_{133},\,\lam_{232}\}$ to be sizable 
\cite{Aad:2015pfa,Khachatryan:2016ovq}, see also \cref{subsubsec:sneutrinoLSPoutsideCMSSM}. However, the excluded 
combination of masses and couplings is not yet competitive 
with the more stringent bounds from the non-observation of $\mu-e$ conversion in nuclei \cite{Faessler:1999jf,Bertl:2006up}. 


\subsubsection{Stau LSP with large $\lam_{ij3}$ or $\lam'_{3jk}$}
This scenario directly corresponds to the case (a) in \cref{sec:stauLSP}, just with a large coupling constant. The 2-body 
stau decays will be prompt and the previous discussion holds.

\subsection{Squark LSPs}
\label{sec:squark-LSPs-CMSSM}

As we see in \cref{tab:LSPs}, there are essentially three cases of squark LSPs in the $\Lambda_{\not R_p}$--CMSSM. 
\begin{enumerate}[leftmargin=*]
\item A first or second generation right-handed down-like squark with a large coupling $\lam''_{212}$, decaying to first or 
second generation quark jets. Experimentally this would lead to four jets, arising from the decay of two on-shell strongly 
interacting particles, \textit{i.e.} these are pair-produced dijet resonances.
\item A right-handed bottom squark decaying via $\lam''_{123,2b3}$, with the same signature as in the previous case.
\item A right-handed top squark decaying via $\lam''_{3jk}$. For $(j,k)=(1,2)$ this is again as in case 1. \!\!above. The 
pair-produced stops lead to two di-jet resonances. For $k=3$ there will be bottom quark jets in the final state.
\end{enumerate}
Of all three, only the third case, namely the top squark LSP, has been directly searched for in the context of RPV, 
assuming the couplings $\lam''_{312}$ \cite{Khachatryan:2014lpa,CMS-PAS-EXO-16-029,ATLAS-CONF-2016-084}, and 
$\lam''_{3b3}$ \cite{Khachatryan:2014lpa,Aad:2016kww,ATLAS-CONF-2015-026}. We assume this is due to the standard 
(RPC) lore that only a top squark can be particularly light. In addition, there has been a considerable effort in the 
theory community to point out interesting signatures, as well as search strategies for the light-stop scenarios, see for 
example Refs.~\cite{Evans:2012bf,Bai:2013xla}, many of which have been adopted by the experimental collaborations. 
The corresponding best lower bounds on the top squark mass are given in \cref{tab:Squark-LSP-searches}.
Included in the table are also two gluino lower mass bounds, obtained in top squark LSP models.

We note that the signature of the other two cases, pair-produced di-jet resonances, is identical to the signature probed in $\lam''_{312}$ \cite{Khachatryan:2014lpa,CMS-PAS-EXO-16-029,ATLAS-CONF-2016-084}. For the case of a very heavy gluino, the production cross section should also be identical, so that the bound can be carried over.

\begin{table*}
\centering
\renewcommand\arraystretch{1.3}
\begin{tabular*}{0.95\textwidth}{l @{\extracolsep{\fill}} l cccc}
\toprule
Particle & Lower Bound [GeV] & RPV Coupling & Simpl. Model & Comment & Reference \\ 
\midrule 
$\tilde \nu_\tau$ & 1280 (3300) & $\lam'_{311}\cdot\lam_{231,132}$   & $d_j\bar d_k\to\tilde\nu_\tau\to e^-\mu^+$ & $\lam'_{311}=\lam_{231,132}=0.01\,(0.2)$&   
    \cite{CMS-PAS-EXO-16-001,Khachatryan:2016ovq}\\
    & 2300 & $\lam'_{311}\cdot\lam_{231,132}$   & $d_j\bar d_k\to\tilde\nu_\tau\to e^-\mu^+$ & $\lam'_{311}=\lam_{231,132}=0.07$& \cite{Aaboud:2016hmk}\\
    & 2200 & $\lam'_{311}\cdot\lam_{133}$   & $d_j\bar d_k\to\tilde\nu_\tau\to e^-\tau^+$ & $\lam'_{311}=\lam_{133}=0.07$& \cite{Aaboud:2016hmk}\\
    & 1900 & $\lam'_{311}\cdot\lam_{233}$   & $d_j\bar d_k\to\tilde\nu_\tau\to \mu^-\tau^+$ & $\lam'_{311}=\lam_{233}=0.07$& \cite{Aaboud:2016hmk}\\
\midrule 
$\tilde u_{Lb}$ &1050$^*$ (1080$^*$) & $\lam'_{1bc}$  ($\lam'_{2bc}$) & $\tilde u_b\to ej\,(\mu j)$ &$\mathrm{Br}=1$, (3,2,{\tiny +}$\frac{2}{3}$)& 
                   \cite{Aad:2015caa,Khachatryan:2015vaa}\\
\midrule
$\tilde d_{Lb}$ & $625^*$   & $\lam^\prime_{ib3} $ & $\tilde d_{Lb}\to b\nu_i$ &  $\mathrm{Br}=1$, (3,2,-$\frac{1}{3}$) & \cite{Aad:2015caa}\\\midrule
$\tilde d_{Rc}$ & $900^*$ ($850^*$)  & $\lam^\prime_{1bc}\ (\lam^\prime_{2bc})$ & $\tilde d_c\to ej/\nu_e j\,(\mu j/\nu_\mu j)$ 
&  $\mathrm{Br}=0.5$ each,      ($\bar 3$,1,-$\frac{1}{3}$)  & \cite{Aad:2015caa}\\
    & $480^*$  & $\lam^\prime_{i3c}$  & $\tilde d_c\to b \nu_i$ & $\mathrm{Br}=0.5$, ($\bar 3$,1,-$\frac{1}{3}$)  & 
    \cite{Aad:2015caa}\\\midrule
$\tilde d_R$    &  650 (450)  & $\lambda''_{313}$  & $\tilde d \to \bar b \bar t$ & $m_{\tilde g}=1.4\,$TeV ($m_{\tilde g}=2\,$TeV)  & 
    \cite{Aaboud:2017dmy}\\
    &  570 (420)  & $\lambda''_{321}$  & $\tilde d \to \bar s \bar t$ & $m_{\tilde g}=1.4\,$TeV ($m_{\tilde g}=2\,$TeV)  & 
    \cite{Aaboud:2017dmy}\\    
    \midrule    
$\tilde t_L$ & $1100$ & $\lam^\prime_{133}\, (\lam^\prime_{233})$ & $\tilde{t}\rightarrow e^+b\, (\mu^+b)$ & & \cite{ATLAS-CONF-2015-015}\\
   & $740$ & $\lam^\prime_{333}$ & $\tilde{t}\rightarrow \tau^+ b$ & & \cite{Khachatryan:2014ura}\\
   & $1010^*\,(1080^*)$ & $\lam^\prime_{132}\ (\lam^\prime_{232})$ & $\tilde{t}\rightarrow e^+ j\,(\mu^+ j)$ & $\mathrm{Br}=1$, (3,2,{\tiny +}$\frac{2}{3}$)
   & \cite{Khachatryan:2015vaa}\\
\midrule
$\tilde b_R$ & $307$ & $\lam^{\prime\prime}_{3b3}$ & $\tilde{b}\to \bar t\bar d_b$ &  & \cite{Khachatryan:2016iqn}\\
   & $560^*$ & $\lam^{\prime}_{333}$ & $\tilde{b}\to t\tau^- $ & Br$=0.5$, ($\bar 3$,1,-$\frac{1}{3}$) & \cite{Khachatryan:2015bsa}\\
\midrule
$\tilde g$ & $650$ & $\lam^{\prime\prime}_{112}$ & $\tilde{g}\to uds$ &  &\cite{Chatrchyan:2013gia}\\
   & $835$ & $\lam^{\prime\prime}_{113}$ ($\lam^{\prime\prime}_{113}$)  & $\tilde{g}\to udb\; (csb)$ &  &\cite{Chatrchyan:2013gia}\\
   & $1360$ & $\lam^{\prime\prime}_{323}$ & $\tilde{g}\to tsb$ & $m_{\tilde q}=5\,$TeV &\cite{CMS-PAS-SUS-16-013} \\
   & $917\, (929)\, [874]$  & $\lam''_{abc} (\lam''_{ab3}) [\lam''_{3b3}]$ &$\tilde g\to 3q$& $m_{\tilde q}=5\,$TeV & \cite{ATLAS-CONF-2013-091}\\
\bottomrule
\end{tabular*}
\caption{Lower mass bounds on supersymmetric particles as the LSP decaying directly via an RPV operator. These are all 
\textit{not} $\Lam_{\not R_P}$--CMSSM scenarios. The bounds marked by an asterisk $^*$ are re-interpreted leptoquark 
scenarios. For the leptoquark searches, we have included the SU(3), SU(2), and U(1)$_{\mathrm{EM}}$ quantum numbers 
in the comment.}
\label{tab:LQ-searches}
\end{table*}


\subsection{Non-Neutralino LSP Outside the CMSSM}
Experimentally there are quite a few searches for sparticles directly decaying through an RPV operator. These 
correspond to scenarios where the sparticle at hand is the LSP. For completeness we briefly collect here the 
cases which are not possible within the CMSSM and which are therefore not listed in \cref{tab:LSPs}.


\subsubsection{Sneutrino LSP}
\label{subsubsec:sneutrinoLSPoutsideCMSSM}
At the LHC it is possible to produce a sneutrino on-resonance via an $L_iQ_j\bar D_k$ operator and for it to decay
via a separate $L_iL_j\bar E_k$ operator. This can lead to spectacular lepton flavor violating signatures 
\cite{Dimopoulos:1988fr,Feng:1997ru,Hewett:1998fu,Dreiner:2000vf,Dreiner:2012np,Moreau:2000bs,Bernhardt:2008jz}
\begin{equation}
d_j \bar d_k\to\tilde\nu_i\to\ell_l\ell_m\,,\qquad L_iQ_j\bar D_k \land L_iL_l\bar E_m\,.
\end{equation}
Experimentally this has been searched for by both \Cms and \Atlas 
\cite{Khachatryan:2016ovq,CMS-PAS-EXO-16-001,Aaboud:2016hmk}. We list the best lower sneutrino mass bounds in 
\cref{tab:LQ-searches}. The production cross section of the sneutrino is proportional to the $\lam'$ coupling squared, 
the bound is correspondingly sensitive. As we see, the bound ranges from 3.3 TeV for a substantial coupling, $\lam'_{311}
=0.2$, to 1.3 TeV for a modest coupling, $\lam'_{311}=0.01$. The bound quickly vanishes when reducing the size of the 
coupling further \cite{Dreiner:2012np}.

The experimental search results are given for a $\tilde\nu_\tau$ \cite{Aaboud:2016hmk}, but the opposite flavor lepton searches apply equally to 
the following coupling combinations which involve a $\tilde\nu_\mu$ propagator 
\begin{eqnarray}
e\mu: &&(\lam'_{211}\cdot\lam_{122})\,, \\
e\tau: &&(\lam'_{211}\cdot\lam_{123}),\,(\lam'_{211}\cdot\lam_{231} )\,,\\
\mu\tau: &&(\lam'_{211}\cdot\lam_{322})\,. 
\end{eqnarray}
We have disregarded the case of an $s$-channel $\tilde \nu_e$ as the required coupling combinations rely on $\lambda'_{111}$, on which 
there are very strict bounds, see \cref{tab:bounds1}.


\subsubsection{Squark LSP}
There have been many searches for squarks decaying directly via RPV operators \cite{Aad:2016kww,Khachatryan:2014ura,Khachatryan:2016iqn,Khachatryan:2015bsa,Aad:2015lea,ATLAS:2016yhq,CMS-PAS-SUS-16-013,Chatrchyan:2013fea,ATLAS-CONF-2015-015,ATLAS:2016mnt}, several focusing on the decay $\tilde{t}\rightarrow bs$ induced by 
the coupling $\lam^{\prime\prime}_{323}$, discussed in \cref{sec:squark-LSPs-CMSSM}. There have also been 
many leptoquark searches which can be directly interpreted as squark production followed by RPV decays via the $L_i
Q_j\bar D_k$ operators, see for example \cite{Aad:2015caa,Khachatryan:2015vaa,Khachatryan:2014ura,Chatrchyan:2012st,Khachatryan:2015bsa}. 
In particular there are three RPV scenarios corresponding directly to leptoquarks:
\begin{eqnarray}
\tilde u_{Lj}&\to& \ell^+_i+d_k\,,\qquad \;\;\;\;\;\;\;\;\;\mathrm{Br}=1\,, \label{eq:sup-L}\\
\tilde d_{Lj}&\to& \bar \nu_i+d_k\,,\qquad \;\;\;\;\;\;\;\;\;\;\mathrm{Br}=1\,, \label{eq:sdown-L}\\
\tilde d_{Rk}&\to& \ell^-_i+u_j/\nu_i+d_j\,,\quad \mathrm{Br}=0.5\,\mathrm{each,} \\
&&\qquad\qquad \qquad\qquad \;\;\mathrm{if\,}j\neq 3\,. \qquad\notag
\end{eqnarray}
If an up-like squark, $\tilde u_{Lj}$, is the LSP, for an $L_iQ_j\bar D_k$ operator it can only decay to a charged lepton 
and a down-like quark with a branching ratio of 1, \textit{cf.} Eq.~(\ref{eq:sup-L}). For a $\bar U_i\bar D_j\bar D_k$ 
operator it will decay to two jets, possibly including one $b$-jet. This latter case has only been considered for top 
quarks, $i=3$. For the former case the best bounds are given in \cref{tab:LQ-searches}, with a lower mass bound 
of about 1\,TeV. 

On the other hand, a left-handed down-like squark, $\tilde d_{Lj}$, can only decay via a neutrino, leading to\ $\ETmissx$, 
\textit{cf.} Eq.~(\ref{eq:sdown-L}). In the table listings we interpret an \Atlas leptoquark search as the decay of a $\tilde d_
{Lb},\,b=1,2$, to a bottom quark resulting in a weaker lower mass bound of 625\,GeV~\cite{Aad:2015caa}. The right-handed 
down-like squark has two possible decay modes, one involving a charged lepton and one involving a neutrino. Combining 
the two often leads to stricter bounds. For jets from the first two generations $\lam'_{1bc,2bc},\,b,c=1,2$, the lower experimental 
bound is about 900\,GeV. For the case $\lam'_{i3c},\,c=1,2,$ the charged lepton mode involves a top quark, which would be a 
separate search. The neutrino mode alone leads to the much weaker bound of only 480\,GeV. These are again all 
reinterpreted leptoquark searches.

The top squark LSP has been more widely considered in the literature, as it is naturally lighter than the other squarks, 
even for RPC, see for example \cite{Evans:2013uwa,Chun:2014jha,Marshall:2014kea}. The decays via $\bar U\bar D
\bar D$ operators  are discussed in \cref{sec:squark-LSPs-CMSSM}. The decays via $LQ\bar D$ operators are 
just special cases of the decay of up-like squarks, Eq.~(\ref{eq:sup-L}), and the bounds are similar, around 1100 GeV. 
The case $\lam'_{333}$ leads to a final-state tau and thus weaker bounds, around 750\,GeV.

In the $\Lam_{\not R_p}$--CMSSM, for $\lam''_{3b3}$, the top squark can be the LSP, but not the bottom squark.
Nevertheless the direct decay of a bottom squark via $\lam''_{3b3}$ with 100\% branching ratio was searched for giving a 
lower mass bound of 307\,GeV \cite{Khachatryan:2016iqn}. A leptoquark search was reinterpreted as the direct decay of a 
right-handed bottom squark via the operator $L_3Q_3\bar D_3$ to a top quark and a tau with a branching ratio of 50\%. 
In this case the lower mass bound is 560\,GeV \cite{Khachatryan:2015bsa}.


\subsubsection{Gluino LSP}
As we saw in \cref{tab:LSPs}, a gluino LSP is not dynamically generated in the $\Lam_{\not R_p}$--CMSSM. All 
the same we briefly discuss this scenario here, as there are several LHC searches for such models. Depending on 
the dominant operator, a gluino LSP decays as
\begin{eqnarray}
\tilde g\to \left\{
\begin{array}{ll}
q\bar q\{ \ell_i\nu_j\bar \ell_k,\,\nu_i\ell_j\bar \ell_k\}\,,\;\;\;\; & L_iL_j\bar E_k\,,\\[1.5mm]
\{ \ell_iu_j\bar d_k,\,\nu_id_j\bar d_k\}\,, & L_iQ_j\bar D_k\,,\\[1.5mm]
u_i d_j d_k \,, & \bar U_i \bar D_j\bar D_k\,.
\end{array} \right.
\label{eq:gluino-LSP-decays}
\end{eqnarray}
The first decay proceeds via a virtual squark and a virtual neutralino. In the second and third case, the gluino decays 
via a virtual squark, which couples directly to the relevant operator. Of these three scenarios only the last one has 
been investigated at the LHC \cite{ATLAS:2012dp,Chatrchyan:2013gia,Chatrchyan:2013fea,Aad:2015lea,Khachatryan:2016iqn,ATLAS-CONF-2016-057,CMS-PAS-SUS-16-013,ATLAS-CONF-2013-091}.

However the first case, $LL\bar E$, leads to identical signatures as in \cref{sec:neutralino-LSP-LLE}, the only 
difference is that now the intermediate neutralino is virtual. The second case, $LQ\bar D$, is novel, although very 
similar to the electroweak production in Section~10 of Ref.~\cite{Khachatryan:2016iqn}. There the pair production of 
neutralinos is investigated, followed by their three-body RPV decay. Here one should consider the pair production 
of gluinos, which has a significantly higher cross section, which should lead to stricter lower mass bounds.

The best bounds for the third case in Eq.~(\ref{eq:gluino-LSP-decays}) are listed in \cref{tab:LQ-searches}. 
The weaker bounds in the first two rows for the gluino were obtained with $\sqrt{s}=8\,$TeV data. Nevertheless 
the $tbs$ search is the most sensitive channel. \Atlas has several searches for this scenario 
\cite{ATLAS:2012dp,Aad:2015lea,ATLAS-CONF-2016-057},  however, they always allow more than one coupling 
to be non-zero, often even all $\lam''_{ijk}$ with equal value. It is again not clear how to interpret the resulting 
bounds.

\section{Testing the RPV--CMSSM with CheckMATE}
\label{sec:checkmate-tests}

For the remainder of this paper we are interested in the sensitivity of the LHC with respect to the $\Lam_{\not R_p}
$--CMSSM, \textit{i.e.} to a complete supersymmetric model. In particular, we are interested in how the presence of 
$R$-parity violating operators affects the well-known results for the $R$-parity conserving CMSSM 
\cite{Bechtle:2015nua,Buchmueller:2013rsa,Aad:2015iea,Chatrchyan:2014goa}. For this we shall use the program 
\Checkmate \cite{Drees:2013wra,Kim:2015wza,Dercks:2016npn}. As we saw in \cref{sec:neutralinoLSP,sec:stauLSP,sec:nonneutralinoLSP}, 
the LHC experiments mainly set bounds on simplified supersymmetric $R$-parity 
violating models. They set little or no bounds on the complete $\Lam_{\not R_p}$--CMSSM model. We here use \Checkmate to recast 
\Atlas and \Cms searches and thus set bounds on the various $\Lam_{\not R_p}$--CMSSM models.

\subsection{Method}

The program \Checkmate automatically determines if a given parameter point of a particular model beyond the 
Standard Model (BSM) is excluded or not by performing the following chain of tasks. First, the Monte Carlo 
generator \Madgraph \cite{Alwall:2014hca} is used to simulate proton proton collisions. The resulting parton 
level events are showered and hadronized using \Pythiaeight \cite{Sjostrand:2014zea}. The fast detector 
simulation \Delphes \cite{deFavereau:2013fsa} applies efficiency functions to determine 
the experimentally accessible final state configuration, including the determination of the jet spectrum using \Fastjet 
\cite{Cacciari:2011ma,Cacciari:2005hq}. Afterwards, various implemented analyses from \Atlas and \Cms designed to identify 
different potentially discriminating final state topologies are used.

Events which pass well-defined sets of constraints are binned in signal regions for which the corresponding prediction for the Standard 
Model and the number of 
experimentally observed events are known. By comparing the predictions of the Standard Model and the user's BSM model of 
interest to the experimental result using the CL$_{\text{S}}$ prescription \cite{Read:2002hq}, \Checkmate{} concludes if the input 
parameter combination is excluded or not at the 95\% confidence level. For more information we refer to 
Refs.~\cite{Drees:2013wra,Kim:2015wza,Dercks:2016npn}.

\subsubsection{Model Setup}
For the proper description of the RPV Feynman rules in \Madgraph{}, we take the model implementation from 
\Sarah{}, which we already used in \cref{sec:rges}, and export it via the UFO format \cite{Degrande:2011ua}. 
For a given set of $\Lam_{\not R_p}$--CMSSM parameters, we make use of the respective \Spheno{} libraries 
created from the same \Sarah{} model used to determine the low energy particle spectrum, the mixing matrices 
and the decay tables. We calculate the SUSY and Higgs masses including RPV-specific two-loop corrections 
\cite{Goodsell:2014bna,Goodsell:2015ira,Goodsell:2016udb} which are particularly important for light stops 
\cite{Dreiner:2014lqa}. As discussed in \cref{sec:RGEinduced}, four-body decays of the stau can be dominant 
and lead to important experimentally accessible final states. In regions where this occurs, see for instance in 
\cref{sec:checkmate:udd}, the four-body stau decays have been determined using \Madgraph.

\subsubsection{Monte Carlo Simulation}
\label{sec:checkmate:montecarlo}
In $\Lam_{\not R_p}$--CMSSM parameter regions where the entire SUSY spectrum is kinematically accessible at the LHC, \textit{i.e.}\ 
with masses  at or below $\mathcal{O}($1 TeV$)$, there exists a plethora of possible final state configurations. To maintain computational 
tractability in our study, we applied the following list of simplifying assumptions:
\begin{itemize}[leftmargin=*]
\item We include only two-body supersymmetric final state production: $p p \to A B+X_\mathrm{soft}$, and require both 
supersymmetric particles, $A$ and $B$, to be produced on-shell. Note that in RPC supersymmetry,  additional hard QCD 
radiation, \textit{i.e.} $pp \to A B j$, is important in parameter regions with highly degenerate spectra due to the resulting 
kinematic boost of the decay products, see e.g. Ref.~\cite{Dreiner:2012gx}. However, due to the instability of the LSP in 
the $\Lam_{\not R_p}$--CMSSM, this additional boost is not needed and therefore this final state is not expected to 
contribute sizably to the final constraining event numbers.
\item We do not consider final state combinations which are strongly suppressed by the  relevant parton density 
distributions and/or which only exist in RPV supersymmetry. Most importantly, this excludes flavor-off-diagonal 
combinations of  ``sea''-squarks (we clarify the meaning below) or squark-slepton combinations. 
\item We include in our simulations production processes, which can only proceed via the electroweak interactions, \textit{i.e.} the 
production of sleptons, electroweak gauginos and the mixed production of electroweak gauginos and squarks or gluinos. However, 
in the case of electroweak gauginos we only include the production of the two lightest neutralinos and the lightest chargino, \textit{i.e.}\ the 
dominantly bino and wino states in a CMSSM setup. We expect no sizeable contributions from the ignored Higgsinos, since these are 
typically significantly heavier and therefore have negligible production rates in comparison with the lighter winos. Similarly, we do not 
include flavor-off-diagonal slepton combinations and mixed electroweak gaugino-squark production with ``sea''-squarks. 
\item With the above considerations, the resulting set of  production channels that we consider are listed below: 

Strong processes:
\begin{itemize}[leftmargin=*]
\item $\tilde g \tilde g$
\item $\tilde g \tilde q_{V}^{(*)}$ 
\item $\tilde q_{V}^{(*)} q_{V}^{(*)}$ (all combinations)
\item $\tilde q_{S}\tilde q_{S}^{*}$ (only flavor-diagonal)
\end{itemize}
Electroweak processes:
\begin{itemize}[leftmargin=*]
\item $\tilde \chi  \tilde \chi$ [all (non-Higgsino) combinations]
\item $(\tilde \ell_{L}, \tilde \ell_{R}, \tilde \nu_{\ell, L})  (\tilde \ell_{L}^* , \tilde \ell_{R}^* , \tilde \nu_{\ell, L}^*)$ (only flavor-diagonal)
\end{itemize}
Mixed processes:
\begin{itemize}[leftmargin=*]
\item $\tilde g \tilde \chi$
\item $\tilde q_{V}^{(*)} \tilde \chi$ (all combinations)
\end{itemize}
Here, $\tilde q_V$ refers to the superpartners of the light quarks: $\tilde u_{L,R}, \tilde d_{L,R}$ and $\tilde s_{L,R}$, 
while $\tilde q_S$ refers to the remaining 
squarks $\tilde c_{L,R}, \tilde b_{1,2}$ and $\tilde t_{1,2}$. Furthermore, $\tilde \chi$ subsumes the two lightest neutralinos $\tilde \chi_{1, 2}
^0$ and the lightest chargino $\tilde \chi_1^\pm$. 
\item Decays of supersymmetric particles are performed within \Pythiaeight{}, using the information from the 
decay table determined in \Spheno{}. This ignores any potential spin-dependent information, which could be 
relevant when performing the proper matrix-element calculation. However, as we do not assume spin-effects 
to be important here, we take the computationally faster approach of using decay tables. 
\item 
To take into account the sizable contributions from higher order QCD effects in the production cross section, we multiply the leading-order production cross-sections taken from \Madgraph{} with the next-to-leading-logarithm K-factors determined by \texttt{NLLFast} 
\cite{Beenakker:1996ch,Beenakker:1997ut,Kulesza:2008jb,Kulesza:2009kq,Beenakker:2009ha,Beenakker:2010nq,Beenakker:2011fu}. This tool interpolates gluino- and squark-mass-dependent higher-order cross sections for all ``strong Processes'' listed above. \texttt{NLLFast} assumes degenerate first- and second-generation squark sector where we use the median of the squark masses for the calculation. Note that this degeneracy is present in models with small $\LamRpV$ but not necessarily if $\LamRpV$ is large.\footnote{For the only benchmark case with large $\LamRpV$ we consider, the squark sector is not degenerate. However, as in this case stop production dominates, we can still use the NLLFast K-factors for rescaling the cross sections.} Stops and sbottoms are always treated separately and obtain individual K-factors. The consideration of higher order effects for the remaining processes is computationally far more involved as tools like {\tt PROSPINO} \cite{Beenakker:1996ed} need to perform the full NLO calculation. These effects are however expected to be significantly smaller compared to the strong production processes and thus we neglect them here.
\end{itemize}

\newcommand{\includeresults}[1]{\includegraphics[width=1.25\columnwidth]{figures/#1}}
\newcommand{\includewideresults}[1]{\onecolumngrid \includegraphics[width=1.99\columnwidth]{figures/#1} \twocolumngrid}
\newcommand{\includetwinresults}[2]{ \onecolumngrid \includegraphics[width=0.45\textwidth]{figures/#1} \includegraphics[width=0.45\textwidth]{figures/#2} \twocolumngrid}
\newcommand{\inlineboldcomment}[1]{  \onecolumngrid  \begin{center}    \textbf{#1}  \end{center}  \twocolumngrid }

\begin{figure*}
\centering
\includewideresults{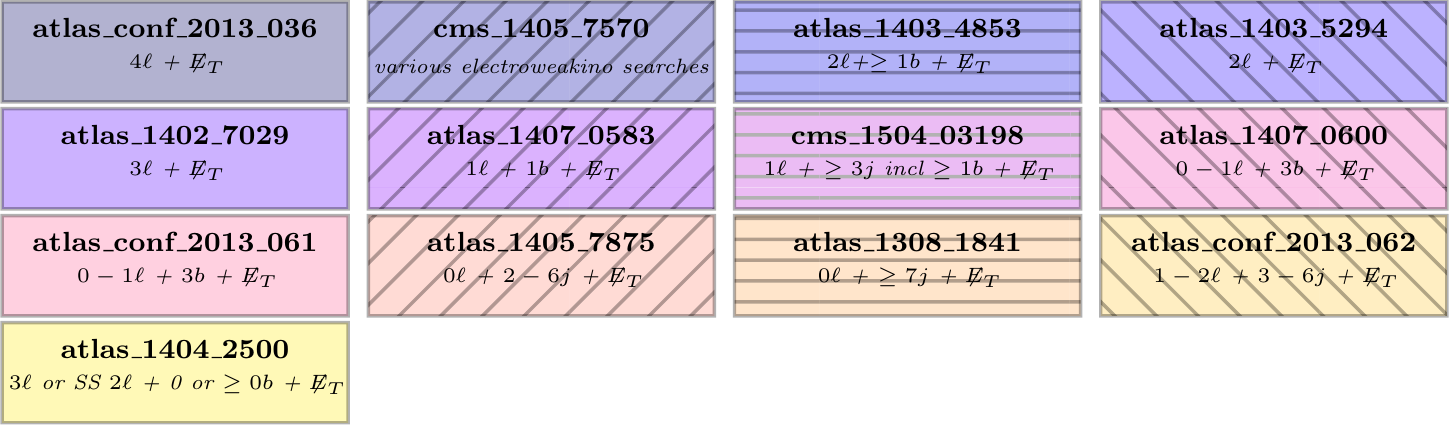}
\caption{This legend shows the different colors/shadings presented on the right-hand side of the following figures. 
Each colored box contains a bold label, which corresponds to the name under which the respective reference is 
listed within \Checkmate{}, furthermore in small type, each box contains a brief description of the analyzed signature.}
\label{fig:checkmate:analysislegend}
\end{figure*}


\subsubsection{Incorporated Analyses}
\Checkmate{} provides a large set of implemented \texttt{ATLAS} and \texttt{CMS} results from both the $\sqrt{s} = 8$ and $13$~TeV runs 
of the LHC. These analyses target a large variety of possible final states which typically appear in theories beyond the Standard Model. The 
vast majority, however, are designed to target RPC supersymmmetry. This implies cuts which require significant amounts 
of missing transverse momentum and/or highly energetic final state objects, namely leptons or jets. As many of the most prominent 
decay chains in the $\Lam_{\not R_p}$--CMSSM indeed correspond to these signatures, it is therefore interesting to determine the relative 
exclusion power of these tailored analyses in comparison to the RPC--CMSSM.

In this study we consider all $\sqrt{s} = 8$ TeV LHC analyses implemented in \texttt{CheckMATE 2.0.1}. For a full, 
detailed list we refer to the documentation in Refs.~\cite{Drees:2013wra,Kim:2015wza,Dercks:2016npn} and the tool's 
website.\footnote{\url{http://checkmate.hepforge.org}} We discuss the target final states of the relevant analyses in 
more detail below, alongside our results. Some final states have been reanalyzed and the corresponding bounds have 
been updated with new LHC results taken at $\sqrt{s} = 13$ TeV center-of-mass energy. For our purpose of comparing 
the \emph{relative} exclusion power when going from an RPC to an RPV scenario, using the $\sqrt{s} = 8$ TeV analysis 
set has the advantage of covering a much larger variety of final states. 

To set the limit, \Checkmate tests all signal regions in all selected analyses, determines the one signal region with the 
largest \emph{expected} sensitivity and checks if the corresponding \emph{observed} result of that signal region is 
excluded at 95\% C.L.\ or not. In the following, the analysis which contains this limit-setting signal region is referred to as 
the ``most sensitive analysis'' for a given 
$\Lam_{\not R_p}$--CMSSM parameter point. Due to the lack of information about correlations in systematic uncertainties between 
different signal regions, \Checkmate is currently incapable of combining information from different signal regions.\footnote{The statistical combination of signal regions for \Checkmate{} is work in progress \cite{demBelknerSebastianSeineArbeit}.}

The list of analyses we employ are shown in \cref{fig:checkmate:analysislegend}. In bold is the name under which the 
analysis is listed in \Checkmate. Underneath in small italics type we briefly denote the physical signature. Here $\ell$ refers to a 
charged lepton,  $\ETmiss$ refers to missing transverse energy.  $j$ refers to a jet in the final state, $b$ specifically a $b$-jet. The references for the 
analyses are given in the \Checkmate documentation. The boxes of the analyses carry different colors and hatchings. This is 
employed in the later exclusion plots, to show which analysis within \Checkmate is the most sensitive.

\begin{figure*}
\centering
\includetwinresults{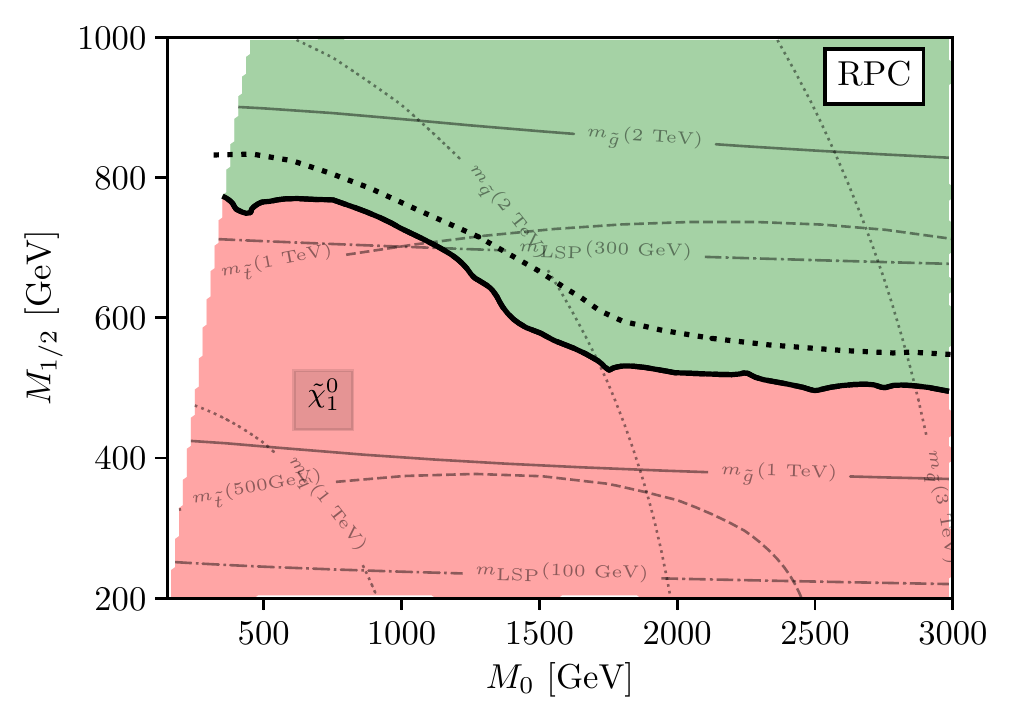}{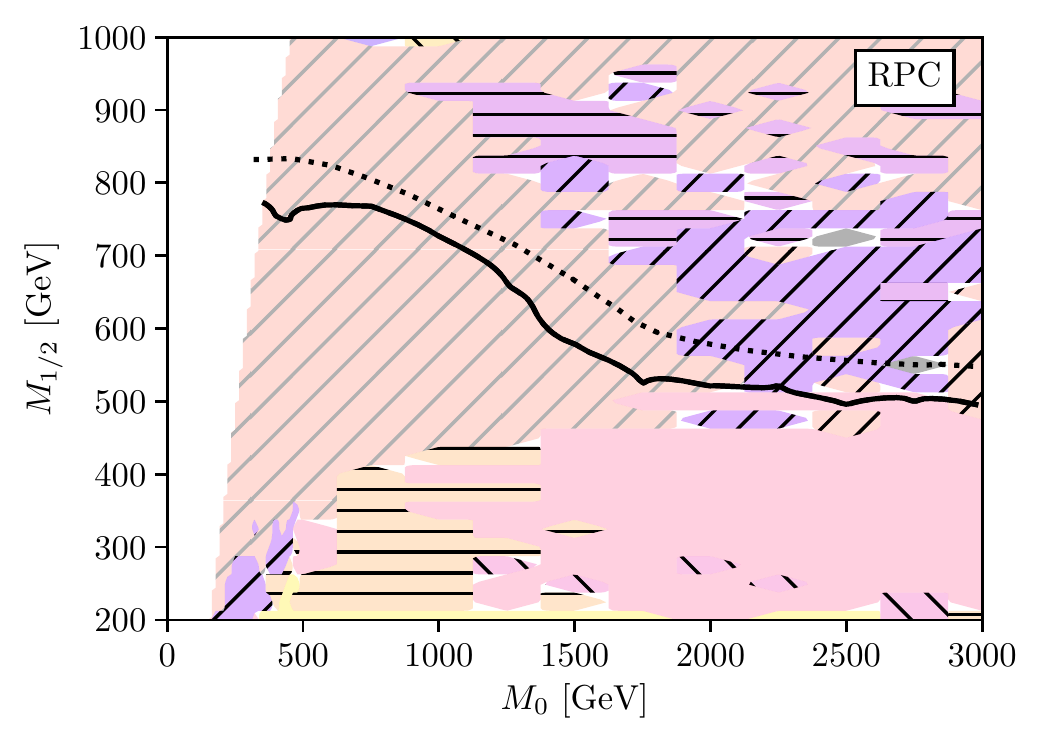}
\caption{LHC exclusion limits in the $M_0$--$M_{1/2}$ plane determined by \Checkmate{} (black solid line) using the 
RPC--CMSSM with the remaining model parameters set to $\tan\beta=30$, $\mu > 0$ and $A_0 = -2 M_0$. On the 
left-hand side, the red region below the solid line is excluded, the green region is allowed by 
\texttt{CheckMATE}. The black dotted line denotes the corresponding \Atlas bound from Ref.~\cite{Aad:2015iea}. The discrepancy is discussed in 
the text. The left-hand figure further contains in light gray the iso-mass contours of the gluino (solid), the squarks 
(dotted), the lightest stop (dashed) and the LSP (dot-dashed), respectively. The white area on the left for which no results are shown 
corresponds to the cosmologically excluded region with a $\tilde \tau$ LSP. On the right-hand side, we show for each 
region the most sensitive LHC analysis, according to the legend in \cref{fig:checkmate:analysislegend}.  
}
\label{fig:checkmate:RPC}
\end{figure*}

\subsubsection{Scanned Parameter Regions}
Even though one of the appealing features of the CMSSM is the  small number of free parameters compared to other 
supersymmetric theories, we still need to fix certain degrees of freedom in order to be able to show results in an 
understandable 2-dimensional parameter plane. As the masses of the supersymmetric particles will be one of the most 
important variables when it comes to the observability of a model realization at the LHC, we  show results in the $M_0$--$
M_{1/2}$ plane. To be specific we scan over the parameter range
\begin{equation}
M_0\in[0,\,3000]\,{\rm GeV},\quad M_{1/2}\in[200,\,1000]\,{\rm GeV}\,.
\end{equation}
For better comparison, we choose the remaining model parameters as in the RPC--CMSSM \texttt{ATLAS} 
analysis in Ref.~\cite{Aad:2015iea} where $\tan \beta$ is fixed to a relatively large value of $30$ while $A_0$ is set via the 
standard formula $A_0 = -2 M_0$ which ensures a realistically large value of the lightest neutral CP-even Higgs boson 
mass. The sign of the $\mu$ parameter is fixed to be positive, to avoid further tension with $(g-2)_\mu$ \cite{Martin:2001st}. 
Allowing any number of RPV-operators to have non-vanishing values would yield an unmanageable set of possible 
scenarios to study. We therefore restrict ourselves to cases where only one of the many operators has a non-zero value at the unification scale. 
In order to directly compare our results to the RPC--CMSSM, we use a small RPV coupling at the GUT scale, $\Lam_{\not 
R_p}|_{\rm GUT}=0.01$. This essentially mimics the RPC mass spectrum, but allows for the prompt decay of the LSP. We 
comment on possible effects of increasing the RPV coupling in \cref{sec:checkmate:other}.


\subsection{Results}
In this section we show the results of the scans. Throughout the analysis of these results, the principle questions 
which we seek to answer are the following:
\begin{enumerate}[leftmargin=*]
\item To what extent do the existing \Atlas and \Cms analyses, which largely focus on RPC 
supersymmetry, exclude the parameter space of the $\Lam_{\not R_p}$--CMSSM?
\item Does breaking $R$-parity weaken the bounds of the RPC--CMSSM due to a gap in the coverage of possible 
final states?
\begin{enumerate}[leftmargin=*]
\item If the answer is \emph{yes}, how could these gaps be closed?
\item Alternatively, if the answer  is \emph{no}, in the cases where the bounds become stronger, which of the effects 
mentioned in previous sections lead to this result?
\end{enumerate}
\end{enumerate}

When presenting our results, we show in the figures for each parameter point, which \Checkmate analysis is the 
most sensitive. We do this by using the color code of \cref{fig:checkmate:analysislegend}.

\subsubsection{$R$-parity Conserving CMSSM}
We start with a short discussion of the RPC--CMSSM in \cref{fig:checkmate:RPC}. On the left we denote in the 
$M_0$--$M_{1/2}$ plane by the thick, solid black line the 95\%-CL exclusion range we obtained using \Checkmate 
for this model. Thus below the curve, in red is the excluded parameter area. Above the curve, in  
green is the allowed area. The remaining CMSSM parameters have been set to $\tan\beta=30$, $\mu > 0$ and $A_0 
=-2 M_0$. In light gray we present supersymmetric mass isocurves for the LSP (dot-dashed), the first two 
generation squarks (dotted), the gluino (solid), and the lightest stop (dashed). The white region on the far left at low $M_0$ results in a 
$\tilde\tau$-LSP, which is not viable phenomenologically if $R$-parity is conserved: as there 
is no possible decay channel for the stau, these regions result in stable charged particles which e.g.\ spoil big bang 
nucleosynthesis \cite{Kohri:2009mi}. This is why here and in the following, we do not show any RPC results in the 
stau-LSP region. We do include them in the RPV cases.
%
%

In the right plot, we show in the $M_0$--$M_{1/2}$ plane, which analysis implemented in \Checkmate is most sensitive at 
a given parameter point or region. We use the color and hash code of \cref{fig:checkmate:analysislegend}. Thus for 
example the point ($M_0=1000\,$GeV, $M_{1/2}=500\,$GeV) is excluded by the analysis denoted 
atlas\texttt{\_}1405\texttt{\_}7875 \cite{Aad:2014wea} in \Checkmate.

The most sensitive analyses target either a 0-lepton multijet ({atlas\texttt{\_}1405\texttt{\_}7875}), or  a $\geq 3b$-jet final 
state ({atlas\texttt{\_}conf\texttt{\_}2013\texttt{\_}061}, Ref.~\cite{ATLAS-CONF-2013-061}), both requiring a significant 
amount of missing transverse momentum. The former final state is especially sensitive when light gluinos decay into jets 
via on-shell squarks and therefore -- as can be seen in our results -- covers the low $M_{0}$ region where the squarks are 
relatively light. On the contrary, the latter targets stop and gluino pair 
production, the dominant modes for large $M_0$, which can produce $4$ $b$-jets due to the resulting top quarks in the final decay chain.

In this particular set of plots, see \cref{fig:checkmate:RPC}, we also show the nominal \Atlas exclusion limit taken from
Ref.~\cite{Aad:2015iea} as an additional, black dotted line. The 
discrepancies 
arise due to 
the  \Atlas result including a statistical combination of the orthogonal sets 
of $0$ and $1\ell$ signal regions which \Checkmate{} cannot perform.
Apart from this 
combination, the detailed, analysis-dependent results given in Ref.~\cite{Aad:2015iea} match our determination of the 
respective most sensitive analysis in this model. 


\subsubsection{$LL\bar E$, \LamCMSSM}
\label{sec:checkmate:lle}

\begin{figure*}
\includetwinresults{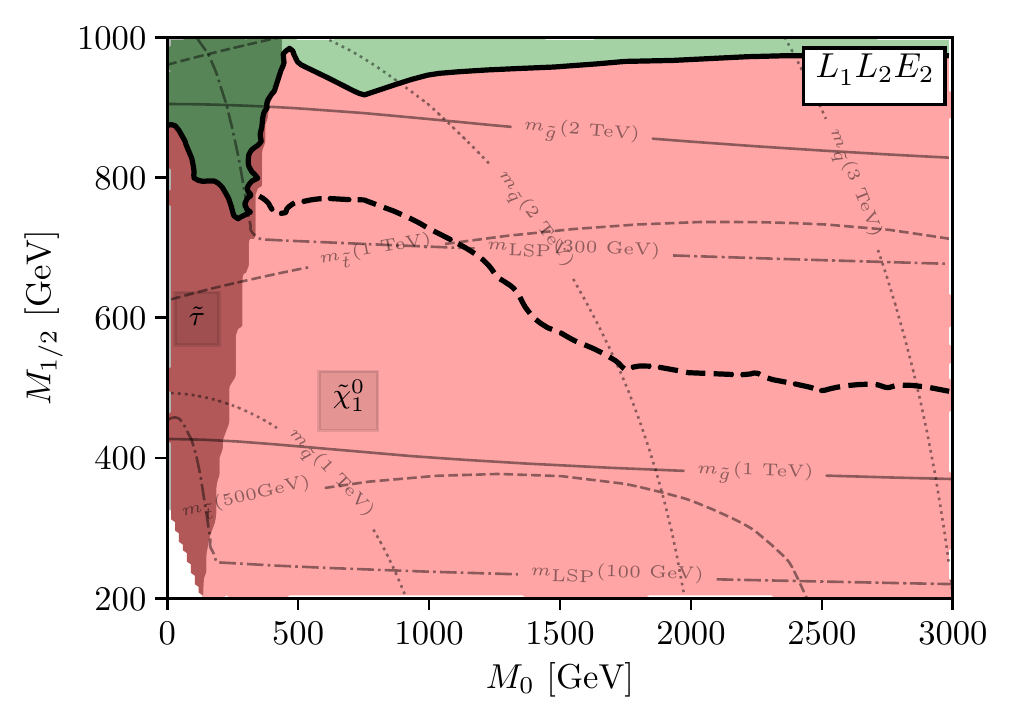}{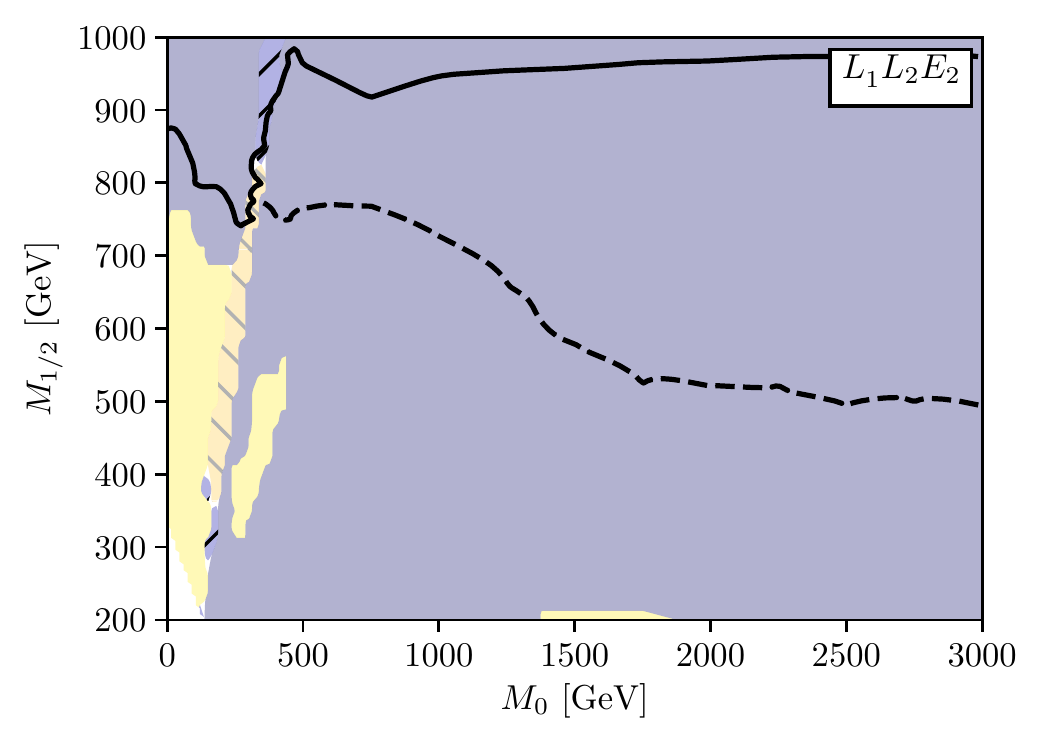}
\includetwinresults{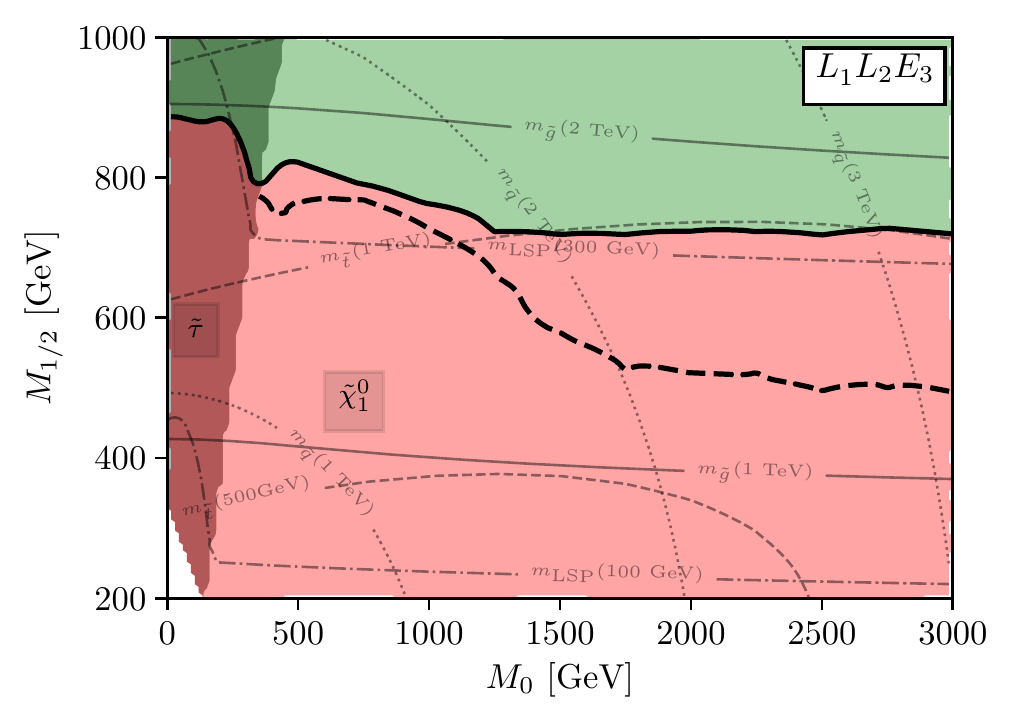}{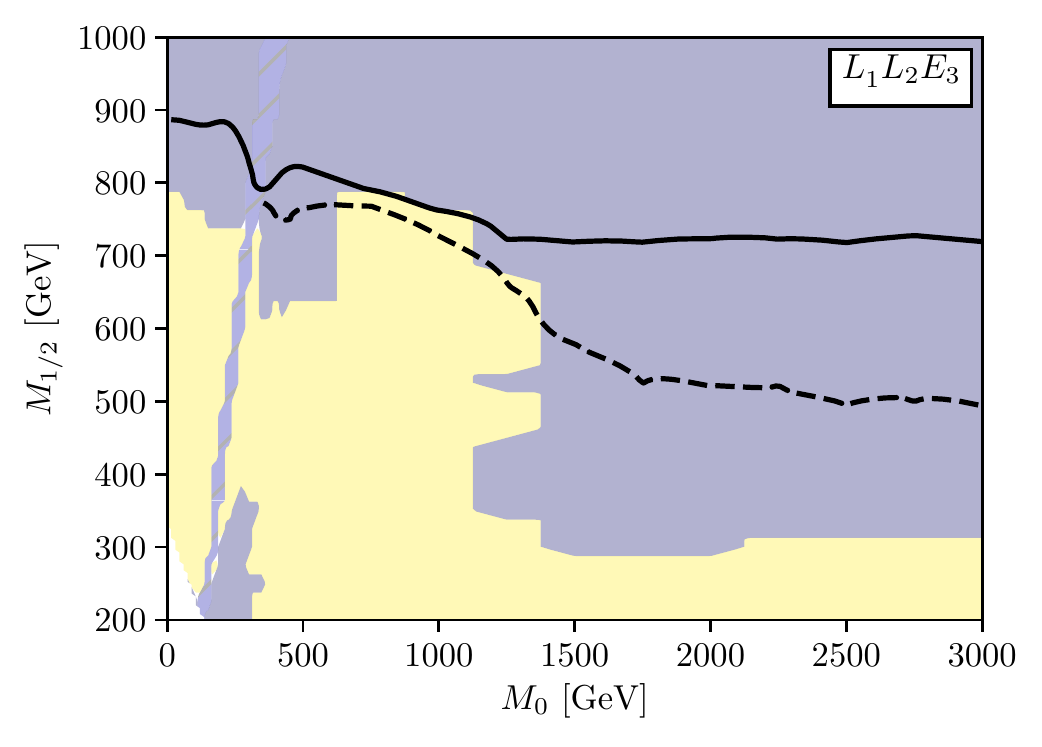}
\caption{Interpreting the LHC experimental searches as exclusion regions of the $LL\bar E$, $\Lam_{\not R_p}$--CMSSM 
in the $M_0$--$M_{1/2}$-plane using \Checkmate, and focusing on the cases $\lam_{122}$ and $\lam_{123}$. The other 
CMSSM parameters are as in \cref{fig:checkmate:RPC}. In the upper left hand plot, corresponding to a $\lam_{122}$ 
model, regions below the thick solid black line are excluded, and colored in red. Parameter regions above the thick black 
solid line are allowed by \Checkmate and colored in green. The dark red and dark green regions have a $\tilde\tau$ LSP, 
which is viable in the $\Lam_{\not R_p}$--CMSSM. The light red and light green regions have a $\tilde\chi^0_1$ LSP.  The 
light gray iso-mass curves are as in \cref{fig:checkmate:RPC}. The RPC--CMSSM exclusion line of 
\cref{fig:checkmate:RPC} (thick solid black curve there) is shown here as a thick dashed black curve for comparison. 
It does not extend into the $\tilde\tau$-LSP region, as that is not viable in the RPC--CMSSM. The figure on the upper right 
shows in thick solid black the \Checkmate exclusion from the upper left plot, as well as the RPC exclusion from 
\cref{fig:checkmate:RPC}. It furthermore shows which LHC analysis implemented in \Checkmate is most sensitive, at a 
given parameter region, using the color code of \cref{fig:checkmate:analysislegend}. The lower two plots are as the 
upper except for turning on $\lam_{123}$ instead of $\lam_{122}$. The small white area at very low $M_0$ and $M_{1/2}$ is where the lightest stau becomes 
tachyonic.}
\label{fig:checkmate:LLE122and123}
\end{figure*}

\begin{figure*}
\includetwinresults{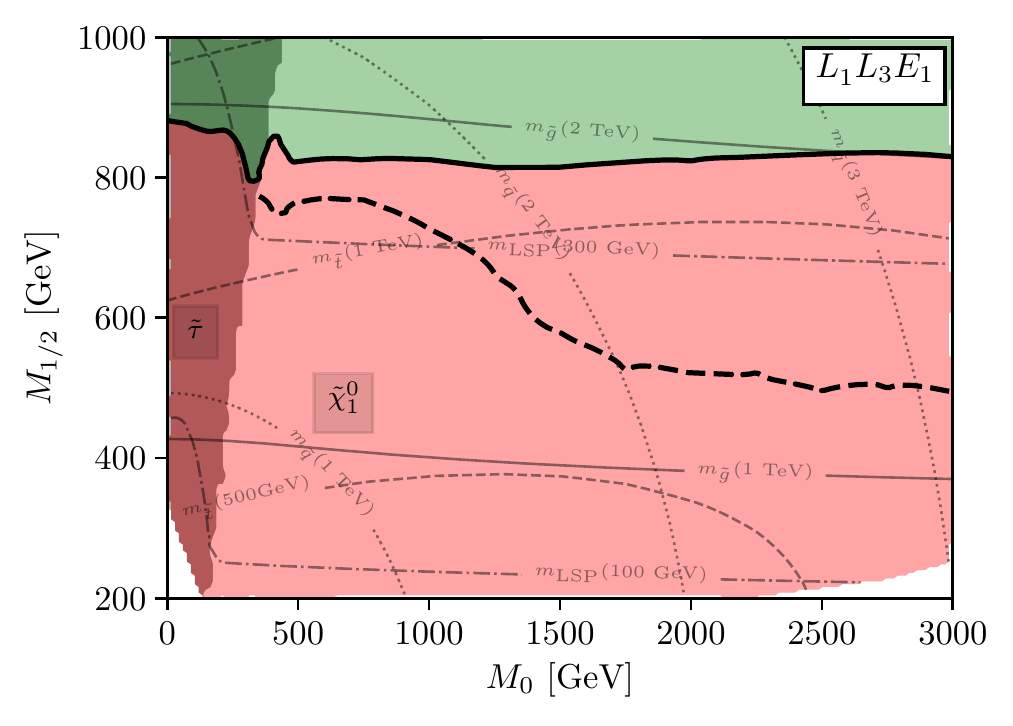}{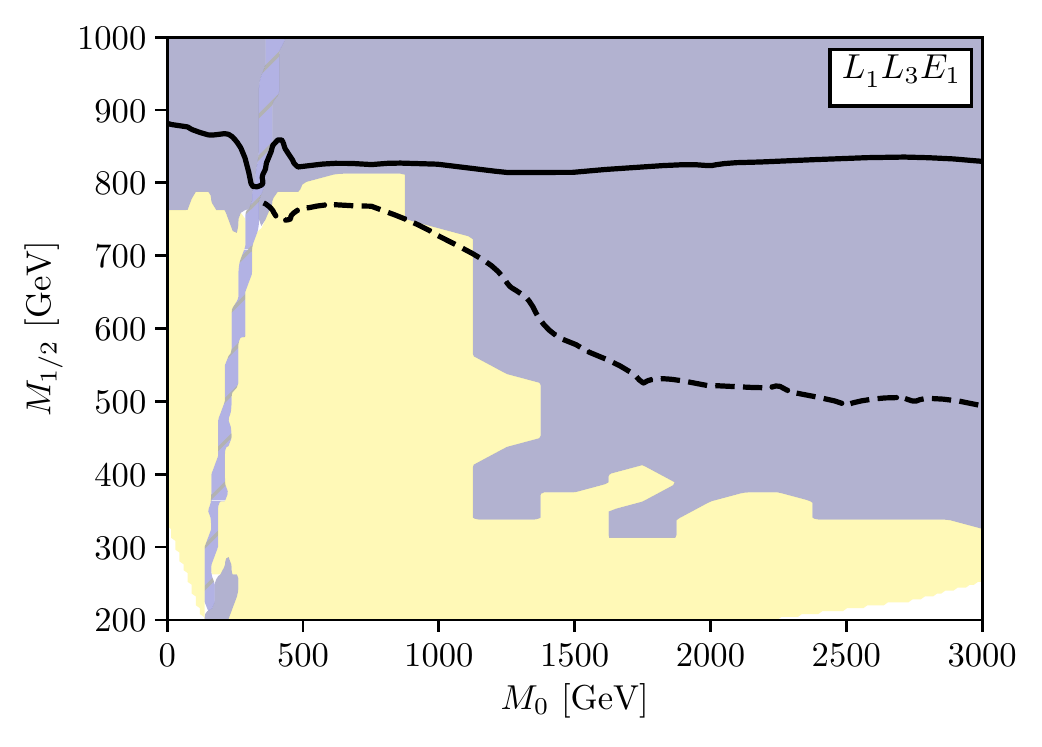}
\includetwinresults{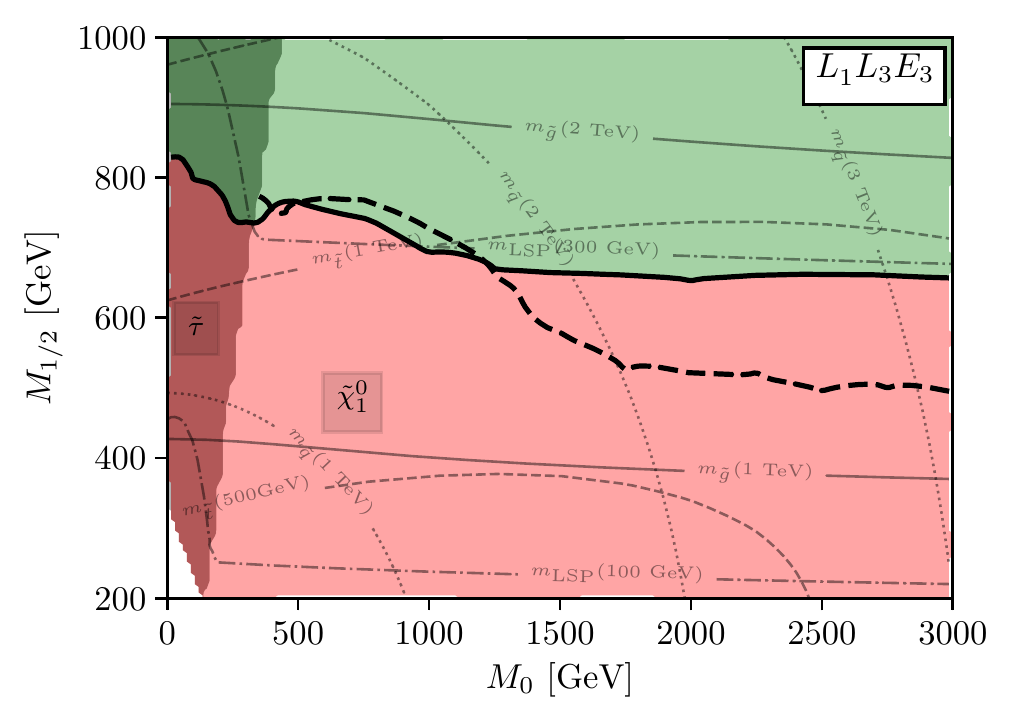}{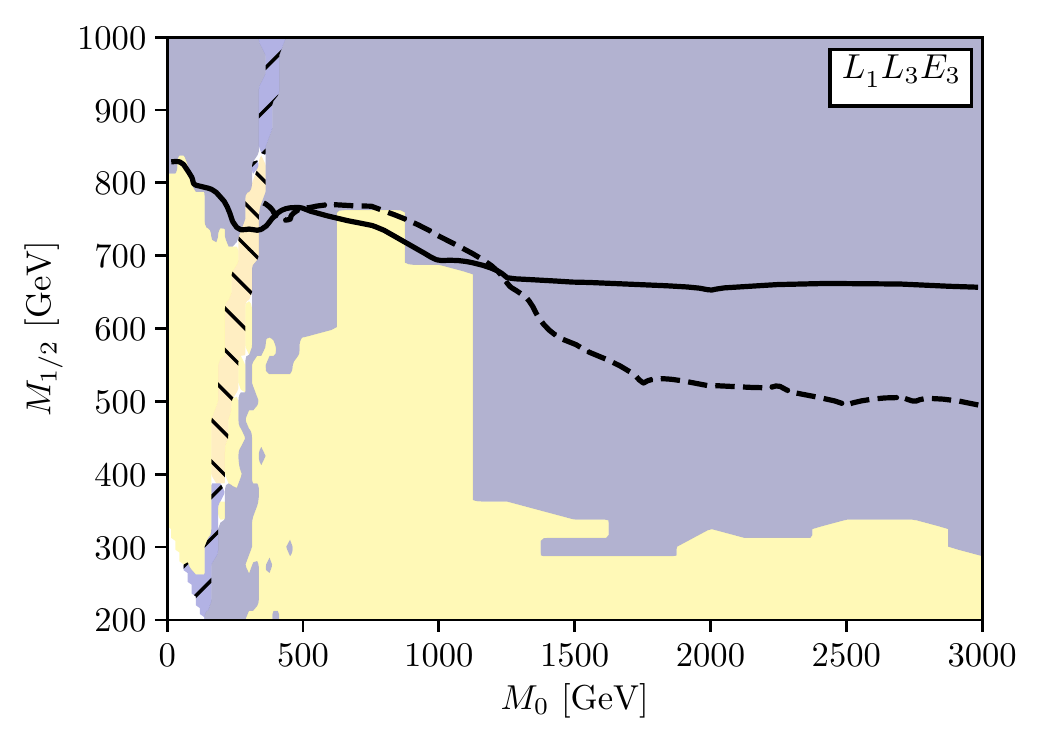}
\caption{Same as for \cref{fig:checkmate:LLE122and123}, but considering the nonzero operators (in descending 
order) $\lam_{131}$ and $\lam_{133}$.}
\label{fig:checkmate:LLE131and133}
\end{figure*}

We now consider the case of a \textit{small} non-zero $LL\bar E$ operator as discussed in 
\cref{sec:neutralino-LSP-LLE,sec:stauLSP}. We determine excluded parameter regions of the corresponding 
$\Lam_{\not R_p}$--CMSSM and compare with the LHC exclusion line obtained in the RPC case. Since we consider 
a small RPV coupling, the particle spectrum remains virtually unchanged with respect to the RPC--CMSSM. However, 
as emphasized before, such an operator leads to the decay of the LSP. Therefore, a neutralino LSP will decay into 
two leptons and one neutrino for a generic $\lam_{ijk}$ coupling, \textit{cf}. Eq.~(\ref{eq:neutralino-LLE-decay}) and 
\cref{tab:lle-signatures}. In parameter regions where the lightest stau is the LSP its possible decay modes are: 
(i)~directly into $\ell_i \nu_j$, if either $i,j$ or $k$ equals $3$, (ii)~via the RGE-generated $\lam_{i33}$ coupling if the 
non-zero RPV coupling at $M_X$ is of form $\lam_{ijj}$ and $\{i,j\}\neq 3$ (see \cref{sec:RGEinduced}) or (iii)~via 
a four-body decay, \textit{cf.} see \cref{sec:stauLSP}, and \cref{tab:stau-LSP-LHC} for the corresponding LHC 
signatures. However, the four-body decay does not happen here due to the large $\tan\beta$ value employed and the consequential 
dominance of the two-body decay through the RGE-generated operators.

The generic LHC searches for RPC supersymmetry look for missing energy in combination with jets and/or leptons. 
They should thus also perform well for the $LL\bar E$ models, due to the many extra leptons from the RPV decay. 
Although the amount of missing energy is in general not as pronounced as for an RPC model, the energy carried 
away by the neutrino in the final decay can still be sizable enough to produce a striking signature, see 
Ref.~\cite{Dreiner:2011wm,Hanussek:2012eh}.  


In the top row of \cref{fig:checkmate:LLE122and123} we show the \Checkmate exclusion in the $M_0$--$M_{1/2}$ 
plane for the $\Lam_{\not R_p}$--CMSSM with $\lam_{122}\neq 0$. The remaining CMSSM parameters and the light 
gray iso-mass curves are  as in \cref{fig:checkmate:RPC}. The excluded 
region is shown below the thick solid black curve in red. The allowed region is shown above this curve in green. The dark 
red/green colored regions correspond to a $\tilde\tau$-LSP, which must be considered in the RPV case. The light red/green 
colored regions correspond to a $\tilde\chi^0_1$-LSP, as in the RPC case. The RPC--CMSSM exclusion line of 
\cref{fig:checkmate:RPC} (thick solid black curve there) is shown here as a thick dashed black curve for comparison. It 
does not extend into the $\tilde\tau$-LSP region, as that is not viable in the RPC--CMSSM. 

The plot on the upper right in \cref{fig:checkmate:LLE122and123} shows in thick solid black the same \Checkmate exclusion from 
the upper left plot, as well as the RPC exclusion from \cref{fig:checkmate:RPC} as a solid dashed line. It furthermore shows which 
LHC analysis implemented in \Checkmate is most sensitive at a given parameter region using the same color code as in
\cref{fig:checkmate:RPC}. When comparing with \cref{fig:checkmate:analysislegend}, we see that most of the parameter range 
is most sensitively covered by atlas\texttt{\_}conf\texttt{\_}2013\texttt{\_}036, Ref.~\cite{ATLAS:2013qla}.

As a result of $\lam_{122}\neq 0$, the neutralino decays will lead to four more charged 1st- or 2nd-generation leptons 
compared to the RPC case, in regions where the neutralino is the LSP, \textit{cf.} \cref{tab:lle-signatures}. 
Consequently, analyses looking for four or more leptons, Ref.~\cite{ATLAS:2013qla}, are very sensitive to this scenario 
and yield a stronger limit than in the RPC case. Thus the solid black curve in the upper left plot is more restrictive than 
the dashed black curve. The search in Ref.~\cite{ATLAS:2013qla} contains separate signal regions designed for both 
RPC and RPV signatures, respectively. It is interesting that, although their signal regions designed for the RPV signatures 
are the ones with the best exclusion power, the RPC signal region performs almost equally well.

When looking more closely at the \Checkmate output we see that it is a specific search region in Ref.~\cite{ATLAS:2013qla}, 
which is most sensitive to the $LL\bar E$ case we are considering here, namely the electroweak pair-production of 
neutralinos and charginos. This production channel is not very promising for RPC models in the CMSSM, as the largest 
electroweak cross-section is usually obtained by the production of a charged and a neutral wino. Within the CMSSM 
boundary conditions, both would decay to the bino by emitting a $W$ and a Higgs/$Z$-boson respectively. This 
comparably small electroweak signal rate is usually not enough for these 
final states to be detected over the background. Thus the RPC--CMSSM is most stringently constrained by gluino pair 
production. In contrast, in the $\Lam_{\not R_p}$--CMSSM the neutralino decays via $\lam_{122}$ lead to a clean signal 
with many charged leptons. 

Specifically, for $M_{0}\gtrsim \SI{500}{\GeV}$, corresponding to regions with a neutralino LSP, we can exclude values of $M_{1/2}
\lesssim\SI{950}{\GeV}$, which feature a bino of $m_{\tilde \chi^0_1}\simeq \SI{400}{\GeV}$ and winos of $m_{\tilde \chi^0_2}\simeq m_{\tilde\chi^\pm_1}\simeq\SI{800}{\GeV}$, as well as gluinos over 2.1\,TeV. Thus within the \LamCMSSM{}, the indirect constraint on
the gluino mass (via the universal gaugino mass) is much stricter than the RPC gluino search can reach.

These \LamCMSSM{} bounds obtained using \Checkmate can, to some degree, be compared to the results from 
Ref.~\cite{Khachatryan:2016iqn},  where bounds on the pair-production of winos decaying via $LL\bar E$ are set. This 
analysis excludes wino masses up to \SI{900}{\GeV} also for $\lam_{122}\neq 0$, when assuming that the neutral wino 
is the LSP. In the case at hand we have a lighter bino to which the wino will decay. This change in kinematics (for instance, 
the final state leptons will be less energetic) with respect to the  simplified model analysis in 
Ref.~\cite{Khachatryan:2016iqn} explains the small differences observed in the bounds on the wino mass.



In the upper left plot of \cref{fig:checkmate:LLE122and123}, the low $M_{0}$ region features a stau LSP and we observe that 
the exclusion power close to the LSP-boundary drops significantly. This occurs due to the produced neutralinos (charginos) decaying into $\tau \tilde 
\tau_1$ $(\nu \tilde \tau_1)$ and $\tilde \tau_1$ decaying via the RGE-induced $\lam_{133}$ coupling. As a result the stau has equal branching 
ratios into both $e\nu$ and $\tau\nu$ final states. Compared to the expected signatures of neutralino-LSP regions explained 
above, several final-state 1st- and 2nd-generation leptons are now replaced by $\tau$ leptons. The expected event rates 
therefore drop by powers of the leptonic tau branching ratio and hence significantly affect the resulting bound from the same 
analysis.

As $M_0$ approaches $0$, the mass of all sleptons and sneutrinos further decreases. This slowly opens further decays of 
the neutralino (chargino) into other $\ell_i \tilde \ell_i$ and $\nu_i \tilde \nu_i$ ($ \nu_i \tilde \ell_i$ and $\ell_i \tilde \nu_i$) combinations, which for $i \neq 3$ lead to the same decay 
signatures via the $\lam_{122}$ as discussed before for the neutralino LSP region. Hence, the exclusion line approaches the 
earlier, stricter bound for $M_0 \rightarrow 0$.


Next, we consider the bottom row of \cref{fig:checkmate:LLE122and123} with a non-zero $\lam_{123}$ coupling. The 
labelling and the included curves are to be understood as for the upper two plots. In the lower right plot we see that again  
atlas\texttt{\_}conf\texttt{\_}2013\texttt{\_}036, Ref.~\cite{ATLAS:2013qla}, is most sensitive over most of the parameter 
region,
however only if $M_0$ is large and the squarks are decoupled. For small $M_0$, analyses which look for jets and like-sign charged leptons, 
for example Ref.~\cite{Aad:2014pda} (denoted {atlas\texttt{\_}1404\texttt{\_}2500} in \cref{fig:checkmate:analysislegend}),  
become more sensitive. In fact, these two analyses are similarly sensitive in this parameter region but the second gets contributions from light squark decays and thus starts dominating the exclusion line for small $M_0$.\footnote{Atlas\texttt{\_}1404\texttt{\_}2500 only becomes the most sensitive analysis if higher order cross sections are used for the strong production modes --- at leading order  the electroweak processes are dominant setting negligibly weaker bounds via analysis atlas\texttt{\_}conf\texttt{\_}2013\texttt{\_}036.}. 
To the right in the lower left plot, the neutralino-LSP decay now always involves tau leptons, \textit{cf.} 
Eq.~(\ref{eq:neutralino-LLE-decay}). The resulting overall bound, the thick solid black curve, is weaker than in the $\lam_
{122}$ case in the neutralino-LSP region, as the multilepton signal is diluted by these taus. However, it is still stricter than 
the RPC case, the thick black dashed curve. We furthermore observe in the lower left plot, that the considered analyses in 
the stau LSP region, \textit{i.e.} for small $M_0$, are \textit{more} sensitive, than in the $\lam_{122}$ case. The search is 
more sensitive, as the stau now decays via $\lam_{123}$ into a neutrino and a 1st- or 2nd-generation charged lepton, rather 
than a tau final state as in the $\lam_{122}$ case, \textit{cf.} the second line in Eq.~(\ref{eq:2bdystau}). 

For completeness we in turn show the results for non-zero values of $\lam_{131}$ and $\lam_{133}$ in \cref{fig:checkmate:LLE131and133}, respectively. 
The notation is as in \cref{fig:checkmate:LLE122and123}.  Again, the differences with respect to 
\cref{fig:checkmate:LLE122and123} can be explained by respectively considering the number of charged 1st- and 
2nd-generation charged leptons versus the number of tau leptons in the relevant final states. For the  $\lam_{131}$ 
coupling, the neutralino will decay with almost equal branching ratios into both $e e\nu$ and $e\tau\nu$ which is why 
the excluded region is larger than for $\lam_{123}$ but smaller than for $\lam_{122}$. For the case of $\lam_{133}$, 
the neutralino decays either into $\tau \tau \nu$ or $e \tau \nu$, which is why the LHC sensitivity is lower compared to all 
previous cases. Remarkably however, it is still more sensitive than the RPC case for most values of $M_0$. Lastly, we 
note there are only minor differences between the exclusion lines if we were to exchange RPV couplings to (s)electrons 
by couplings to (s)muons which is due to the comparable identification efficiency between electrons and muons at both 
\Atlas and \Cms. All other $LL\bar E$ couplings are obtained by exchanging flavor indices $1\leftrightarrow 2$ and 
the respective bounds can therefore be inferred from the scenarios shown above.

For all of the $LL\bar E$ cases, we see that the $4\ell$+$\mathrm{MET}$ search of {atlas\texttt{\_}conf\texttt{\_}2013\texttt{\_}036}, 
Ref.~\cite{ATLAS:2013qla} 
and the jets$+$SS$\ell$ search of {atlas\texttt{\_}1404\texttt{\_}2500}, Ref.~\cite{Aad:2014pda}, are 
the most sensitive in \Checkmate over most of the parameter range. The most important 
message from looking at the different $LL\bar E$ operators and comparing to the RPC case is, however, that within the CMSSM, 
as a complete supersymmetric model, the LHC is actually more sensitive to scenarios in which $R$-parity is violated via an 
$LL\bar E$ operator than if $R$-parity is conserved. This statement holds even if the signal regions designed for RPV in 
Ref.~\cite{ATLAS:2013qla} are disregarded.


\begin{figure*}[htbp]
\includetwinresults{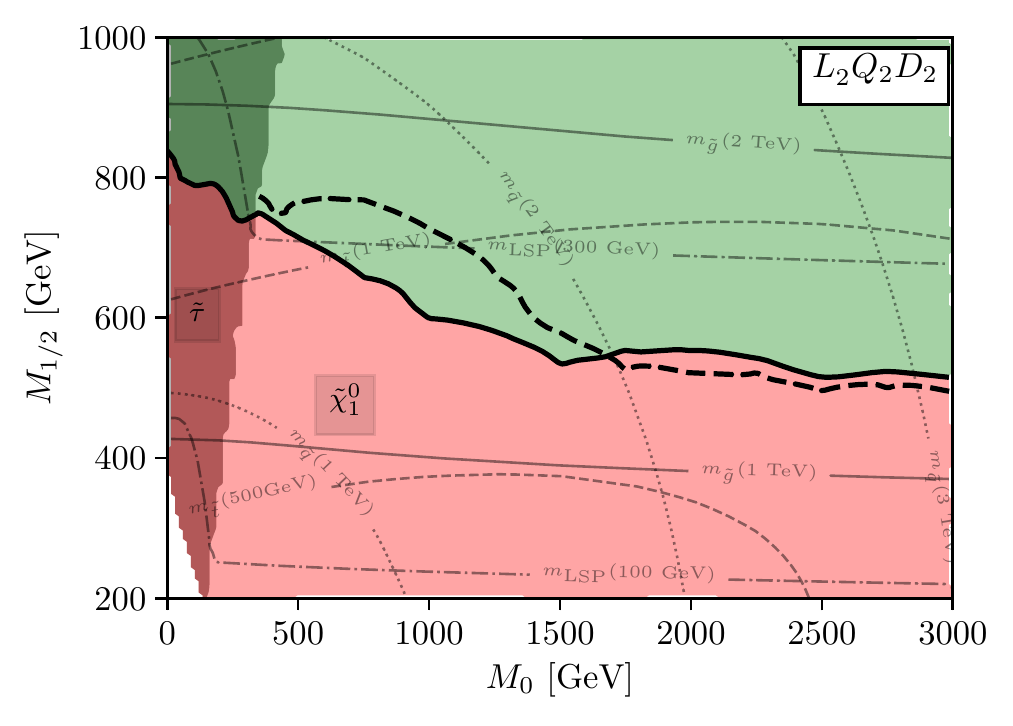}{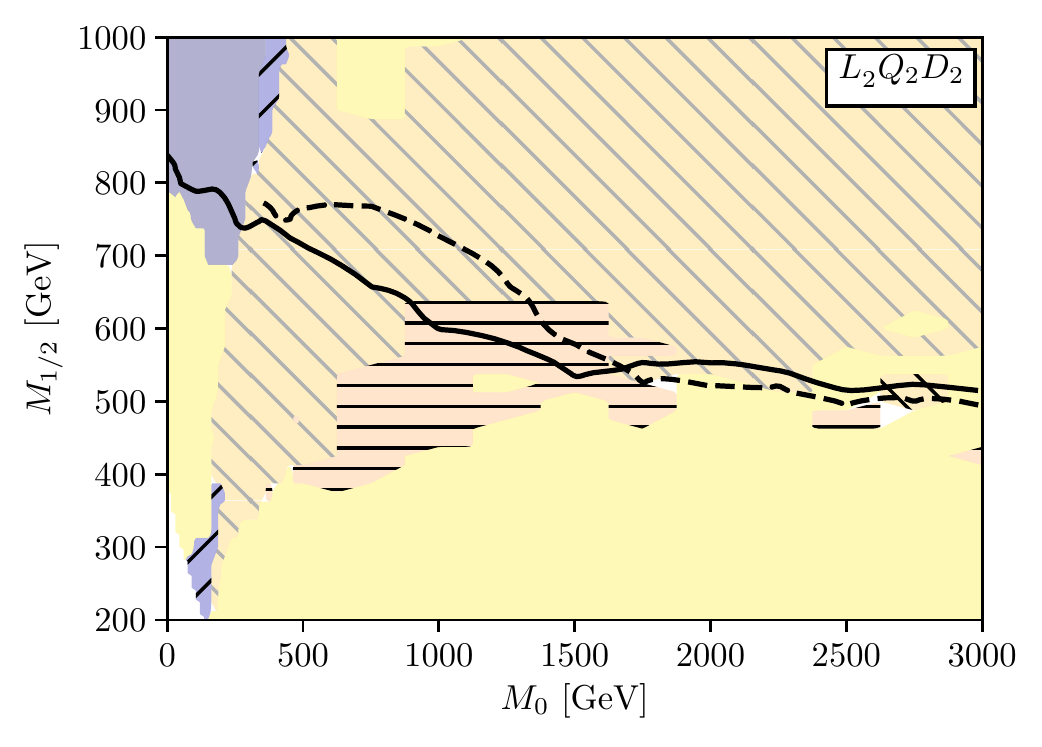} 
\includetwinresults{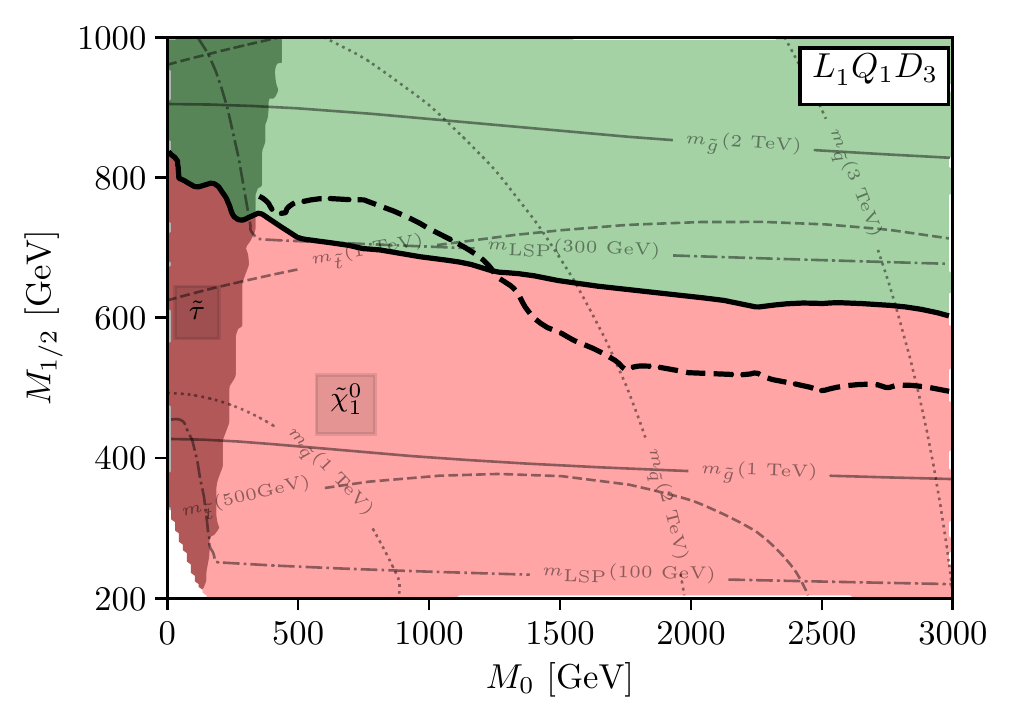}{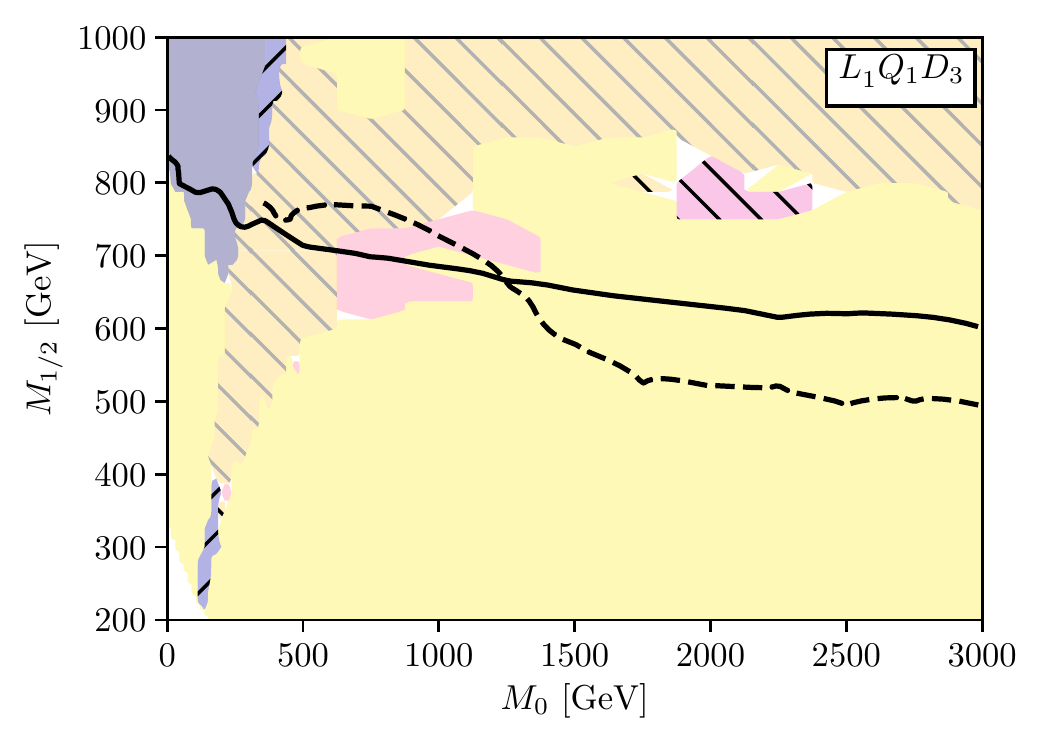}
\includetwinresults{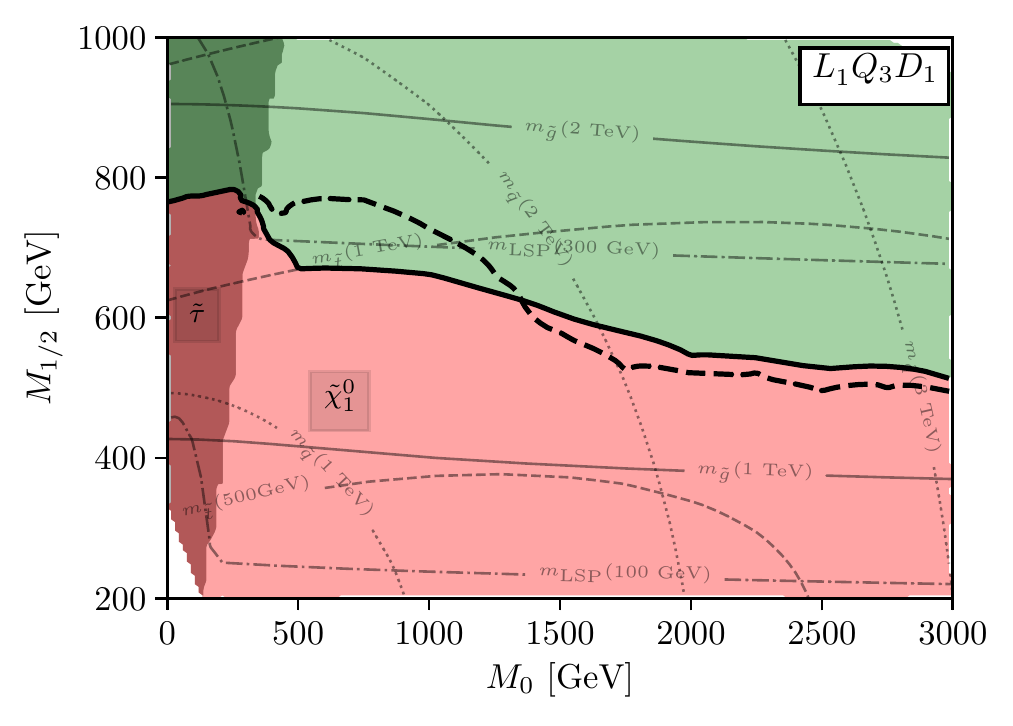}{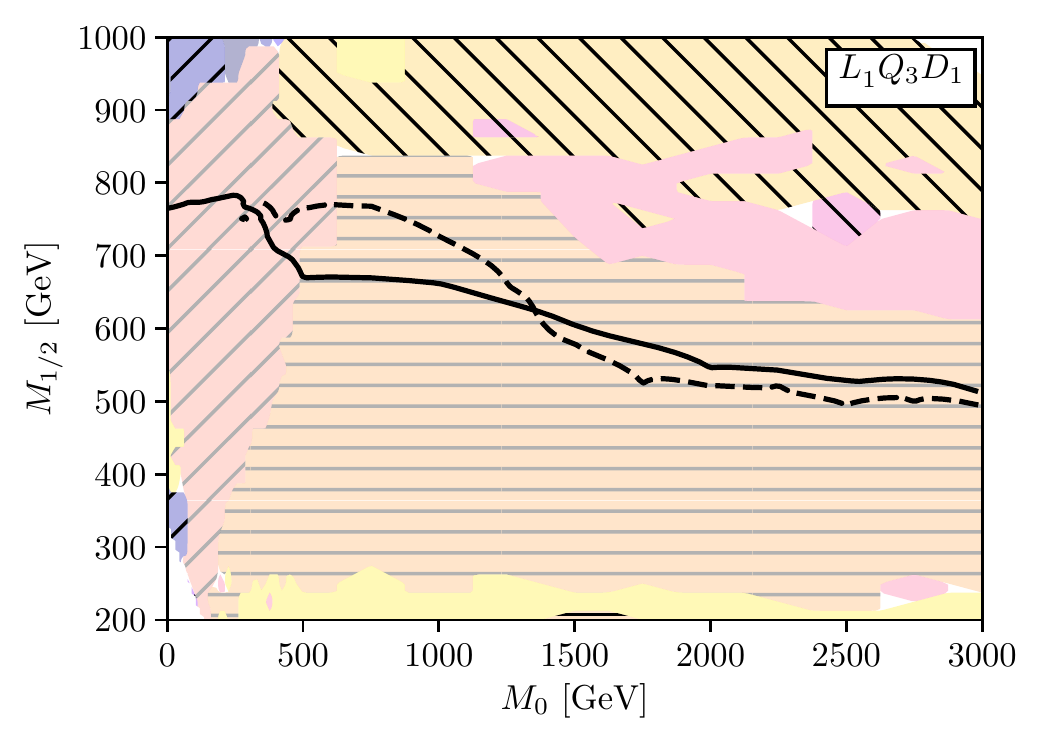}
\includetwinresults{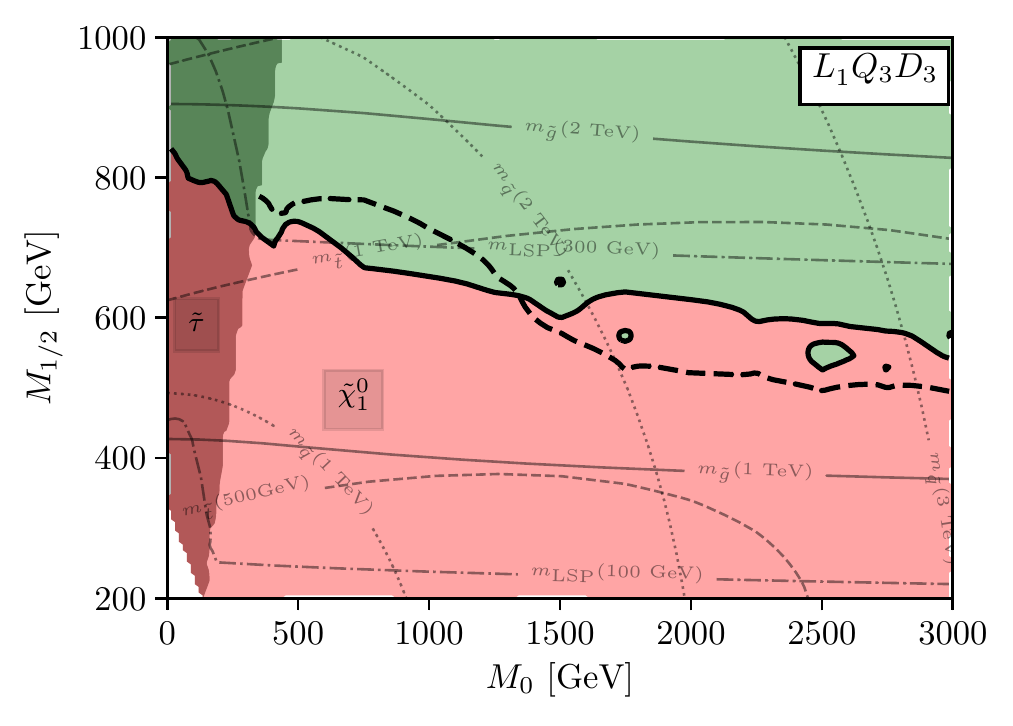}{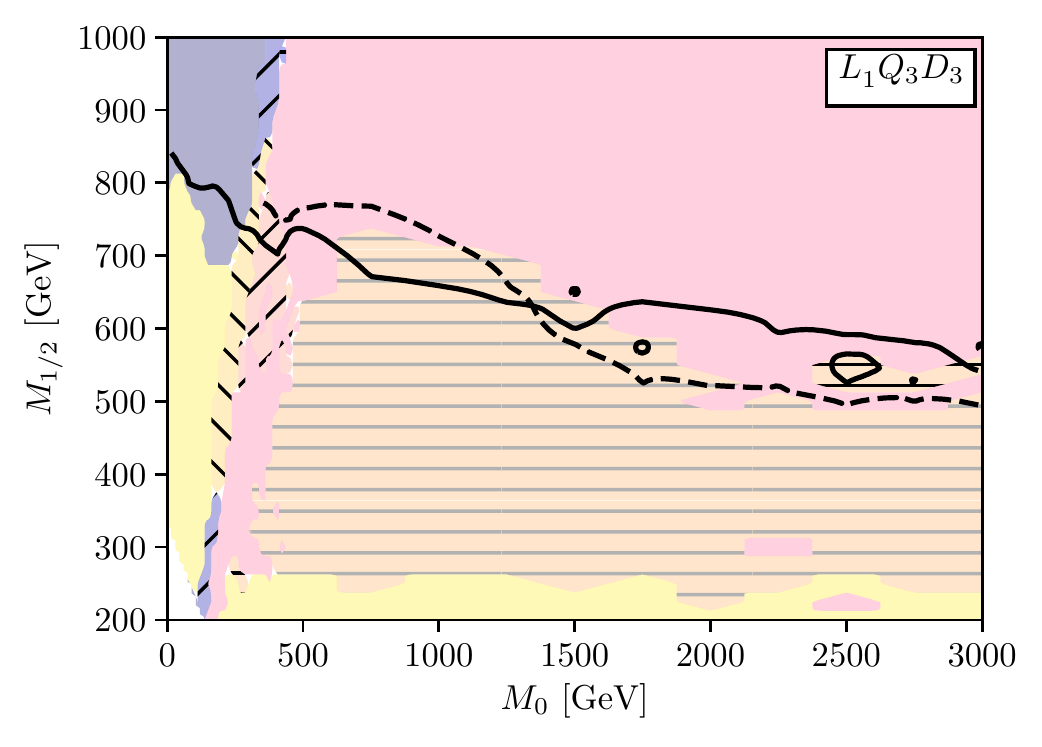}
\caption{Same as for \cref{fig:checkmate:LLE122and123}, but considering the nonzero operators (in descending order)  $\lam'_{222},\lam'_{113},\lam'_{131}$ and $\lam'_{133}$.}
\label{fig:checkmate:LQD222}
\end{figure*}

\subsubsection{$LQ\bar D$, \LamCMSSM}
\label{sec:checkmate:lqd}

We now turn to the discussion of the $LQ\bar D$ operator. In general, when compared to the previous $LL\bar E$ case, it is 
clear that the LHC sensitivity is reduced,  as we have to replace either a neutrino and a charged lepton or even two charged leptons by two quarks in the final
RPV decay of a neutralino LSP, see Eq.~(\ref{eq:neutralino-decay-LQD}) and \cref{sec:NeutralinoLQD}. Stau 
LSPs, in turn, will mostly decay either into a pair of quarks or via a four-body decay into a tau, a charged lepton or neutrino, 
and two quarks, see Eq.~(\ref{eq:2bdystau}) and \cref{sec:stauLSP}.

Furthermore, we generally observe that the electroweak gaugino-pair-production is no longer relevant in the case of a decay via 
$LQ\bar D$. Due to the hadronic decay products of the neutralino- or stau-LSP the efficiency in the electroweak gaugino 
case is no longer significantly higher than in the strong production case. The latter then wins due to the significantly higher
production cross section.

Let us discuss the results for the individual couplings. The first row in \cref{fig:checkmate:LQD222} shows the case 
of a nonzero $\lam'_{222}$ operator. As just mentioned, the overall exclusion sensitivity is significantly lower than in the 
$LL\bar E$ case, and is comparable to the RPC--CMSSM case, shown here as the thick black dashed line. In a small 
region around $M_0=750\,$GeV, the RPC
is even stricter. In the neutralino-LSP region with high $M_0$, we  find that analyses which look for jets and like-sign charged leptons, 
for example Ref.~\cite{Aad:2014pda} (denoted {atlas\texttt{\_}1404\texttt{\_}2500} in \cref{fig:checkmate:analysislegend}),  
are most sensitive. See the right-hand plot. In this region the first and second generation squarks are relatively heavy. Therefore
gluino and stop pair-production are the most dominant production modes. These produce final states with many $b$-jets, lower 
quark generation jets, and leptons and hence populate the $3b$ signal region of a ``$2$ same-sign $\ell$ or $3 \ell\,$'' analysis
({atlas\texttt{\_}1404\texttt{\_}2500}), for 
which the Standard Model background is nearly 
zero. Here, the high final state multiplicity induced by the $LQ\bar D$ decay results in a 
slightly increased sensitivity when compared to the RPC case.

In regions with lower $M_0$ where gluino-squark associated production and squark pair production become relevant, generic 
squark-gluino searches like {atlas\texttt{\_}conf\texttt{\_}2013\texttt{\_}062}, Ref.~\cite{ATLAS-CONF-2013-062}, which look for 
jets, leptons and missing transverse momentum dominate. For these, the increased final state multiplicity via the 
additional $LQ\bar D$-induced decays results in a worse bound than for the RPC case. This is due to the signal regions setting 
strong cuts on the required momentum of the final state objects and the missing transverse momentum of an event. These are 
necessary to sufficiently reduce the Standard Model background contribution, especially from multiboson production, 
which also produces final states with high jet and lepton multiplicity and some missing transverse momentum. Since the 
expected missing transverse momentum of the event is significantly larger in RPC models for which the LSP does not decay, 
breaking $R$-parity weakens the bounds in these regions, \textit{cf.} Ref.~\cite{Dreiner:2011wm,Hanussek:2012eh}.

Within the stau LSP region, the wedge at low $M_0$, the stau will undergo two-body decays due to the RGE-generated 
$\lam_{233}$ operator for large $\tan \beta$, see also \cref{fig:staudecayTB}. Therefore, this scenario mimics the results 
from the stau LSP region in the case where $\lam_{233}$ is already present at $M_X$, see our discussion of the 
phenomenologically almost identical $\lam_{133}$ operator, in \cref{sec:checkmate:lle}.

\begin{figure*}[h!]
\includetwinresults{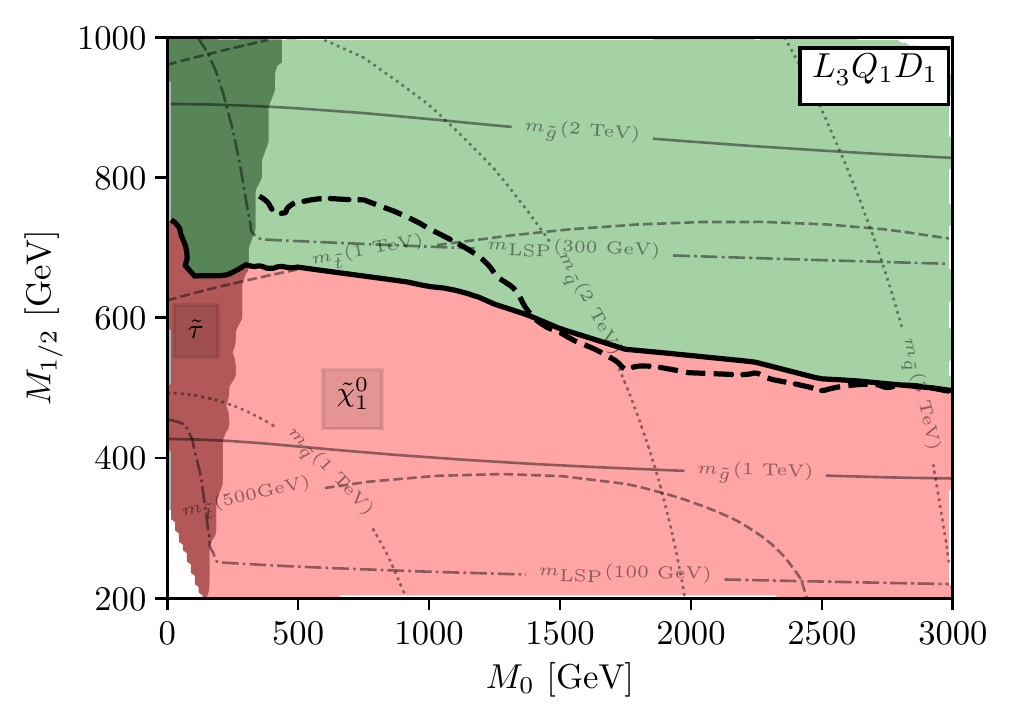}{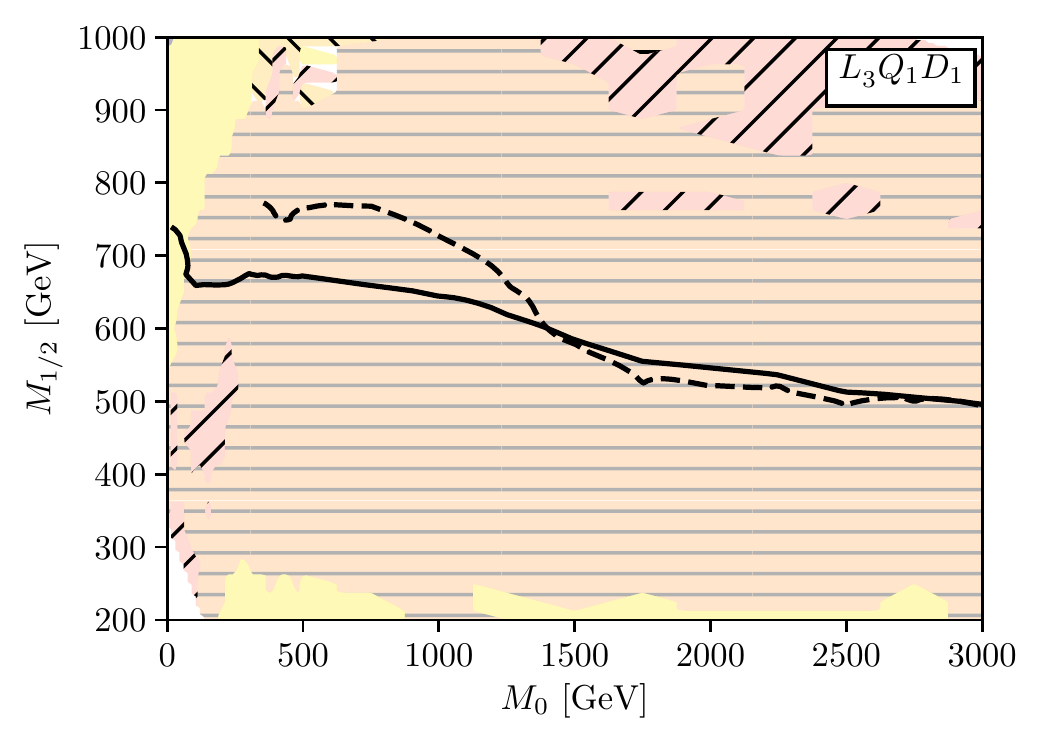}
\includetwinresults{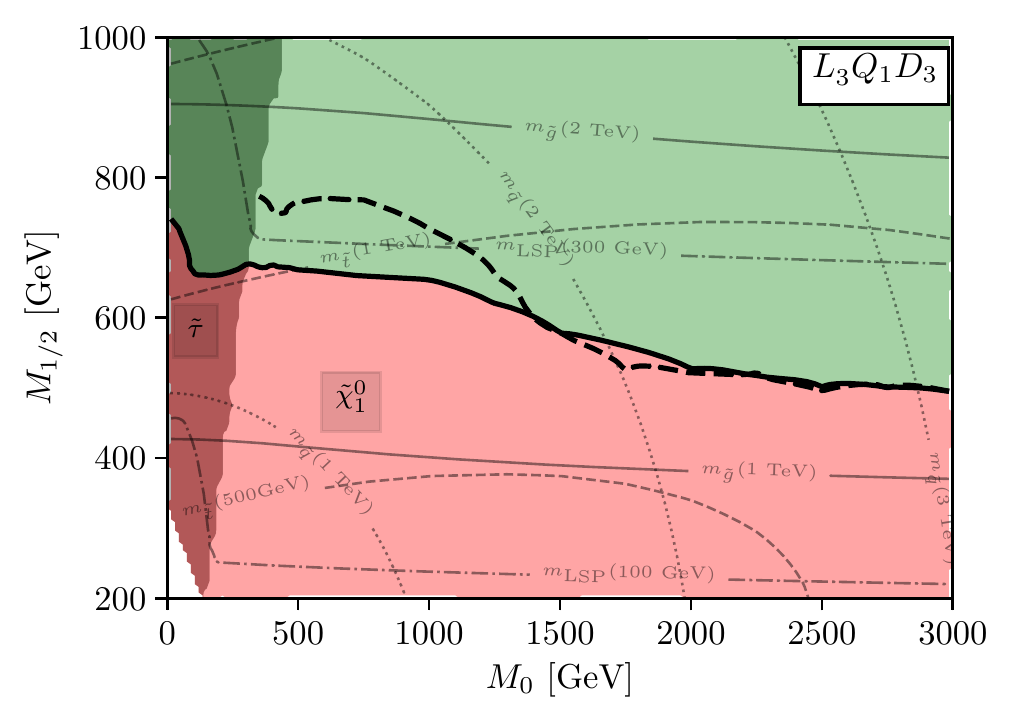}{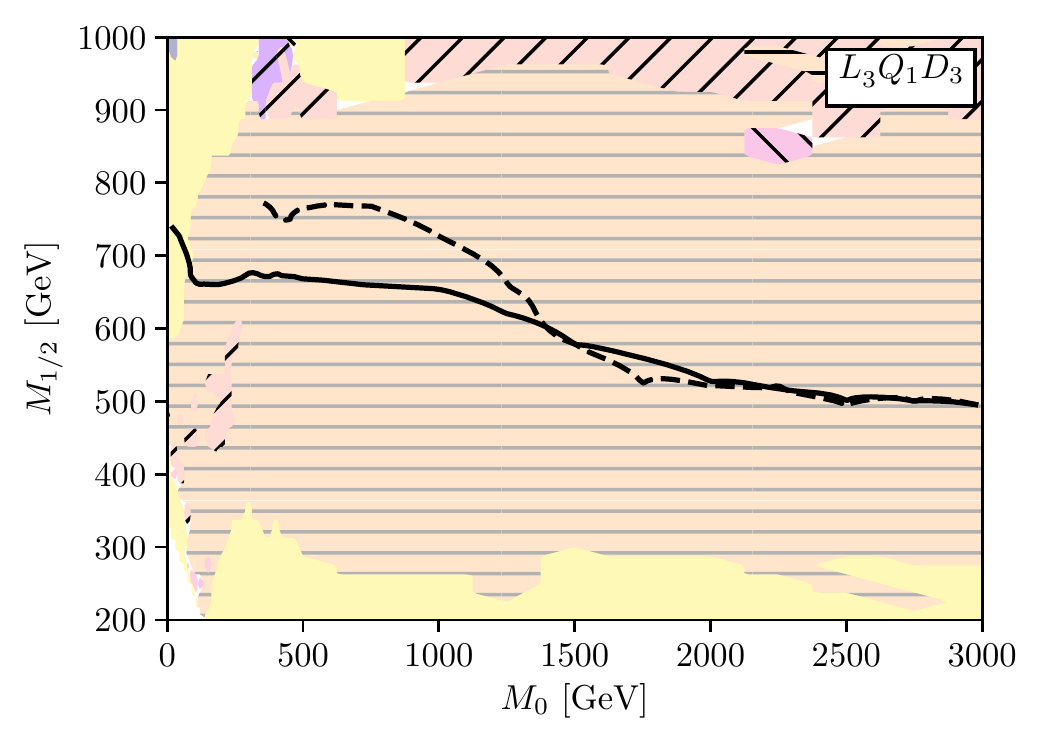}
\caption{Same as for \cref{fig:checkmate:LLE122and123}, but considering the nonzero operators (in descending order)  
$\lam'_{311}$ and $\lam'_{313}$, which all violate tau lepton number.
}
\label{fig:checkmate:LQD3ij}
\end{figure*}

We continue with the discussion of $\lam'_{113}$, with the only phenomenologically relevant difference that $\bar D_2$ is 
replaced by $\bar D_3$. Hence, in the neutralino LSP case, the only phenomenological difference is that two $b$-jets 
replace two normal jets. (We found that the sensitivity in the $\lam'_{222}$ and $\lam'_{112}$ cases are almost identical, 
since the experimental efficiencies for muons and electrons are similar.) Due to the good $b$-jet tagging efficiency, this 
clearly improves the distinguishability with respect to the Standard Model background and results in an increase in 
sensitivity. This effect is most prominent for large values of $M_0$. The same analysis as in the previous $\lam'_{222}$ 
case, see {atlas\texttt{\_}1404\texttt{\_}2500}, Ref.~\cite{Aad:2014pda}, provides the most stringent bounds as it contains 
special signal regions which tag additional $b$-jets. For smaller values of $M_0$ barely any change in sensitivity 
is visible in comparison to before. 

In the $\tilde \tau$-LSP region of the $\lam'_{113}$ case, the stau will almost always undergo a four-body decay, thereby decaying 
into both $\tau e b j$ and $\tau \nu b j$ at approximately equal rates. The increase in sensitivity with respect to the neutralino-LSP 
region comes from the additional tau leptons in the final state.

We continue with the cases $\lam'_{131}$ and $\lam'_{133}$ in the lower two rows in \cref{fig:checkmate:LQD222} and focus 
on the neutralino LSP region first. Here, the top quark in the decay products does not improve the sensitivity when 
compared to the $\lam'_{222}$ case. When
comparing to the $\lam'_{113}$ case, we see the sensitivity also goes down. On the one side we no longer have the bottom quark 
jet in every decay and on the other hand the operator $\lam'_{13i}$ in principle allows for neutralino decays into both $t+\ell+j_i$ 
and $b + \nu+j_i$. However, the mass of the LSP is so low in the relevant parameter range, that the decay into the top quark 
is kinematically suppressed. Hence, most of the neutralinos will decay via the neutrino mode and as such do not produce the final 
state leptons which are required for the aforementioned ``$2$ same-sign $\ell$ or $3 \ell\,$'' analysis to be sensitive. Instead, the 
most relevant analysis in the large-$M_0$ region turns out to be a search looking for events with more than seven jets plus missing 
energy, see  {atlas\texttt{\_}1308\texttt{\_}1841}, Ref.~\cite{Aad:2013wta}. The high jet multiplicity for $\lam'_{131,133}$ arises 
from hadronically decaying tops, produced from the standard $\tilde g \rightarrow t \tilde t, \tilde t \rightarrow t\tilde\chi^0_1$ decay chains 
in this parameter region, as well as jets from the final neutralino decay $\tilde \chi^0_1\rightarrow jj\ETmiss$.

Comparing $\lam'_{133}$ to $\lam'_{131}$, the two additional $b$-jets in the final state result in a slightly improved exclusion power 
via the multi-$b$ analysis in {atlas\texttt{\_}conf{\_}2013\texttt{\_}061}, Ref.~\cite{ATLAS-CONF-2013-061}.

In the $\tilde \tau$-LSP region, the case $\lam'_{133}$ is analogous to the $\lam'_{222}$ case in that the RGE-generated $\lam_
{133}$ operator determines the $\tilde \tau$ decay, leading to similar bounds. In the $\lam'_{131}$ case, the situation is similar to 
$\lam'_{113}\neq 0$ in that the four-body decay dominates. However, as the final state including the top quark, $\tilde \tau \to \tau 
e t j$, is kinematically suppressed, the stau almost exclusively decays into $\tau \nu b j$. This scenario therefore exhibits the worst 
LHC measurement prospects of all the $\lam'_{aij} \neq 0$, $a=1,2$, $\tilde \tau$-LSP scenarios.

Turning to the $\lam'_{311}$ scenario shown in the top row of \cref{fig:checkmate:LQD3ij}, all differences with respect to 
the former $\lam^\prime_{222}$ case can be explained by the exchange of muons by taus in the final state, which reduces 
the overall final state identification efficiency. As a result, the searches looking for leptons lose sensitivity and, similar to the 
above $\lam^\prime_{i3i}$ 
cases, the best constraints are instead provided by the high jet multiplicity analysis described in {atlas\texttt{\_}1308\texttt{\_}1841}, 
Ref.~\cite{Aad:2013wta}. Whilst in the $\lam^\prime_{222}$ scenario the lower $M_0$ region was most constrained by the 
squark-gluino searches in {atlas\texttt{\_}conf{\_}2013\texttt{\_}062}, Ref.~\cite{ATLAS-CONF-2013-062}, here this region is 
again covered by the high multiplicity jet analyses. A closer look at the event rates however reveals that these two analyses are 
almost equally sensitive and hence the resulting bounds are nearly the same. 

In the $\lam'_{311}$ scenario, the LHC sensitivity does not change significantly when traversing from the neutralino- into the 
stau-LSP region since the stau itself decays directly into light quark jets. Hence, only the kinematics change when the LSP 
crossover occurs, while the final state signatures stay the same. This is why we see exactly the same behavior for the other $i=3$ 
cases $\lam'_{313}$, $\lam'_{331}$ and $\lam'_{333}\neq 0$, the former of which we show 
in the second row of  \cref{fig:checkmate:LQD3ij}. 
Consequently, the additional $b$-tagging in these scenarios does not noticeably improve the exclusion power of {atlas\texttt{\_}1308\texttt{\_}1841}, Ref.~\cite{Aad:2013wta}.

Here we have considered all  distinct types of nonzero $LQ\bar D$ operators, which in principle have differing LHC 
phenomenology. In the region where the neutralino is the LSP and $M_0\lesssim 1.2~$TeV, the corresponding LHC bounds 
that we obtain using \Checkmate{} are slightly weaker compared to the $R$-parity-conserving CMSSM. This corresponds to 
the region where squark pair-production dominates and the additional decay of the neutralino LSP reduces the $\ETmiss$. 
In the parameter region where gluino and stop pair production dominates, \textit{i.e.} for large $M_0$, we instead find most $LQ
\bar D$ scenarios are as constrained as the RPC--CMSSM because of the equally good performance of the multijet searches 
preferred by RPV and the multi-$b$ searches sensitive to RPC. The special cases $L_i Q_j \bar D_3$ with $i, j \in \{1, 2\}$ are 
significantly more constrained in this region of the $R$-parity violating CMSSM, due to the additional extra leptons and $b$-jets 
in the final state. In all cases, regions with stau LSP are well covered by either multilepton or combined lepton+jet searches and 
yield comparable bounds as in parameter regions with a neutralino LSP.

\subsubsection{$\bar U \bar D \bar D$}
\label{sec:checkmate:udd}

\begin{figure*}
\includetwinresults{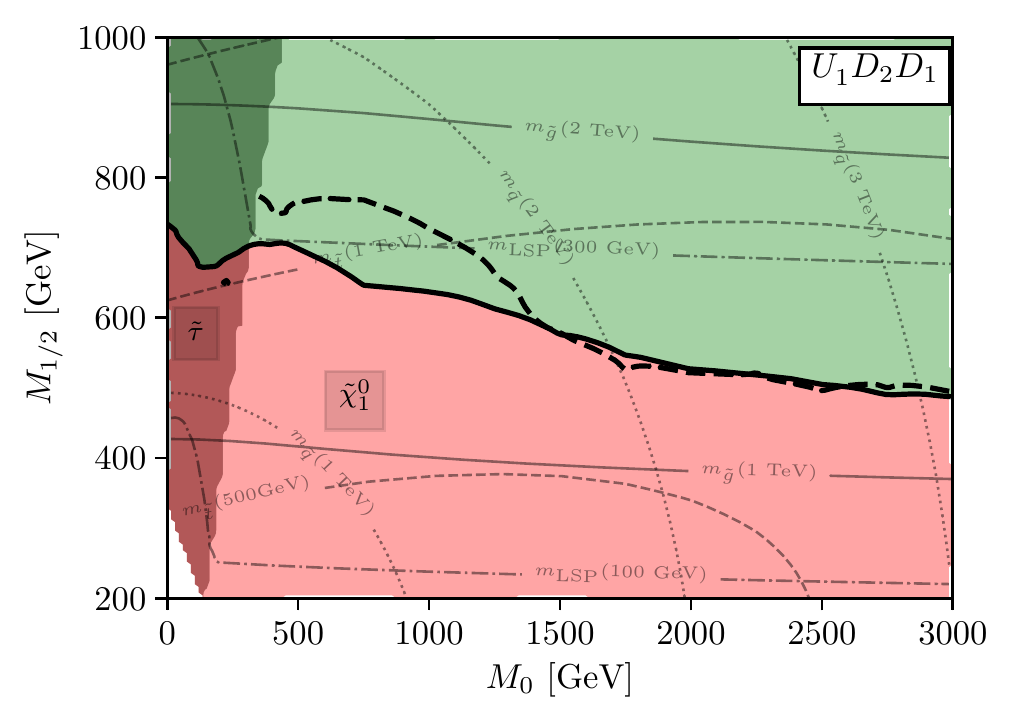}{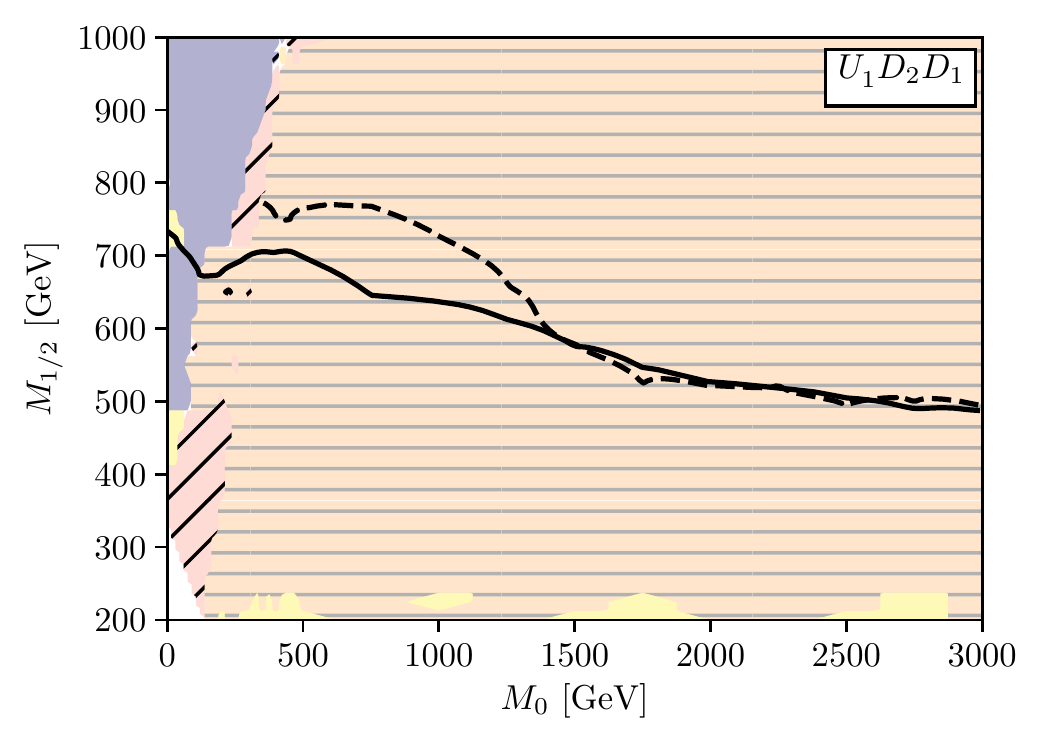}
\includetwinresults{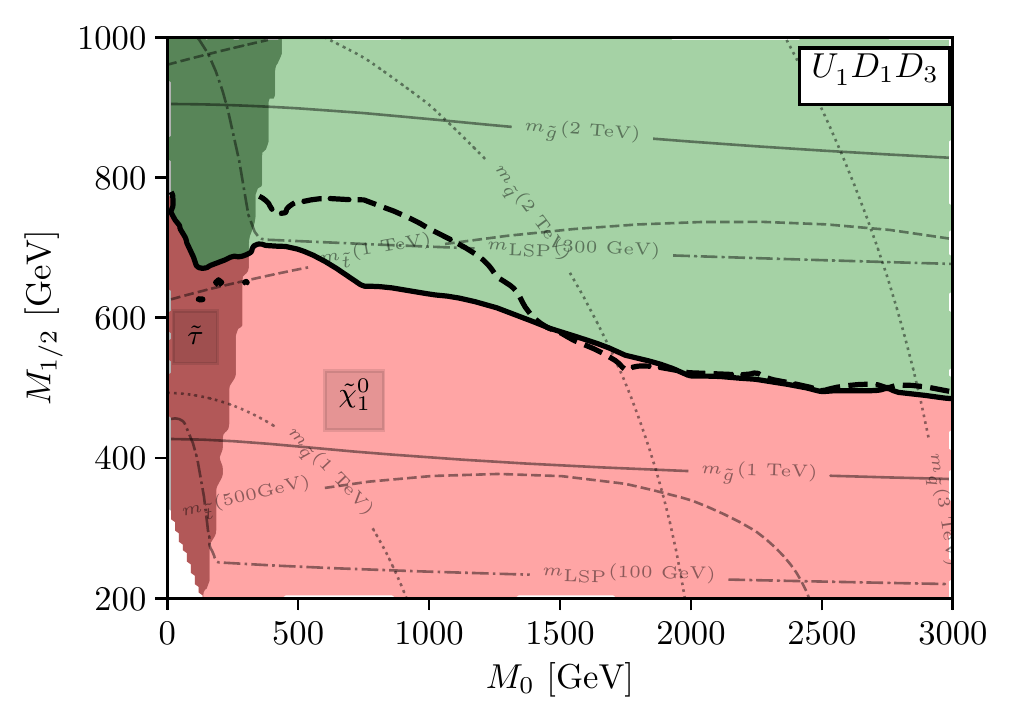}{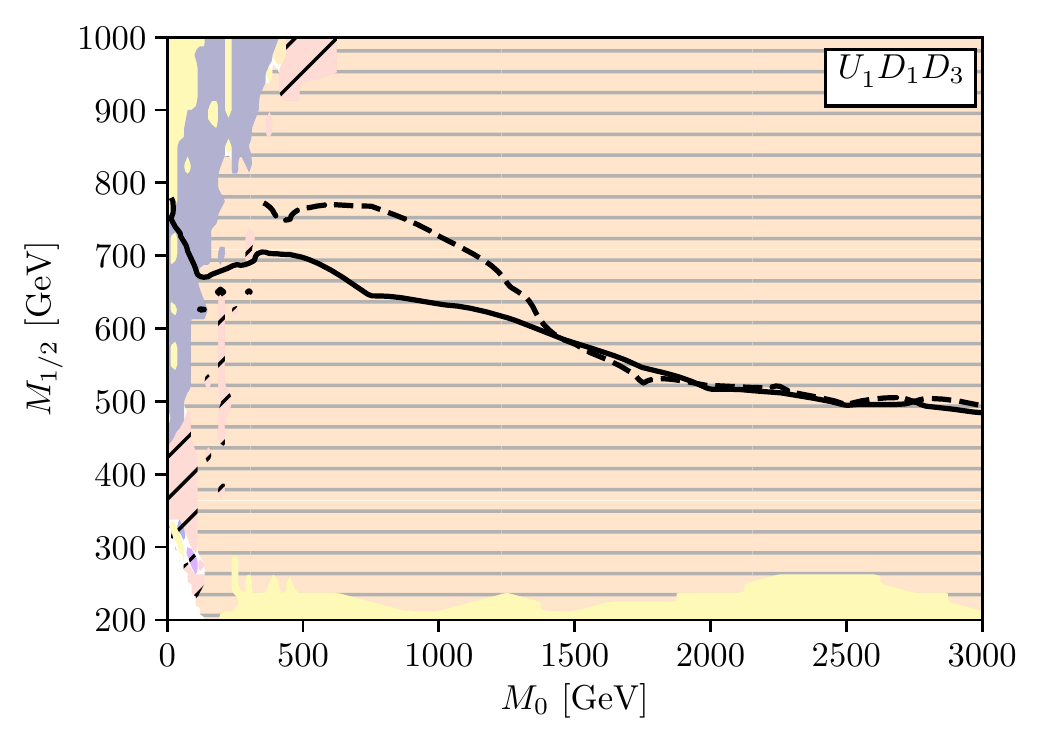}
\includetwinresults{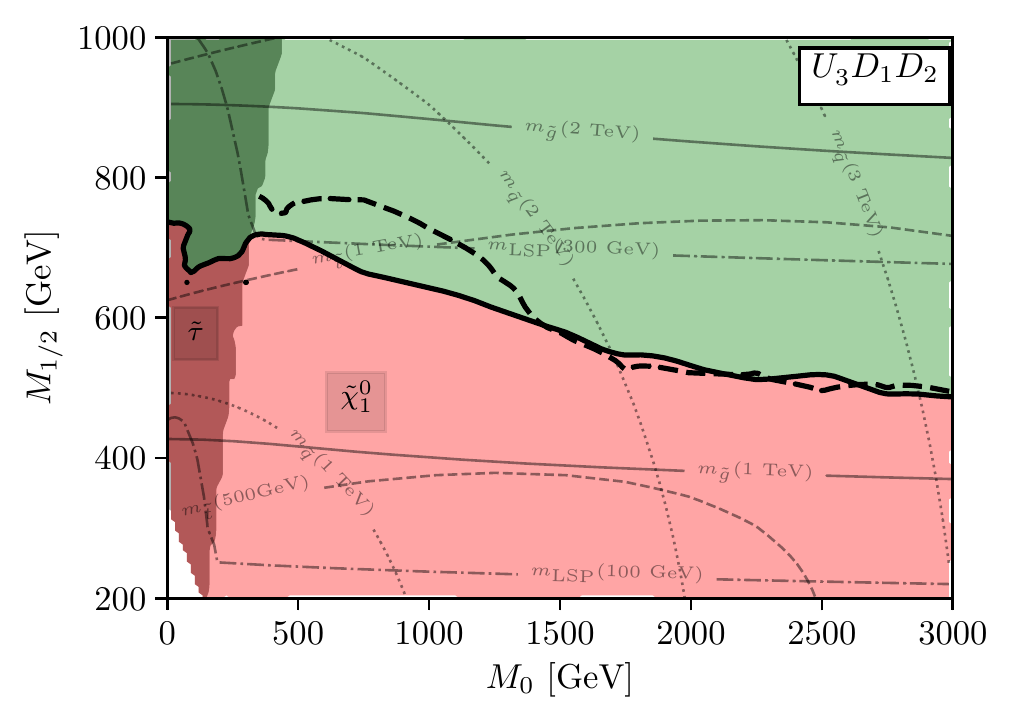}{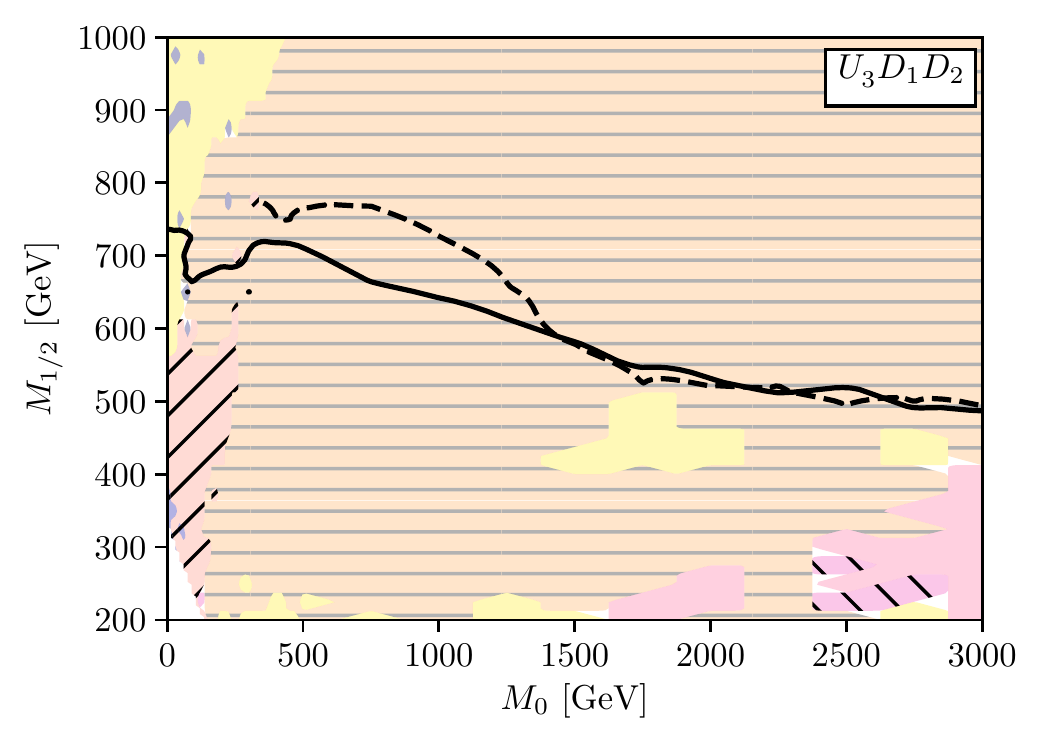}
\includetwinresults{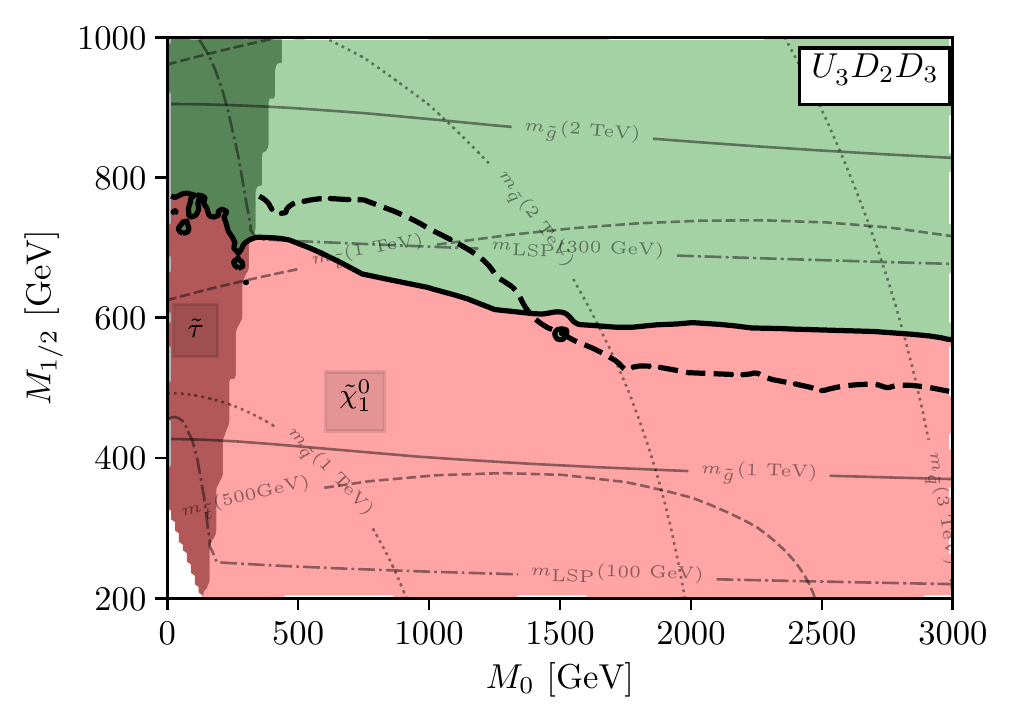}{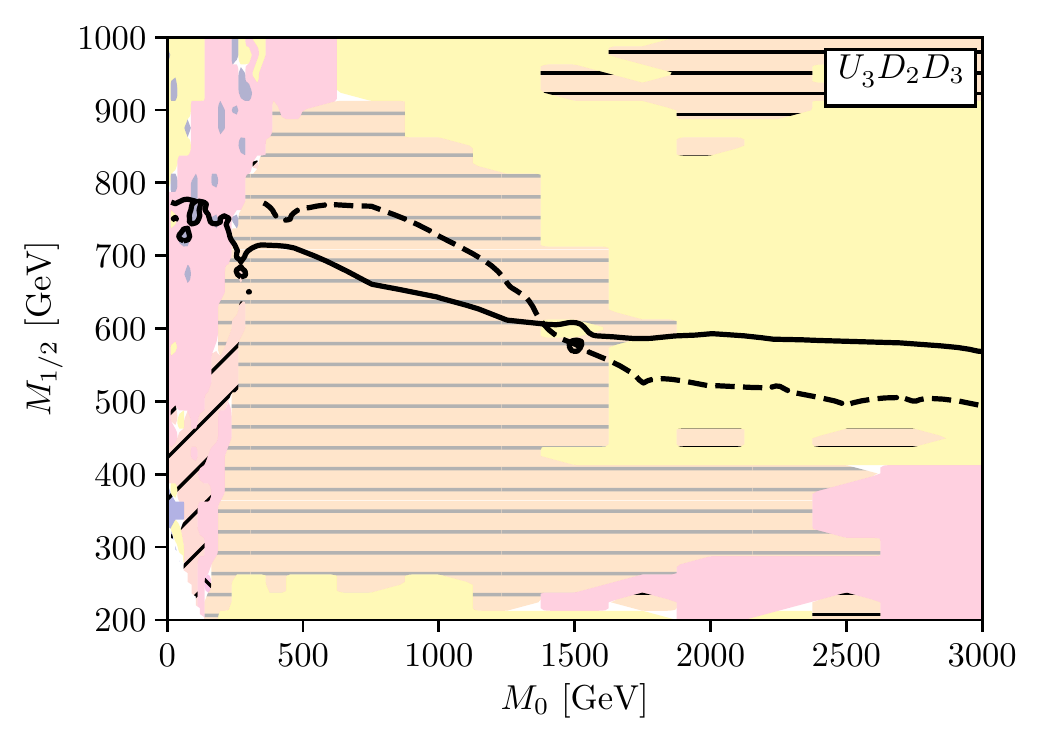}
\caption{Same as for \cref{fig:checkmate:LLE122and123}, but considering the nonzero operators (in descending order)  $\lam''_
{121},\lam''_{131},\lam''_{312}$ and $\lam''_{323}$.}
\label{fig:checkmate:UDD121and323} 
\end{figure*}

Here we discuss the $\bar U \bar D \bar D$ operator for which one typically expects the weakest LHC bounds as there is no striking 
missing energy signal nor any additional leptons, see for example Ref.~\cite{Allanach:2001xz,Allanach:2001if,Evans:2013uwa}.

In \cref{fig:checkmate:UDD121and323}, we show the results in analogy with the previous subsections. We first consider 
the case of $\lam''_{121}$, the top row, where the neutralino LSP decays into three light jets. Therefore multi-jet searches 
should yield the most stringent limits for such scenarios. Indeed, as can be seen in the top right plot of 
\cref{fig:checkmate:UDD121and323},  the analysis in {atlas\texttt{\_}1308\texttt{\_}1841}, Ref.~\cite{Aad:2013wta}  provides 
the best exclusion power for the entire neutralino-LSP region. Interestingly, the bounds on the parameter space which we 
obtain are almost as strong as the bounds on the RPC scenario, the thick black dashed line, \textit{cf.} 
\cref{fig:checkmate:RPC}. This can be regarded as an impressive success for the experimental groups, since multi-jet 
analyses belong to the most challenging signatures at a hadron collider.

In the large $M_0$ region where the exclusion lines from RPC and RPV are very similar, gluino pair production has the 
highest cross section. The gluinos then decay down to a top quark and a stop which itself decays to a top and a neutralino. 
The dominant $\tilde g \to t \tilde t$ decay occurs because of the large stop mixing in this region which significantly 
reduces the $\tilde t_1$ mass with respect to the other squark masses. The neutralino then eventually undergoes a 
three-body decay into three light jets. As a result, the signal region looking for $\geq 10$\;jets and missing energy (denoted 
``10j50'' in Ref.~\cite{Aad:2013wta}) provides the best constraints. This is somewhat surprising as the analysis vetoes 
against isolated leptons while requiring missing energy. Naively, one would have expected searches for $b$-jets, missing 
energy and leptons to dominate. However, we find that only the next-best analysis looks for that, Ref.~\cite{Aad:2014lra}, 
with the best applicable signal region ``SR-1$\ell$-6j-C'' looking for one lepton, more than six jets and missing energy. 
Furthermore note that also in Ref.~\cite{Aad:2013wta}, a RPV interpretation has been performed, assuming 
gluino pair-production, which decay into $\tilde t \bar t$, with $\tilde t\to bs$, obtaining bounds of $m_{\tilde g}\gtrsim 
1\,$TeV. Translating the bounds we obtain for the CMSSM-like scenario to gluino mass bounds, we obtain even stricter 
mass limits, which is due to the higher jet multiplicity in the final state from including the intermediate neutralino in the 
decay chain.

In the lower $M_0$ regions where the exclusion in RPC parameter space is stricter, squark-gluino associated production is dominant. 
Therefore, while the RPC--CMSSM provides a large missing energy signal and is therefore probed by analyses like 
{atlas\texttt{\_}1405\texttt{\_}7875}, Ref.~\cite{Aad:2014wea}, the RPV counterpart is still best covered by the 10-jet signal region of \cite{Aad:2013wta}. 

We once more want to emphasize that the bounds we obtain on the parameter space rely on the boundary conditions which 
we impose at the high scale and the corresponding (s)particle spectrum. In particular, as seen in the large $M_0$ region, 
the presence of top quarks in the decay is of major importance. In Ref.~\cite{Buckley:2016kvr}, a re-interpretation of LHC 
results in a natural SUSY context and $\lam_{212}'' \neq 0$ has been performed. In their scenario the stop and gluino 
masses are varied independently and a Higgsino LSP decaying into three light jets is assumed. The results show that in this 
case, gluino and stop masses are generically less constrained when compared to the RPC analogue.

In the stau LSP region, at low $M_0$, the decay into one tau lepton and three light jets dominates for $\lam''_{121}$ and into 
one tau, one $b$-jet and two light jets for $\lam''_{113}$. Interestingly, most of this area is best covered by the 4-lepton 
analysis {atlas\texttt{\_}conf\texttt{\_}2013\texttt{\_}036} \cite{ATLAS:2013qla}, which requires, in case of a $\tau$-tag, at least 
three additional light leptons. This means that the other three $\tau$-leptons can only be identified  through their leptonic decay, 
reducing the overall acceptance by a factor $[{\rm BR}(\tau \to \ell \nu \nu)]^3 \simeq 0.044$. In addition, we find that 
wino pair-production is important in this part of parameter space and that the charged wino state decays into $e/\mu +\tilde 
\nu$ in up to 30\% of the cases, further contributing a charged lepton in the final state.

Turning to the cases where third-generation quarks are among the LSP decays, namely the bottom three rows in 
\cref{fig:checkmate:UDD121and323}, we see that the bounds in the low $M_0$ region are similar to the $\lam''_{121}$ case, 
and that again the multi-jet search is most sensitive. In the high $M_0$ region, where gluino and stop pair-production becomes 
relevant, searches for same-sign leptons, 
{atlas\texttt{\_}1404\texttt{\_}2500} \cite{Aad:2014pda}, become 
effective for $\lam''_{3ij}$. This is due to leptonically decaying top quarks in the final state and has already been 
analysed in detail in Ref.~\cite{Bardhan:2016gui}. Note that in comparison to the $LQ\bar D$ operator $\lam'_{i3j}$, 
where the top in the final state was phase-space suppressed and as such the alternative neutrino decay mode was 
favored, there is no other comparable decay mode for the neutralino in the case of $\bar U \bar D \bar D$. As such, it will 
always decay into a top quark whenever kinematically accessible, which is the case for $M_{1/2}\gtrsim 400~$GeV. 
Therefore, both gluino and stop pair production can lead to same-sign leptons from the leptonic top decay modes, rendering 
this scenario slightly more constrained than the $\lam''_{aij}$, $a=1,2$, cases. Interestingly, we find that in addition electroweak
gaugino production is even more important for the limit setting in the large-$M_0$ area than stop pair-production.

For the $\lam''_{323}$ case, the additional possibility of tagging more $b$-jets further improves sensitivity, such that 
the large-$M_0$ region is considerably more constrained than the RPC analogue. 

For $\lam''_{312}$, in regions 
with the stau being the LSP, the kinematical suppression of final states with top quarks results in the most abundant decay chains 
going via off-shell charginos into a neutrino, a bottom quark and two light-flavor jets, \textit{cf.} \cref{eq:staudec}. Therefore, 
searches for missing energy and several 
jets, \textit{e.g.}\ {atlas\texttt{\_}1308\texttt{\_}1841} and {atlas\texttt{\_}1405\texttt{\_}7875}, Refs.~\cite{Aad:2013wta,Aad:2014wea}, provide a good coverage. At very low $M_0$, 
where the mass difference $m_{\tilde\chi^0_1} - m_{\tilde\tau_1}$ is largest, even the search for same-sign leptons, 
{atlas\texttt{\_}1404\texttt{\_}2500} \cite{Aad:2014pda}, which is sensitive to the leptonically 
decaying taus from $\tilde \chi^0 \to \tilde \tau \tau$, becomes effective enough to exclude the area below $M_{1/2}\lesssim 730~
$GeV. However, the sensitivity of this analysis to the scenario at hand quickly drops off with decreasing $m_{\tilde\chi^0_1}-m_
{\tilde\tau_1}$, as can be seen in the $\lam''_{312}$ case of \cref{fig:checkmate:UDD121and323}. For $\lam''_{323} \neq 0$, which 
features at least four $b$-jets in the final state, the search for large missing transverse momentum and at least three $b$-jets, 
{atlas\texttt{\_}conf\texttt{\_}2013\texttt{\_}061} \cite{ATLAS-CONF-2013-061}, is furthermore able to exclude the rest of the 
$\tilde \tau$-LSP parameter space below around $M_{1/2}\simeq 760~$GeV.

Summarizing,  $\bar U \bar D \bar D$ couplings within the CMSSM are almost as well covered by LHC analyses as the RPC counterpart.
Similarly to the $LQ\bar D$ case, regions with low $M_0$ are harder to detect at the LHC than the RPC scenario. For large $M_0$ the 
searches for many jets and missing energy are very sensitive, leading to bounds as strong as in the RPC--CMSSM, while in the case of 
a $\lambda''_{3i3}$ coupling the bounds are even stricter. We stress again that these results are, in particular in the large $M_0$ region, 
specific to the CMSSM boundary conditions. For instance, if the stops were heavier than the gluinos, the bounds which one could set on 
the corresponding scenario would be considerably weaker \cite{Buckley:2016kvr}. In the stau LSP region, searches for several leptons 
provide the best constraints whereas for $\lambda''_{323}$, multi-$b$-jet analyses are even more sensitive.

Finally, a comment is in order. Much of the considered parameter space can be excluded or detected in the near future due to the decay products of intermediate top quarks in the final state. This is a consequence of the CMSSM boundary conditions where the stops often appear in either the production or decay channels. At the LHC, 
there are two methods to identify top quarks. The first method involves reconstructing the individual decay products of the top quark. The second, referred to as top-tagging, involves reconstructing the top-quark decay products inside a single fat-jet. This is possible by analysing the jet substructure if the top is boosted enough, and  tagging is in principle already possible if $p_{T,t} \gtrsim 150\,$GeV \cite{Anders:2013oga}. 
While top quarks can be produced directly from squark or gluino decays in $LQ\bar D$ and $\bar U \bar D \bar D$ scenarios, this does not happen in the considered scenarios because of the small couplings and the lighter neutralinos to which each coloured sparticle will decay first. Moreover, the lightest  neutralino decays to a top final states in $\lambda''_{3ij}$ scenarios, but boosted tops would require much heavier neutralinos than what is accessible in the near future (in a CMSSM context). Hence the only possibility for boosted tops is through the stop decays into $\tilde \chi^0_1 t$ which happens for $\tilde t_R$ mainly (while $\tilde t_L$ would decay to the wino first).
A naive estimate shows that $p_{T,t}$ of $\mathcal O(150\,{\rm GeV})$ is possible in all the remaining parameter space which has not yet been excluded in the figures above. Requiring, however, a significant boost of $\mathcal O(400\,{\rm GeV})$ for which the tagging efficiency is greatly improved \cite{Abdesselam:2010pt}, then this occurs only  in the upper 
part of the $M_0-M_{1/2}$ plane, as this requires a significant mass splitting between the stop, the top and the neutralino. 
The associated region is not yet accessible with stop pair production, in particular not at the 8\,TeV LHC,  since it requires stop masses beyond a TeV, featuring a production cross section of sub-fb. 
This region will, however, be accessible with more accumulated data at the 13\,TeV LHC.

\begin{figure*}[htbp]
\includetwinresults{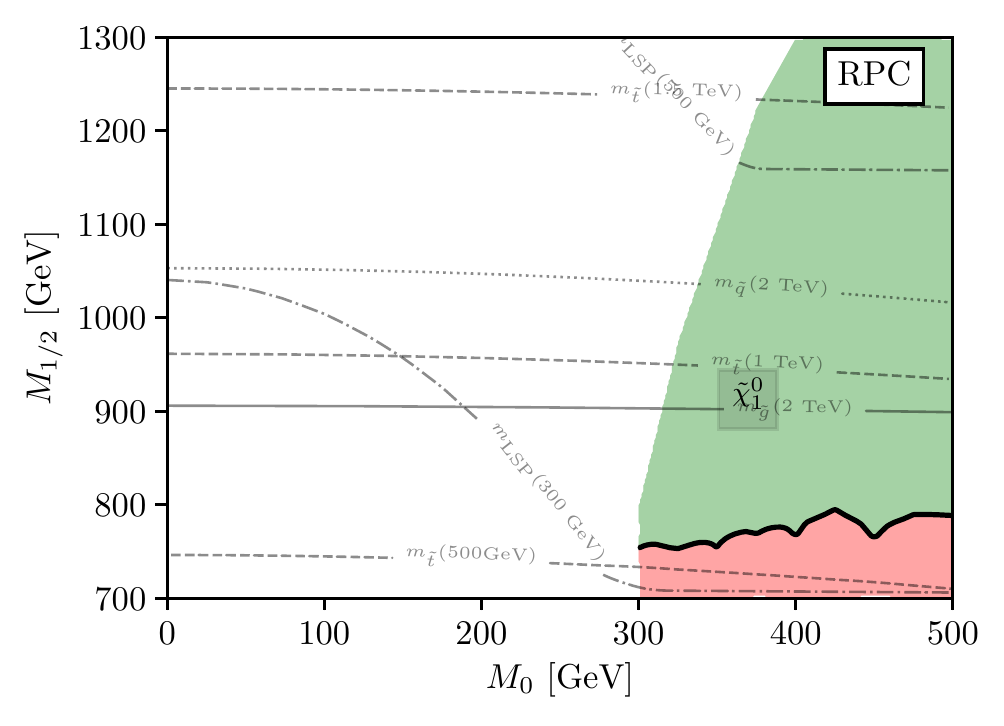}{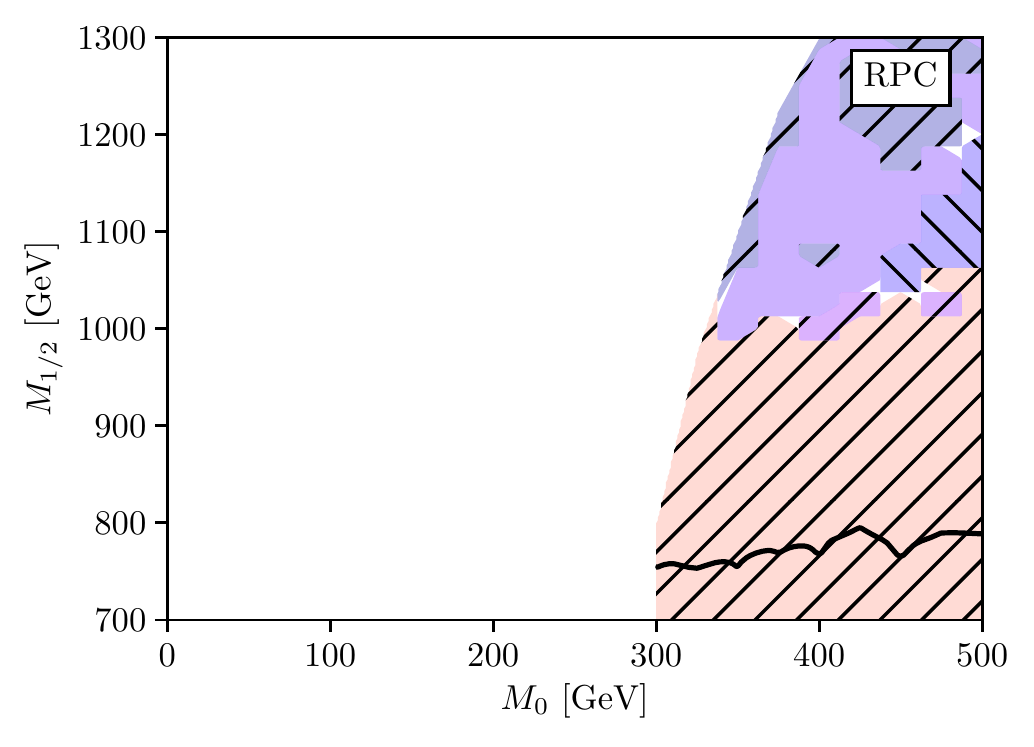}
\includetwinresults{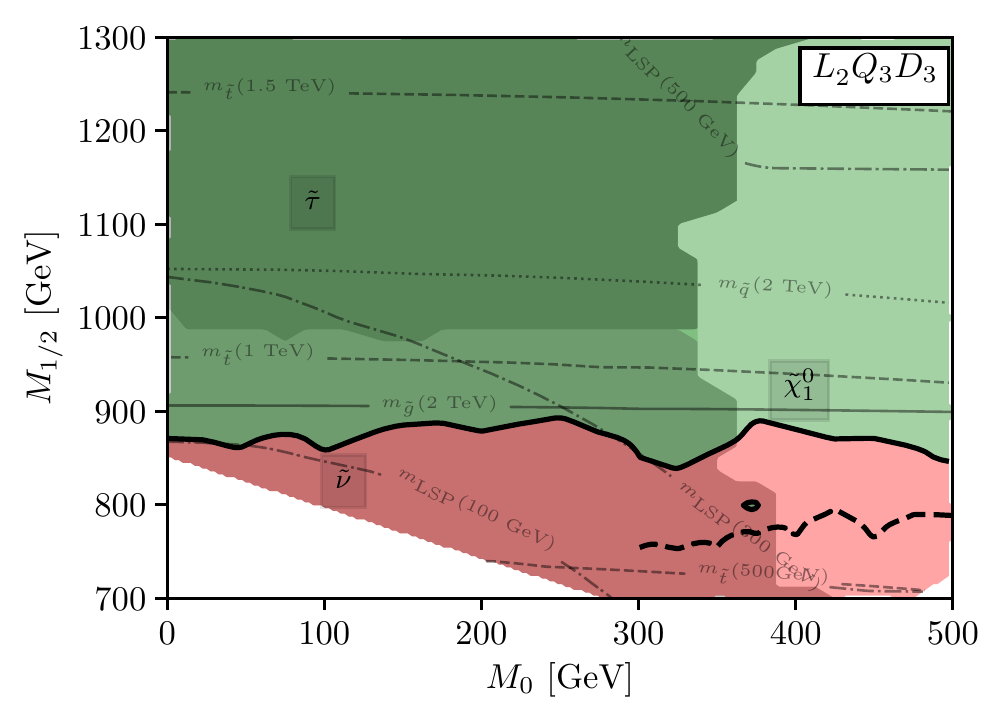}{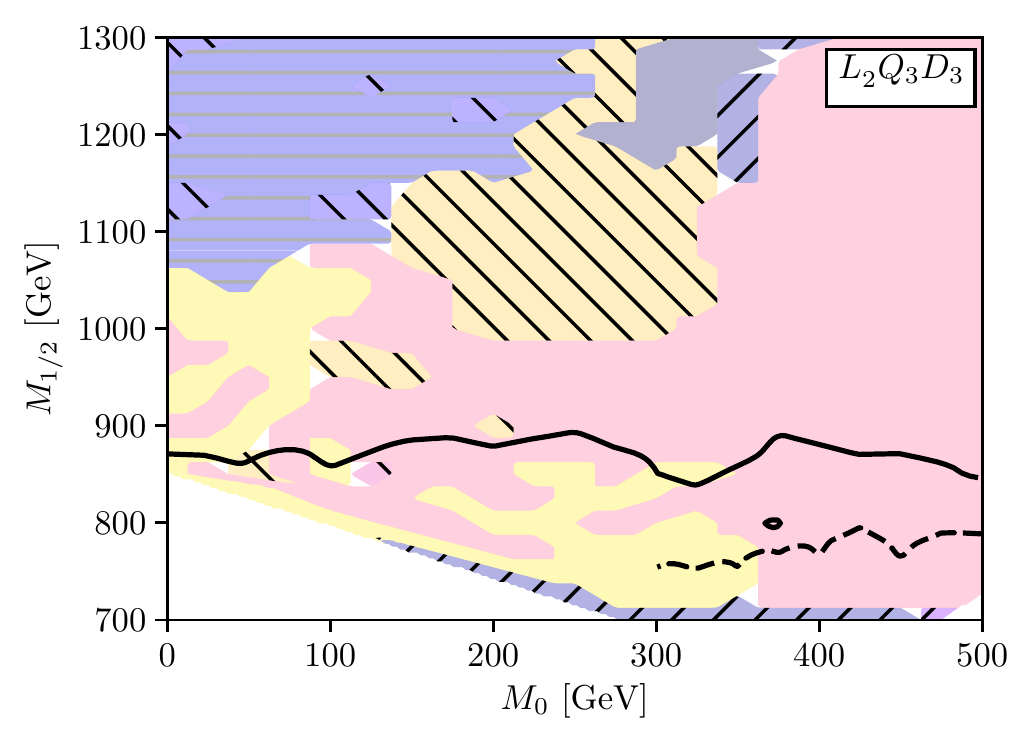}
\caption{Same as for \cref{fig:checkmate:LLE122and123}, but this time using the  parameter choices as in \cref{fig:sneutrino-LSP}, 
\textit{i.e.} with $A_0=-2800\,$GeV fixed, and $\lam'_{233}=0.08$. Note also the change in $M_0$ and $M_{1/2}$ parameter ranges 
compared to Figs.~\ref{fig:checkmate:RPC}-\ref{fig:checkmate:UDD121and323}. This makes the $\tilde\nu$-LSP region more readily visible. Contrarily to the previous figures, we also show the mass contours in the forbidden charged LSP region in the top left figure. This is however only for the purpose of improving the readability of the individual contour labels.}
\label{fig:checkmate:largeLQD}
\end{figure*}


\section{Large $\Lam_{\not R_p}$: other LSP scenarios}
\label{sec:checkmate:other}

Here, we briefly comment on scenarios with a large $\Lam_{\not R_p}$ coupling, as discussed in \cref{sec:nonneutralinoLSP}. 
In this case, the mass spectrum changes with respect to the RPC--CMSSM so that a direct comparison is no longer
meaningful. However, qualitative changes only occur if not only the spectrum but also the relative hierarchy of particle 
masses is altered. This can lead to (i)~changes in the final state signature and/or (ii)~changes in the kinematic distributions. 

A drastic example of case (i) is the squark LSP scenario, which we envisage for large $\bar U \bar D \bar D$ couplings, 
\textit{cf.} \cref{tab:LSPs}. Then, as discussed in \cref{sec:nonneutralinoLSP},  squark LSPs will be 
pair-produced and decay directly into pairs of dijet resonances via the (large) $\lam''$ coupling. Even though we operate 
within CMSSM boundary conditions here, the possible squark LSP scenarios correspond quite closely to the simplified 
models employed in the experimental analyses searching for this exact scenario, as the pair-production of the comparably 
light squarks will dominate over all other production modes. We therefore refer to the analyses summarized in 
\cref{sec:nonneutralinoLSP} for the respective bounds on the squark masses. 

We cannot perform a recast of the bounds on the squark LSP scenarios with the current version of \Checkmate{} since 
the respective analyses containing 4-jet final states of which two combine to a dijet resonance are not yet handled by 
this tool. We leave the inclusion of these results for future work.

In the case of a slepton or sneutrino LSP, the change in the final-state signature is milder as the pair-production of the LSP is 
suppressed with respect to squark/gluino and also wino production. There will therefore at most be changes in the intermediate 
cascade decays as well as the kinematics of the final state particles; we do not expect drastic changes. 

It is nevertheless instructive to check this statement with our tools at hand. In \cref{fig:checkmate:largeLQD} we show an 
example of parameter space with either neutralino or sneutrino LSP, corresponding to the scenario shown earlier in 
\cref{fig:sneutrino-LSP}. In the top plot on the left we have turned off the $R$-parity violating coupling, $\lam'_{233}=0$, 
and show only the $\tilde\chi^0_1$-LSP region, as is appropriate in the RPC--CMSSM. This agrees with 
\cref{fig:checkmate:RPC}. On the right we see that the most sensitive signature in 
the exclusion region, the lower right-hand corner of the plot, is the $0\ell$, 2-6$j$+$\ETmiss$ search of 
atlas\texttt{\_}1405\texttt{\_}7875 \cite{Aad:2014wea}.

In the lower two plots in \cref{fig:checkmate:largeLQD} we have $\lam'_{233}=0.08$ at $M_X$ (corresponding to 
$\lam'_{233}\simeq 0.19$ at the weak scale).  Comparing the lower left plot to the upper left plot, we have now 
included the $\tilde\tau$- and the $\tilde\nu$-LSP regions (lower left and upper left regions in the plots in the lower row). We see that the $\tilde\chi^0_1$-LSP region has slightly 
shrunk, compared to the RPC case in the upper left plot. This is the effect of the RPV coupling on the RGE running 
of the masses.

\begin{table*}
\renewcommand\arraystretch{1.40}
\begin{tabularx}{\linewidth}{l X X X X X | X X X X X X }
\toprule
                & \multicolumn{5}{c}  {$\tilde \chi^0_1$ LSP region } & \multicolumn{6}{|c}   {$\tilde \tau_1$ LSP region }\\ \midrule

Coupling        & $m_{\tilde g}$  & $m_{\tilde t_1}$ & $m_{\tilde q_{\rm 1st/2nd}}$ & $m_{\tilde \chi^0_1}$ & $m_{\tilde \chi^\pm_1}$ & 
 $m_{\tilde g}$        & $m_{\tilde t_1}$ & $m_{\tilde q_{\rm 1st/2nd}}$ & $m_{\tilde \chi^0_1}$ & $m_{\tilde \chi^\pm_1}$ & $m_{\tilde \tau_1}$ \\ \midrule
RPC  &   1280  &   710  &   1560  &   220  &   430  &   --  &   --  &   --  &   --  &   --  &   -- \\ \midrule
$\lam_{122}$  &   2070  &   1320  &   1960  &   400  &   750  &      1690  &   1140  &   1520  &   320  &   600  &   230 \\ 
$\lam_{123}$  &   1700  &   980  &   1630  &   310  &   600  &   1790  &   1220  &   1620  &   340  &   640  &   260 \\ 
$\lam_{131}$  &   1850  &   1120  &   1700  &   350  &   670  &  1740  &   1180  &   1580  &   330  &   620  &   260 \\ 
$\lam_{133}$  &   1590  &   920  &   1540  &   290  &   560  &   1690  &   1140  &   1520  &   320  &   600  &   230 \\ \midrule
$\lam'_{111}$  &   1220  &   700  &   1520  &   210  &   410  &      1690  &   1140  &   1520  &   320  &   600  &   230 \\ 
$\lam'_{113}$  &   1480  &   850  &   1530  &   260  &   510  &      1690  &   1140  &   1520  &   320  &   600  &   230 \\ 
$\lam'_{131}$  &   1310  &   750  &   1450  &   230  &   440  &   1690  &   1150  &   1520  &   320  &   600  &   220 \\ 
$\lam'_{133}$  &   1310  &   750  &   1470  &   220  &   440  &   1690  &   1140  &   1520  &   320  &   600  &   230 \\ 
$\lam'_{311}$  &   1250  &   750  &   1400  &   210  &   420  &   1530  &   1040  &   1360  &   280  &   530  &   190 \\ 
$\lam'_{313}$  &   1290  &   730  &   1410  &   220  &   440  &   1530  &   1040  &   1360  &   280  &   530  &   190 \\ 
$\lam'_{323}$  &   1280  &   720  &   1400  &   220  &   430  &   1530  &   1040  &   1370  &   280  &   540  &   200 \\ 
$\lam'_{331}$  &   1330  &   750  &   1440  &   230  &   450  &   1580  &   1080  &   1420  &   290  &   560  &   210 \\ 
$\lam'_{333}$  &   1350  &   770  &   1420  &   240  &   470  & 1620  &   1060  &   1460  &   310  &   600  &   240 \\ \midrule
$\lam''_{113}$  &   1250  &   720  &   1350  &   210  &   420  &  1420  &   970  &   1270  &   260  &   490  &   180 \\ 
$\lam''_{121}$  &   1260  &   730  &   1350  &   210  &   420  &  1480  &   1010  &   1330  &   270  &   520  &   200 \\
$\lam''_{312}$  &   1250  &   730  &   1350  &   210  &   420  &  1430  &   960  &   1290  &   260  &   500  &   180 \\ 
$\lam''_{323}$  &   1400  &   780  &   1350  &   250  &   480  &  1530  &   1040  &   1360  &   280  &   530  &   190 \\ \bottomrule
\end{tabularx}
\caption{Lower mass bounds on the particle spectrum in GeV. The mass bounds for each RPV coupling are obtained from the the 
most conservative points that appear in the $M_0$--$M_{1/2}$ planes of Figs.~\ref{fig:checkmate:RPC} - \ref{fig:checkmate:UDD121and323} 
from \cref{sec:checkmate-tests}. In the RPC case there are no $\tilde\tau$-LSP regions.
}
\label{tab:checkmateboundstable}
\end{table*}

In the RPV case, we now want to compare the bounds on $M_{1/2}$ in the sneutrino-LSP and the neutralino-LSP 
region, respectively. In both cases, the production of stop squarks dominates the LHC supersymmetric production 
cross section. Despite the large $\lam'_{233}$ coupling invoked, the stop mainly decays into a top quark and a bino. 
In the region on the right-hand side of the figure, the bino-LSP has a dominant three-body decay. While in the left-hand 
region, the bino is \textit{not} the LSP and it first decays into $\nu_\mu\tilde\nu_\mu$ or $\mu\tilde\mu$, with the nearly 
degenerate $\tilde\nu$ or $\tilde\mu$ on-shell. The latter then decay further via the $\lam'$ coupling. The dominant final 
state is therefore the same in the $\tilde\chi^0_1$-LSP and the $\tilde\nu$-LSP scenarios, and only the final state 
particles' kinematics differ. As expected, we thus observe that, except for a small dip in the cross-over region  ($|m_{\tilde
\chi^0_1}-m_{\tilde\nu_1}|$ is small) the bounds in the sneutrino-LSP and the neutralino-LSP region are comparable. 

Interestingly, in some of the sneutrino-LSP region, the cross section of the pair-production of sneutrinos and smuons is 
even comparable to the stop pair-production. However, due to the additional top quark in the final state, the latter 
provides a better discrimination against Standard Model background and sneutrino/smuon pair-production does not 
provide any mentionable constraints on the parameter space by itself.

From this, we conclude that even though the mass hierarchies of the lightest supersymmetric particles may be affected 
for larger RPV couplings, the resulting bounds are hardly dependent on the details of this hierarchy, as long as both LSP 
and NLSP are only electroweakly interacting. Small, fine tuned parameter regions with degenerate LSP-NLSP masses 
form a mild exception as here the soft decay kinematics of the NLSP-to-LSP decay weaken the resulting bounds.

\section{Absolute lower mass bounds on RPV-CMSSM scenarios}
\label{sec:checkmateboundstable}

In this last section, we present  a set of lower supersymmetric mass bounds within the CMSSM. These thus assume a complete 
supersymmetric model, with possibly involved cascade decay chains. This is unlike the experimental bounds in 
\cref{tab:best-bounds-lle,tab:best-bounds-lqd,tab:best-bounds-udd},
which are based on simplified models. 

The bounds here in \cref{tab:checkmateboundstable} are the result of the analyses of \cref{sec:checkmate-tests}, which 
lead to the Figs.~\ref{fig:checkmate:RPC} - \ref{fig:checkmate:UDD121and323}. For each case the allowed parameter range (green) 
is scanned and the lightest respective sparticle mass is determined. We list separately the bounds for the case of a neutralino LSP 
(left) and for a stau LSP (right). In both cases we give the lower mass bounds for: the gluino,  the lightest stop, the first/second 
generation squarks, the lightest neutralino and the lightest chargino. These are the particles which are also directly produced. 
Bounds on other particles also exist, but are always indirect, and obtained only through the CMSSM boundary conditions.  They inform
us about the $\LamRPV$--CMSSM, not necessarily the sensitivity of the LHC. For the stau LSP scenario we include the lower bound 
on the lightest stau, as this is an essential parameter of these models. We emphasize that all the bounds in the stau-LSP case are 
new, as such bounds do not yet exist in the literature.

Looking at the bounds more closely, for example in the upper left plot of Fig.~\ref{fig:checkmate:LLE122and123} for $\lam_{122}\not
=0$, we would expect the lightest allowed gluino mass in the $\tilde\chi^0_1$-LSP case to correspond to the dip in the exclusion 
curve near $(M_0,\,M_{1/2})\simeq(800\,\mathrm{GeV},\,920\,\mathrm{GeV})$. Looking at the light gray dot-dashed gluino mass 
iso-curve, we therefore expect a lower mass exclusion bound of just over 2000\,GeV. In \cref{tab:checkmateboundstable} in the
row for $\lam_{122}$, we see the lower gluino mass bound is indeed 2070\,GeV. Correspondingly in the lower left plot on the left in 
\cref{fig:checkmate:LLE122and123}, for $\lam_{123}\not=0$, the exclusion curve is significantly lower, about  three quarters of the way between 
the 1~TeV and 2~TeV gluino iso-mass curves. It also has no marked dip. The bound in \cref{tab:checkmateboundstable} is 
1700\,GeV. In these cases, the strong bounds on the gluino mass are caused by direct bounds on the electroweak 
gaugino sector through multi-lepton searches which translate into bounds on the gluino mass via the CMSSM boundary conditions. 
Since the cross section for gluino production at such high masses is of order $\mathcal{O}($ab$)$, direct measurements of 
gluino-induced topologies cannot provide competetive bounds.

Turning to the case $\lam'_{113}\not=0$ shown in \cref{fig:checkmate:LQD222}, we see the exclusion curve sloping 
downwards for large $M_0$, thus the bound is obtained at the limit of our scan region, $M_0=3000\,$GeV. This is similar to the 
extended simplified models considered in Ref.~\cite{CMS-PAS-SUS-12-027}, where both the squarks and the gluinos were 
kinematically accessible. The lower mass bounds on the gluino/squarks shown in the second to last row of the gluino section and the 
last row of the squark section in \cref{tab:best-bounds-lqd} strongly depend upon the chosen squark/gluino masses, \textit{i.e.} 
the gluino mass bound depends upon the assumed or allowed squark masses and vice versa. 

The remaining gluino mass bounds in the $\tilde\chi^0_1$-LSP scenario for $\lam'$ and $\lam''$ are all very similar, mainly 
around 1300\,GeV. They are determined by the limit of the scanning region and  rest on the production cross sections at 
$M_0=3000\,$ GeV. Contrarily to the above $LL\bar E$ discussion, the most sensitive signatures require the production of 
gluinos and hence set a comparably weaker bound. In all these cases a lighter gluino should be possible for completely 
decoupled squarks.

The lower mass bounds on the lightest top squark are typically significantly weaker, in the range of 700 to 800\,GeV over all couplings 
in the $\tilde\chi^0_1$-LSP scenario. This is similar to the RPC CMSSM bound of 710\,GeV. The exception are the cases $\lam_{122}\,
(\lam_{131})$ with $m_{\tilde t_1}> 1320$\,GeV (1120\,GeV). The reason is that, because of the employed relation $A_0=-2\,M_0$, 
the stop mass splitting increases with increasing $M_0$. Thus the lowest stop mass bounds come from the bounds at large $M_0$. 
As is seen in \cref{fig:checkmate:LLE122and123,fig:checkmate:LLE131and133}, the cases $\lam_{122,131}$ are the only ones where 
the lower bound on $M_{1/2}$ is as severe or even stricter for large $M_0$ as for lower $M_0$.
For the 1st/2nd generation squarks the lower mass bounds come from the low-$M_0$ region and
are consistently around 1400-1600\,GeV, as in the RPC case. They are only markedly stricter in the $LL\bar E$ scenarios.

For the stau LSP scenarios the lower gluino mass bound is typically obtained in the dip region along the neutralino-LSP--stau-LSPcross 
over or close by. For $\lam_{122}$ this is clearly significantly lower, for $\lam_{123,131,133}$ it is comparable to the neutralino-LSP 
case and for $\lam'_{111,113,131,133,311,313},\,\lam''_{121,113,312,323}$ the lower gluino mass bound in the stau-LSP case should be
considerably stricter than in the neutralino LSP case. This is confirmed  in \cref{tab:checkmateboundstable}. For the $LL\bar E$ cases 
these gluino mass lower bounds are due to electroweak gaugino production. For $LQ\bar D$ and $\bar U\bar D\bar D$ the
production process leading to the most sensitive limits is gluino and/or squark production, possibly involving top squarks. The other mass 
bounds are then determined indirectly via the CMSSM boundary conditions. Thus the gluino and lightest neutralino mass bounds can be 
roughly understood as the mass ratio $M_1/M_3\simeq 1/6$. The chargino mass bounds are also due to direct electroweak gaugino 
production for the $LL\bar E$ case but are otherwise also derived quantities in the RPV-CMSSM. The stop and lightest generation squark 
bounds are a mixture, sometimes derived, but sometimes also obtained via direct production. The stau mass bounds in the stau-LSP 
scenarios are all derived quantities. Overall the stau-LSP parameter range is very narrow and thus the lower mass bounds are very
similar across all couplings.

To emphasize these bounds are the result of using \Checkmate and therefore contain all the same deficiencies as discussed in 
\cref{sec:checkmate:montecarlo}. 
 Also, experimental bounds are typically interpreted within simplified
 models whereas CMSSM-based scenarios like ours have various potentially interesting decay signatures which appear simultaneously. Lacking a statistical combination of the numerous search channels then leads to a significant dilution of the bounds that can be derived compared to a single simplified model.

Despite being conservative our RPV-CMSSM gluino mass bounds are stricter in the $LL\bar E$ $\tilde\chi^0_1$-LSP case than in
\cref{tab:best-bounds-lle}. This is because they are in fact due to electroweak gaugino production, which can be reinterpreted within 
the $\LamRPV$-CMSSM.  In the $LQ\bar D$ case the bounds are largely similar, except for those from 
Ref.~\cite{Aaboud:2017dmy,Aaboud:2017faq} which are based on $\sqrt{s}=13\,$TeV data. In the $\bar U\bar D\bar D$ case the bounds
are also similar.

\section{Summary}
\label{sec:summary}
We have performed a systematic appraisal of the LHC coverage of $R$-parity violating supersymmetric models. We have 
mainly focused on the $\LamRPV$-CMSSM, with only a single non-zero RPV coupling at the unification scale. We have 
obtained the following results

\textbf{(1)}  In \cref{sec:rges}, starting from the small set of $\LamRPV$-CMSSM parameters at $M_X$ in 
Eq.~(\ref{eq:rpv-cmssm}), we have dynamically determined the possible LSPs at the weak scale, taking in particular the 
Higgs mass constraint into account. This is an update of Ref.~\cite{Dreiner:2008ca}. The results are presented in 
\cref{tab:LSPs}. We find an extensive parameter range with either a neutralino or a stau LSP. For special large RPV 
couplings we can also have one of $\{\tilde e_R,\,\tilde\mu_R,\,\tilde\nu_{e,\mu},\,\tilde s_R,\,\tilde d_R,\,\tilde b_1,\,\tilde 
t_1\}$ as the LSP.

\medskip

\textbf{(2a)} In \cref{sec:neutralinoLSP}, we focussed first on the $\tilde\chi^0_1$-LSP scenarios. For the various possible 
dominant operators, we have compiled tables detailing all possible LHC signatures: Tabs.~\ref{tab:lle-signatures} ($LL\bar E$), 
\ref{tab:lqd-signatures} ($LQ\bar D$), and \ref{tab:udd-signatures} ($\bar U\bar D\bar D$). We have then compiled {\it all} relevant 
LHC analyses by \Atlas and \Cms, and have presented the resulting bounds on the simplified supersymmetric mass spectra in 
Tabs.~\ref{tab:best-bounds-lle}~($LL\bar E$), \ref{tab:best-bounds-lqd} ($LQ\bar D$), and \ref{tab:best-bounds-udd} ($\bar U\bar 
D\bar D$), again depending on the nature of the dominant RPV operator. These bounds are  independent of the assumption of 
CMSSM-like boundary constraints and can thus be applied to all RPV models, provided the appropriate branching ratios are 
implemented.  Comparing the two sets of tables we can thus determine the coverage of these models at the LHC. We have 
observed the following: 

\begin{itemize}[leftmargin=*]
\item {$\tilde\chi^0_1$-LSP, $L L \bar E$:} These scenarios are very well covered via LHC analyses looking for 4 leptons 
(including a number of taus) plus missing transverse energy, \textit{cf.} Ref.~\cite{Aad:2014iza}. We also note that electroweak 
gaugino production can play an important role due to the large number of additional leptons in the finals states, significantly 
boosting the efficiencies. 

\item {$\tilde\chi^0_1$-LSP, $L Q \bar D$:} The typical signatures containing charged leptons with a number of jets, $b$-tagged 
or otherwise, are well covered. However, final states lacking charged leptons and instead containing hadronically decaying taus 
and or missing transverse energy are not completely covered, \textit{cf.} cases IIg-k in \cref{tab:lqd-signatures}. Most existing 
analyses focus on high jet multiplicity plus missing transverse energy which provide some sensitivity to the above scenarios. Note 
the recent analysis in Ref.~\cite{Aaboud:2017faq} tags an isolated electron or muon, further requiring high jet multiplicity 
with no veto on missing transverse energy. This search is sensitive to many of the RPV scenarios beyond just the $L Q \bar D$ 
operators.

\item {$\tilde\chi^0_1$-LSP, $U\bar D \bar D$:} Many scenarios are well covered, especially in the case that top quarks are 
produced in the cascade decay chain, yielding leptons in the final states. Searches for only jets with and without $b$-jets are in
principle sensitive to all possible final states. However  in Ref.~\cite{ATLAS-CONF-2016-057} all $\bar U\bar D \bar D$ couplings 
where simultaneously switched on. This makes it very difficult to reinterpret the particular analysis in comparison to single operator 
dominance. Finally based on the simplified analyses with single operator dominance, $\lam''_{312}$ coverage is lacking. 
\end{itemize}

\textbf{(2b)} In \cref{sec:stauLSP}, we next considered the stau-LSP scenarios in detail. The list of LHC signatures has been 
presented in Ref.~\cite{Desch:2010gi}. However, these models, irrespective of the dominant RPV coupling, are not explicitly 
searched for at the LHC. The only exception is a CMSSM model with non-zero $\lam_{121}$ \cite{ATLAS:2012kr}. This analysis 
only uses $\sqrt{s}=7\,$TeV data and makes a number of assumptions about the four-body decay channels, completely ignoring 
two-body decay channels, which become relevant for large $\tan\beta$. We thus do not present a list of lower mass bounds on the 
supersymmetric particles in these scenarios. We are however encouraged by the recent search for multiple leptons with up to two 
hadronic tau candidates, motivated by electroweak gaugino production in $R$-parity conserving models \cite{CMS:2017fdz}. These 
signatures are also highly relevant for many stau-LSP scenarios.  We further encourage experimentalists to perform dedicated 
analyses looking for  final states with high tau, charged lepton and jet multiplicities, \textit{cf.} \cref{tab:stau-LSP-LHC}. 

\medskip

\textbf{(2c)} In \cref{sec:nonneutralinoLSP} we summarize the experimental LHC bounds on the non-standard LSP 
scenarios, \textit{i.e.} those listed in \cref{tab:LSPs}, which are obtained for large RPV couplings. The results are summarized 
in \cref{tab:Squark-LSP-searches,tab:LQ-searches}. There are typically no direct searches for these scenarios at
the LHC, except in the $\tilde t_1$-LSP scenario. However in most cases the $\tilde\chi^0_1$ is the NLSP and for example the 
gluino cascade decay proceeds through the same chain as in the corresponding RPV $\tilde\chi^0_1$-LSP case. The only 
difference is that the neutralino decay is now two-body instead of three-body. Thus the final state kinematic distributions should 
be slightly different. We expect this to only moderately affect the search sensitivities. In special cases entire decay modes can be 
kinematically blocked due to the heavy top quark for example, which in turn can affect bounds more significantly.

We have also collected a set of related searches which involve non-neutralino/stau LSPs, which do not arise in the $\LamRPV
$--CMSSM, but for which there are experimental searches. Here we have a sneutrino LSP, a squark LSP, and a gluino LSP.
These are compiled in \cref{tab:LQ-searches}. Here we have also included reinterpreted leptoquark searches.

\medskip

\textbf{(3)} In \cref{sec:checkmate-tests} we performed collider studies using the program \Checkmate to assess the 
coverage at the LHC of RPV models in comparison to the CMSSM with $R$-parity conserved. We consistently only use
analyses implemented in \Checkmate for the $\sqrt{s}=8\,$TeV data. We also only considered all supersymmetric production
cross sections at leading order for both the $R$-parity violating and the $R$-parity conserving case, multiplied with K-factors determined by \texttt{NLLFast} for strongly produced final states.. We found that the LHC 
constraints on RPV models are, for most regions of parameter space, at least comparable to the RPC case, while for the 
$LL\bar E$ operator the constraints are significantly stronger (\textit{cf.} \cref{fig:checkmate:LLE122and123}). The main 
caveats are $\bar U \bar D \bar D$ and $L Q \bar D$ operators with $M_0$ in the range 300 to 1000\,GeV, \textit{cf.} 
\cref{fig:checkmate:LQD222,fig:checkmate:LQD3ij,fig:checkmate:UDD121and323}. 

In the RPC case, the most 
sensitive analyses in these regions are searches looking for high-$p_T$ jets plus missing transverse momentum. Including 
these RPV operators decreases sensitivity through both the reduction of the missing transverse energy and the distribution of the 
jet $p_T$ over many jets. Therefore searches for many jets and a moderate amount of missing energy are usually most sensitive here, 
but not competitive to the RPC sensitivity. One should note that the analysis in Ref.~\cite{Aaboud:2017faq} is not currently available in \Checkmate. We 
expect this search to be far more sensitive in these parameter regions, especially as it does not trigger on missing 
transverse energy. 

The increased sensitivity outside of this $M_0$ range occurs as many RPV operators can lead to the production of particles in 
the final state which are not only easier to detect experimentally but can also lead to greatly reduced SM background 
contamination. 
For example $\bar U\bar D \bar D$ operators which involve couplings to the top squarks, $\lambda''_{3ij}$, 
can lead to final states with jets (including $b$-jets) and two like-sign leptons or even three or more leptons, which is what analyses like 
Ref.~\cite{Aad:2014pda} have been designed for. Because of these rather special lepton signatures, these final states can be better 
discriminated against the SM background compared to the typical signatures arising from the RPC-CMSSM in the large-$M_0$ region,
which are one lepton, $b$-jets and missing transverse momentum.

We stress that we have used all available \SI{8}{\TeV} analyses implemented in 
\Checkmate, however, this does not yet include a large number of \Atlas and \Cms analyses optimized for RPV models. We find 
that many of these missing analyses are essential when one considers RPV couplings which are large enough to directly affect the 
mass spectrum of the model. Since \Checkmate{} is currently restricted to cut-and-count based analyses, no resonance searches could be considered which might also be relevant for certain RPV scenarios, as discussed in the main text. 

\medskip

\textbf{(4)} In \cref{sec:checkmateboundstable} we have collected the resulting mass bounds of the RPV--CMSSM \Checkmate
analysis of \cref{sec:checkmate-tests}. We present explicit lower mass bounds for the gluino, the first/second generation squarks,
the lightest stop, the lightest neutralino and the lightest chargino. We present these separately for the case of a $\tilde\chi^0_1$-LSP
and for a $\tilde\tau$-LSP. We compare the bounds with the corresponding RPC-bounds obtained also with \Checkmate, as well as with
the experimental bounds collected here in \cref{sec:neutralinoLSP}. The lower mass bounds in the stau-LSP case are all new,
as both \Atlas and \Cms have not yet determined any lower mass bounds in this case.

\medskip

Overall $R$-parity violating models have been searched for, but we strongly encourage the experimental collaborations to increase
the effort to systematically cover all possible models.

\section*{Acknowledgments}
We thank Santiago Folgueras for interesting discussions about multi-lepton searches at 
\Cms, as well as Florian Staub for technical support with \texttt{SARAH}. One of us, H.K.D. thanks the organizers of the conference
`Is SUSY Alive and Well?' held at the IFT UA Madrid in Sept. 2016, which partially stimulated this work. We also thank Howie Haber, 
Steve Martin and Tim Stefaniak for discussions. D.D. acknowledges the support of the Collaborative Research Center SFB 676 
``Particles, Strings and the Early Universe'' of the DFG. M.E.K. thanks the DFG for financial support through the Research Unit 2239 
``New Physics at the LHC'' and the IFIC Valencia for hospitality while part of this work was completed. T.O. is supported by the 
SFB--Transregio TR33 ``The Dark Universe''. A.R. thanks the Cusanuswerk for funding, and Tel Aviv University and the Weizmann 
Institute for hospitality while part of this work was completed. H.K.D. thanks Nikhef and MITP Mainz for hospitality while part of this 
work was completed. 

\medskip

\appendix



\section{RpV Bounds}
\label{rpv-bounds}
Here we summarize the current status of the bounds on the $R$-parity violating trilinear Yukawa couplings. In 
Tab.~\ref{tab:bounds1} we present the constraints on single couplings. These results are adopted from \cite{Allanach:1999ic} \
and are based on indirect decays and perturbativity. 
\begin{table*}
\centering
\renewcommand\arraystretch{1.4}
\begin{tabular*}{\textwidth}{l @{\extracolsep{\fill}} ccccc}
\toprule
ijk &\(\lambda_{ijk}(M_W)\)&\(\lambda^\prime_{ijk}(M_W)\)&\(\lambda^{\prime\prime}_{ijk}(M_W)\)\\
 &&\\[-5mm]
\hline &&\\[-2mm]
111 & - & \(5.2\times 10^{-4}\times \left(\frac{m_{\tilde{e}}}{\SI{100}{GeV}}\right)^2\times\sqrt{\frac{m_{\tilde{\chi}^0}}{\SI{100}{GeV}}}\) & -\\
112 & - & \(0.021\times \frac{m_{\tilde{s}_{R}}}{\SI{100}{GeV}}\)& \(10^{-15}\times\left(\frac{m_{\tilde{q}}}{\tilde{\Lambda}\SI{}{GeV}}\right)^{5/2}\)\\
113 & - & \(0.021\times \frac{m_{\tilde{b}_{R}}}{\SI{100}{GeV}}\)& \(10^{-4}\)\\ &&\\[-2.5mm]
\hline &&\\[-2mm]
121 & \(0.049\times\frac{m_{\tilde{e}_R}}{\SI{100}{GeV}}\) & \(0.043\times \frac{m_{\tilde{d}_{R}}}{\SI{100}{GeV}}\)& \(10^{-15}\times\left(\frac{m_{\tilde{q}}}{\tilde{\Lambda}\SI{}{GeV}}\right)^{5/2}\)\\
122 & \(0.049\times\frac{m_{\tilde{\mu}_R}}{\SI{100}{GeV}}\) & \(0.043\times \frac{m_{\tilde{s}_{R}}}{\SI{100}{GeV}}\)& - \\
123 & \(0.049\times\frac{m_{\tilde{\tau}_R}}{\SI{100}{GeV}}\) & \(0.043\times \frac{m_{\tilde{b}_{R}}}{\SI{100}{GeV}}\)& \((1.23)\)\\
&&\\[-2.5mm]
\hline&&\\[-2mm]
131 \footnote{The constraint on $\lambda^\prime_{131}$ is at the $3\sigma$ level, since the data disagree with the standard model prediction.} & \(0.062\times\frac{m_{\tilde{e}_R}}{\SI{100}{GeV}}\) & \(0.019\times \frac{m_{\tilde{t}_{L}}}{\SI{100}{GeV}}\)& \(10^{-4}\)\\
132 & \(0.062\times\frac{m_{\tilde{\mu}_R}}{\SI{100}{GeV}}\) & \(0.28\times \frac{m_{\tilde{t}_{L}}}{\SI{100}{GeV}}\ (1.04)\)& \((1.23)\)\\
133 & \(0.0060\times\sqrt{\frac{m_{\tilde{\tau}}}{\SI{100}{GeV}}}\) & \(0.0014\times\sqrt{ \frac{m_{\tilde{b}}}{\SI{100}{GeV}}}\)& - \\
&&\\[-2.5mm]
\hline &&\\[-2mm]
211 & \(0.049\times\frac{m_{\tilde{e}_R}}{\SI{100}{GeV}}\) & \(0.059\times \frac{m_{\tilde{d}_{R}}}{\SI{100}{GeV}}\)& - \\
212 & \(0.049\times\frac{m_{\tilde{\mu}_R}}{\SI{100}{GeV}}\) & \(0.059\times \frac{m_{\tilde{s}_{R}}}{\SI{100}{GeV}}\)& (1.23) \\
213 & \(0.049\times\frac{m_{\tilde{\tau}_R}}{\SI{100}{GeV}}\) & \(0.059\times \frac{m_{\tilde{b}_{R}}}{\SI{100}{GeV}}\)& (1.23) \\
&&\\[-2.5mm]
\hline &&\\[-2mm]
221 & - & \(0.18\times \frac{m_{\tilde{s}_{R}}}{\SI{100}{GeV}}\ (1.12)\)& (1.23) \\
222 & - & \(0.21\times \frac{m_{\tilde{s}_{R}}}{\SI{100}{GeV}}\ (1.12)\)& - \\
223 & - & \(0.21\times \frac{m_{\tilde{b}_{R}}}{\SI{100}{GeV}}\ (1.12)\)& (1.23) \\ &&\\[-2.5mm]
\hline &&\\[-2mm]
231 & \(0.070\times\frac{m_{\tilde{e}_R}}{\SI{100}{GeV}}\) & \(0.18\times \frac{m_{\tilde{b_L}}}{\SI{100}{GeV}}\)& (1.23) \\
232 & \(0.070\times\frac{m_{\tilde{\mu}_R}}{\SI{100}{GeV}}\) & \(0.56\ (1.04)\)& (1.23) \\
233 & \(0.070\times\frac{m_{\tilde{\tau}_R}}{\SI{100}{GeV}}\) & \(0.15\times\sqrt{ \frac{m_{\tilde{b}}}{\SI{100}{GeV}}}\)& - \\
&&\\[-2.5mm]
\hline &&\\[-2mm]
311 & \(0.062\times\frac{m_{\tilde{e}_R}}{\SI{100}{GeV}}\) & \(0.11\times \frac{m_{\tilde{d}_{R}}}{\SI{100}{GeV}}\ (1.12)\)& - \\
312 & \(0.062\times\frac{m_{\tilde{\mu}_R}}{\SI{100}{GeV}}\) & \(0.11\times \frac{m_{\tilde{s}_{R}}}{\SI{100}{GeV}}\ (1.12)\)& 0.50\ (1.00) \\
313 & \(0.0060\times\sqrt{\frac{m_{\tilde{\tau}}}{\SI{100}{GeV}}}\) & \(0.11\times \frac{m_{\tilde{b}_{R}}}{\SI{100}{GeV}}\ (1.12)\)& 0.50\ (1.00) \\ &&\\[-2.5mm]
\hline &&\\[-2mm]
321 & \(0.070\times\frac{m_{\tilde{e}_R}}{\SI{100}{GeV}}\) & \(0.52\times \frac{m_{\tilde{d}_{R}}}{\SI{100}{GeV}}\ (1.12)\)& 0.50\ (1.00) \\
322 & \(0.070\times\frac{m_{\tilde{\mu}_R}}{\SI{100}{GeV}}\) & \(0.52\times \frac{m_{\tilde{s}_{R}}}{\SI{100}{GeV}}\ (1.12)\)& - \\
321 & \(0.070\times\frac{m_{\tilde{\tau}_R}}{\SI{100}{GeV}}\) & \(0.52\times \frac{m_{\tilde{b}_{R}}}{\SI{100}{GeV}}\ (1.12)\)& 0.50\ (1.00) \\ &&\\[-2.5mm]
\hline &&\\[-2mm]
331 & - & \(0.45\ (1.04)\)& 0.50\ (1.00) \\
332 & - & \(0.45\ (1.04)\)& 0.50\ (1.00) \\
333 & - & \(0.45\ (1.04)\)& - \\
\bottomrule
\end{tabular*}
\caption{Upper bounds on the magnitude of $R$-partiy violating couplings at the $2\sigma$ confidence level, taken from 
\cite{Allanach:1999ic}. The constraints arise from indirect decays. The concrete processes are described in detail in the 
original paper. Additionally the perturbativity constraints are shown in parentheses in case they are more stringent than
for $m_{\tilde{q},\tilde{l}}=\SI{1}{TeV}$. The numbers where no mass-dependence is specified where derived assuming 
a degenerate mass spectrum of $\SI{100}{GeV}$. The constraints on $\lambda^{\prime\prime}_{112,121}$ are derived 
from double nucleon decay and depend on a hadronic scale $\tilde{\Lambda}$ that can be varied from $0.003$ to $1$.}
\label{tab:bounds1}
\end{table*}

\bibliography{RpV}

\end{document}